\documentclass[aps,twocolumn,superscriptaddress]{revtex4-1}
\usepackage{amssymb, amsmath, bm}
\usepackage{graphicx,onlyamsmath}

\usepackage{natbib}
\usepackage{xcolor}
\usepackage[normalem]{ulem}  

\begin{document}

\title{Relativistic collapse of Landau levels of Kane fermions in crossed electric and magnetic fields}

\author{Sergey S.~Krishtopenko}
\email[]{sergey.krishtopenko@gmail.com}
\affiliation{CENTERA Laboratories, Institute of High Pressure Physics, Polish Academy of Sciences, PL-01-142 Warsaw, Poland}

\author{Fr\'{e}d\'{e}ric Teppe}
\affiliation{CENTERA Laboratories, Institute of High Pressure Physics, Polish Academy of Sciences, PL-01-142 Warsaw, Poland}
\affiliation{Laboratoire Charles Coulomb (L2C), UMR 5221 CNRS-Universit\'{e} de Montpellier, F- 34095 Montpellier, France}
\date{\today}

\begin{abstract}
Using an elegant model involving only $\Gamma_{6c}$ and $\Gamma_{8v}$ bands, massless Kane fermions were defined as the particles associated with the peculiar band structure of gapless HgCdTe crystals. Although their dispersion relation resembles that of a pseudo-spin-1 Dirac semimetal, these particles were originally considered to be hybrids of pseudospin-1 and -1/2 fermions. Here we unequivocally find that by considering an additional $\Gamma_{7c}$ conduction band inherent in HgCdTe crystals, the Kane fermions are ultimately two nested Dirac particles. This observation allows the direct application of Lorentz transformations to describe the relativistic behavior of these particles in crossed electric and magnetic fields. By studying the relativistic collapse of their Landau levels at different orientations between the crossed fields and the main crystallographic axes, we demonstrate that the Kane fermions strikingly decay into two independent Dirac particles with increasing of electric field. Our results provide new insight into semi-relativistic effects in narrow-gap semiconductors in crossed electric and magnetic fields.
\end{abstract}

\pacs{73.21.Fg, 73.43.Lp, 73.61.Ey, 75.30.Ds, 75.70.Tj, 76.60.-k} 
\keywords{}
\maketitle

\section{\label{Sec:Int} Introduction}
According to the solutions of the Dirac equation, essential in high energy physics, relativistic particles are classified into Dirac, Majorana or Weyl fermions.
In solids, the interaction of electrons with the periodic potential of the crystal lattice
can give rise to dispersion relations at high symmetry points of the Brillouin zone, which mimic the relativistic particles.

This relativistic analogy is usually used for materials,
in which conduction and valence bands are well separated from the other bands, considered as the remote bands. In the early 1960s, Keldysh~\cite{Flds1} and Wolff~\cite{Flds2} independently discovered that the two-band description of the bismuth, PbTe, PbSe and PbS-type semiconductors was equivalent to the Dirac equation for free particles. If the band-gap vanishes, the band structure mimics either by massless Dirac fermions or Weyl fermions. Solid state analogues of massless relativistic particles were demonstrated in graphene~\cite{Flds6}, semiconductor quantum wells~\cite{Flds7,Flds8,Flds9b,Flds9,Flds10}, topological insulators~\cite{Flds11,Flds12,Flds13}, Dirac~\cite{Flds14,Flds15,Flds16,Flds17} and Weyl~\cite{Flds18,Flds19,Flds20} semimetals.

One of the relativistic properties of spin-1/2 fermions, which can also be probed in ``non-relativistic'' solids, arises in the presence of perpendicular electric and magnetic fields. Aronov and Pikus~\cite{Flds3,Flds4} on the one hand and Zawadzki and Lax~\cite{Flds5,Flds5b} on the other hand utilized this relativistic analogy to describe the inter-Landau level transitions within a two-band approximation in crossed electric $\mathcal{E}$ and magnetic $\mathcal{B}$ fields. Particularly, Aronov and Pikus~\cite{Flds3,Flds4} were the first who proposed to apply the Lorentz transformation with ``an effective speed of the light'' to eliminate the magnetic field when $\mathcal{E}>\mathcal{B}$ in the moving coordinate system, and the electric field when $\mathcal{E}<\mathcal{B}$. In the former case, the particle motion is infinite and no quantization takes place, while in the latter case the Landau quantization remains until its collapse when the drift velocity $V_d=c\mathcal{E}/\mathcal{B}$ (where $c$ is the speed of light in vacuum) reaches the ``effective speed of the light'' $\tilde{c}$ in the system. Forty years later, similar results were obtained for graphene~\cite{Flds25}. Note that the ``effective speed of the light'' in solids represents the maximum particle velocity in the system -- for instance, the Fermi velocity in graphene~\cite{Flds6}.

A fundamental difference between electrons in solids and those at high energy is that
Lorentz invariance is not always preserved in condensed matter physics.
By generalizing the Dirac equation, one can find many other free fermionic excitations that have no high-energy analogs~\cite{Flds21,Flds22,Flds23,Flds24}. Massless ``pseudo-spin-1 and -3/2'' Dirac and Weyl fermions are a remarkable example of such particles, whose low-energy Hamiltonian has the form of $\mathbf{k}\cdot\mathbf{S}$, where $\mathbf{S}$ is the vector of spin-1 or -3/2 matrices.

Massless Kane fermions, observed in gapless HgCdTe crystals~\cite{s1,s2}, represent another type of fermionic excitations formed by the crossing of three bands. Its unique band structure is characterized by doubly degenerate conical bands intersected at the vertex by an additional flat band, closely resembling a pseudo-spin-1 Dirac semimetal~\cite{Flds24}. However, unlike the pseudo-spin-1 Dirac fermions, Kane fermions are not protected by symmetry or topology, and their band structure can be set at will~\cite{s2}. The discussion of the nature of Kane fermions has become more confusing after Malcolm and Nicol~\cite{Flds26}, who noted that the three-band Kane Hamiltonian involving the $\Gamma_{8v}$ and $\Gamma_{6c}$ bands maps onto an intermediate value ($\alpha=1/\sqrt{3}$) of the $\alpha$-$\mathcal{T}_3$ model~\cite{Flds27}. Since the $\alpha$-$\mathcal{T}_3$ model interpolates between the spin-1/2 (graphene, $\alpha=0$) and pseudo-spin-1 (dice or $\mathcal{T}_3$ lattice, $\alpha=1$) Dirac-Weyl systems, it was concluded that the Kane fermions are hybrids of pseudo-spin-1 and -1/2 Dirac fermions.

In this work, by taking into account an additional conduction $\Gamma_{7c}$ band resulting in a finite curvature of the heavy-hole band in HgCdTe crystals, we unequivocally identify the Kane fermions as spin-1/2 particles. Furthermore, using this extended (3+1)-band model, we show that the Kane fermion is composed of two \emph{mutually hybridized} Dirac particles. This representation makes it possible to directly apply the Lorentz transformation for the calculation of the Landau levels in crossed electric and magnetic fields. We have found that increasing of the electric field first leads to the decay of the Kane fermion into two independent Dirac particles. Then, the Landau levels of the Dirac particles collapse when their drift velocities $V_d$ reach the value $V_d^{*}$ lying between \emph{half} and \emph{full} value of the ``effective speed of the light''. This is a distinctive feature of Kane fermions, since the Landau levels of conventional Dirac fermions normally collapse as $V_d$ approaches the "effective speed of light"~\cite{Flds3,Flds4,Flds5,Flds5b,Flds25}. In addition, this limiting speed depends on the orientation of the electric and magnetic fields with respect to the main crystallographic axes of the zinc-blende crystals.

The paper is organized as follows. In Sec.~\ref{Sec:Problem}, we briefly review the existing three-band description of Kane fermions in crossed electric and magnetic fields. In Sec.~\ref{Sec:Lorentz}, we describe the main consequences of taking into account the additional $|\Gamma_{7c},{\pm}1/2\rangle$ band. Subsequently, we consider in detail the application of the Lorentz transformation allowing to eliminate the electric field in the moving reference frame for two orientations of the magnetic field ($\mathcal{B}\parallel[001]$ and $\mathcal{B}\parallel[010]$) resulting in various conditions of Landau level collapse.
Section~\ref{Sec:RandD} provides the comparison between Dirac and Kane fermions in crossed electric and magnetic fields and a discussion on the Landau level collapse in HgCdTe bulk crystals with inverted and non-inverted band structure. Finally, the main results are summarized in Sec.~\ref{Sec:Sum}.

\section{\label{Sec:Problem} Brief overview of the problem}
Let us first briefly overview the three-band description of Kane fermions~\cite{s1,s2}, which until now has allowed them to be considered as hybrids of pseudospin-1 and -1/2 fermions. Up to first-order in \textbf{k$\cdot$p} theory, the three-band Hamiltonian involving the $\Gamma_{6c}$ and $\Gamma_{8v}$ bands can be presented in the form~\cite{q6}:
\begin{equation}
\label{Keq:1}
\mathcal{\hat{H}}_{6\times6}(\mathbf{k},\theta)=\begin{pmatrix}
\hat{\mathcal{H}}_{0}(k_x,k_y,\theta) & \hat{\mathcal{H}}_{z}(k_z) \\ \hat{\mathcal{H}}_{z}(k_z)^{\dag} & \hat{\mathcal{H}}_{0}^{*}(-k_x,-k_y,\theta)\end{pmatrix},
\end{equation}
where the asterisk stands for complex conjugation and ``$\dag$'' corresponds to Hermitian conjugation. The $3\times3$ blocks $\hat{H}_{0}({k}_x,{k}_y,{k}_z)$ and $\hat{H}_{z}({k}_z)$ in Eq.~(\ref{Keq:1}) are written as
\begin{equation}
\label{Keq:2}
\hat{H}_{0}=\begin{pmatrix}
C_{0}+M_{0} & V{k}_{+}\sin\theta & V{k}_{-}\cos\theta \\
V{k}_{-}\sin\theta & C_{0}-M_{0} & 0\\
V{k}_{+}\cos\theta & 0 & C_{0}-M_{0}\end{pmatrix}
\end{equation}
and
\begin{equation}
\label{Keq:3}
\hat{H}_{z}=\begin{pmatrix}
0 & 0 & V{k}_{z} \\
0 & 0 & 0\\
V{k}_{z} & 0 & 0\end{pmatrix},
\end{equation}
where $k_{\pm}=k_x\pm i k_y$ with $k_x$, $k_y$, $k_z$ being momentum operators, $C_0$ is an energy reference; $V/\hbar$ is the ``effective speed of the light'' $\tilde{c}$ related with the momentum matrix element $P$ between the $\Gamma_{6c}$ and $\Gamma_{8v}$ bands~\cite{s1}; and $M_0$ is the mass parameter describing the ordering of $\Gamma_{6c}$ and $\Gamma_{8v}$ bands~\cite{s2}. In Eq.~(\ref{Keq:2}), we have additionally introduced a ``hybridization angle'' $\theta$ describing the mixing between light- and heavy-holes at finite quasimomentum~\cite{q6}. For the Kane Fermions, $\theta=\pi/3$. Since $\mathcal{\hat{H}}_{6\times6}(\mathbf{k},\theta)$, $\mathcal{\hat{H}}_{6\times6}(\mathbf{k},-\theta)$ and $\mathcal{\hat{H}}_{6\times6}(\mathbf{k},\pi/2\pm\theta)$ in the absence of magnetic field are all related by unitary transformation, one can only consider the $\theta$ values within the range $[0,\pi/2)$.

As first noted by Malcolm and Nicol~\cite{Flds26}, the Hamiltonian in Eq.~(\ref{Keq:1}) at $k_z=0$ and $M_0=0$ is nothing but two copies of the $\alpha$-$\mathcal{T}_3$ model~\cite{Flds27} interpolating between the spin-1/2 and pseudo-spin-1 Dirac-Weyl systems. In our notation, $\mathcal{\hat{H}}_{6\times6}(\mathbf{k},\theta)$ describes the physics of graphene (dice lattice) at $\theta=0$ ($\theta=\pi/4$), where two diagonal blocks pertain to the distinct chiral centers $K$ and $K'$, respectively~\cite{Flds27}. In the 3D case, non-zero $k_z$ takes the role of connecting two otherwise independent chiral blocks.

An inherent feature of $\mathcal{\hat{H}}_{6\times6}(\mathbf{k},\theta)$ is the independence of its eigenvalues from $\theta$, that makes the Kane fermion energy dispersion ($\theta=\pi/3$) the same as the one of Dirac fermion ($\theta=0$) with independent flat band:
\begin{equation}
\label{Keq:4}
E_{\pm}(\mathbf{k})=C_0\pm\sqrt{M_0^2+V^2k^2},~~~~~E_0=C_0-M_0,
\end{equation}
where $k^2=k_x^2+k_y^2+k_z^2$. However, the difference between Kane and Dirac fermions arises in the presence of a magnetic field. Assuming the orientation of magnetic field $\mathcal{B}$ in the $z$ direction, the Landau level energies are written in the form
\begin{eqnarray}
\label{Keq:5}
E^{(\sigma)}_{\pm}=C_0\pm\sqrt{M_0^2+V^2k_z^2+\dfrac{2V^{2}}{a_B^2}(n+1+\sigma\sin^2\theta)},\nonumber\\
E_0=C_0-M_0,~~~~~~~~~~~~~~~~~~~~~~
\end{eqnarray}
where $n\geq{-1}$ is the Landau level index, $\sigma=-1$ for $n\geq{0}$, while $\sigma=+1$ for $n\geq{-1}$, and $a_B$ is the magnetic length ($a_B^2=c\hbar/e\mathcal{B}$, $c$ is the speed of light in vacuum). As seen from Eq.~(\ref{Keq:5}), Landau levels of Dirac fermion are doubly degenerate at $n\geq{0}$, while the degeneracy is lifted for the Kane fermion case.

The first attempt to solve the problem of particle motion in crossed fields within the three-band approximation was made by Zawadzki~\emph{et~al.}~\cite{Flds29,Flds30} in the context of InSb semiconductor ($M_0>0$). In their work, \emph{neglecting} the electric-field-induced inter-band terms arising for the squared Hamiltonian $\mathcal{\hat{H}}_{6\times6}(\mathbf{k},\theta)$, the Landau level energies for the $\Gamma_6$ band were written as
\begin{multline}
\label{Keq:6}
E_{\Gamma_6}=C_0+\hbar{V_d}k_y+\sqrt{1-\delta^2}\biggl[M_0^2+V^2k_z^2+\\
+\sqrt{1-\delta^2}\dfrac{2V^{2}}{a_B^2}\left(n'+\frac{1}{2}+\frac{\sigma'}{4}\right)\biggr]^{1/2},
\end{multline}
where $V_d$ is a drift velocity ($V_d=c\mathcal{E}/\mathcal{B}$), $n'\geq{0}$ is the Landau level index, $\sigma'={\pm}1$ for all values of $n'$. In order to obtain this expression in our notations from those of Ref.~\cite{Flds29}, one has to use the following expressions for the band gap $\epsilon_g$, the effective mass $m_0^{*}$ and the spin-splitting factor $g_0^{*}$ at the band edge~\cite{Flds30}:
\begin{equation}
\label{Keq:7}
\epsilon_g=2M_0,~~~~~m_0^{*}=\dfrac{\hbar^{2}M_0}{V^2},~~~~~g_0^{*}=-\dfrac{m_0}{m_0^{*}},
\end{equation}
where $m_0$ is the free-electron mass.

The most important quantity in Eq.~(\ref{Keq:6}) is $\delta$ defined as $\delta={\hbar}V_d/V=(c\hbar\mathcal{E})/(V\mathcal{B})$. In deriving Eq.~(\ref{Keq:6}), it was assumed that the electric field $\mathcal{E}$ is oriented along the $x$ axis. As seen, the results of Zawadzki~\emph{et~al.}~\cite{Flds29,Flds30} are very similar to the case of Dirac fermions~\cite{Flds3,Flds4,Flds5,Flds5b,Flds25} -- the Landau levels collapses when $V_d$ approaches the ``effective speed of the light'', $\tilde{c}=V/\hbar$. By redefining $n'$ and $\sigma'$, $E_{\Gamma_6}$ in Eq.~(\ref{Keq:6}) in the absence of electric field is reduced to $E^{(\sigma)}_{+}$ in Eq.~(\ref{Keq:5}) at $\theta=\pi/3$.

Let us now show that the assumptions made by Zawadzki~\emph{et~al.} in the derivation of Eq.~(\ref{Keq:6}) are equivalent to the semiclassical quantization rule for $E_{\pm}(\mathbf{k})$ in Eq.~(\ref{Keq:4}). According to the generalization of Lifshitz-Onzager quantization rule made by Lifshitz and Kaganov~\cite{Flds31}, semiclassical Landau levels in crossed electric and magnetic fields can be found from the relation:
\begin{equation}
\label{Keq:8}
\mathcal{S}(\varepsilon^{*},k_z)=\dfrac{2\pi}{a_B^2}(n+\gamma),
\end{equation}
where $\gamma$ is constant and $\mathcal{S}(\varepsilon^{*},k_z)$ is the cross-section  of the surface $E(\mathbf{k})-{\hbar}V_{d}k_y=\varepsilon^{*}$ in the momentum space. This generalization is due to the fact that in the crossed fields, instead of the energy, the quantity $E(\mathbf{k})-{\hbar}V_{d}k_y=$ is conserved. The latter is nothing but an ellipse in the momentum space
\begin{multline}
\label{Keq:9}
V^2k_x^2+\left(V^2-\hbar^2V_d^2\right)\left[k_y-\dfrac{{\hbar}V_d}{V^2-\hbar^2V_d^2}\left(\varepsilon^{*}-C_0\right)\right]^2=\\
=\dfrac{V^2}{V^2-\hbar^2V_d^2}\left(\varepsilon^{*}-C_0\right)^2-M^2-V^2k_z^2.
\end{multline}
By calculation the ellipse cross-section, the energies of the semiclassical Landau levels are written as
\begin{multline}
\label{Keq:10}
E_{\pm}=C_0+\hbar{V_d}k_y\pm\sqrt{1-\delta^2}\biggl[M_0^2+V^2k_z^2+\\
+\sqrt{1-\delta^2}\dfrac{2V^{2}}{a_B^2}\left(n+\gamma\right)\biggr]^{1/2},
\end{multline}
where $\gamma$ can be defined from the comparison of Eq.~(\ref{Keq:10}) at $\delta=0$ with Eq.~(\ref{Keq:5}). The latter makes $E_{+}$ to be equivalent to $E_{\Gamma_6}$ in Eq.~(\ref{Keq:6}).

The main problem of the three-band approximation in crossed electric and magnetic fields manifests itself when one cannot neglect the inter-band coupling. For the $\Gamma_{6c}$ band, this arises when $M_0^2$ becomes comparable with $\beta{\hbar}^2V^2\cos\theta/a_B^2$ and $\beta{\hbar}^2V^2\sin\theta/a_B^2$, which represent the coupling strength with the light- and heavy-hole bands, respectively. On the contrary, for the $\Gamma_{8v}$ band, the inter-band coupling is always relevant due to the proximity of the light- and heavy-hole bands. Thus, the impossibility of neglecting the inter-band coupling in materials such as HgCdTe makes it impossible to clearly understand the behavior of Landau levels at relatively high electric fields. The problem is however less challenging beyond the three-band approximation, in which the heavy hole band is flat.

\section{\label{Sec:Lorentz} Lorentz boost in (3+1)-band model}
To overcome the problems resulting from the three-band approximation, we propose to consider an additional conduction band, whose \textbf{k$\cdot$p} interaction induces the non-zero curvature of the heavy hole band. For the zinc-blende materials, this is the second conduction $|\Gamma_{7c},{\pm}1/2\rangle$ band~\cite{q1,q2a,q2}. Further, we briefly outline the main consequences of taking into account the $|\Gamma_{7c},{\pm}1/2\rangle$ band, while a more rigorous derivation of their proper (3+1)-band Hamiltonian from the 14$\times$14 \textbf{k$\cdot$p} model~\cite{q1,q2a,q2} of cubic semiconductors can be found in Appendix~\ref{sec:A2}.

\begin{figure}
\includegraphics [width=1.0\columnwidth, keepaspectratio] {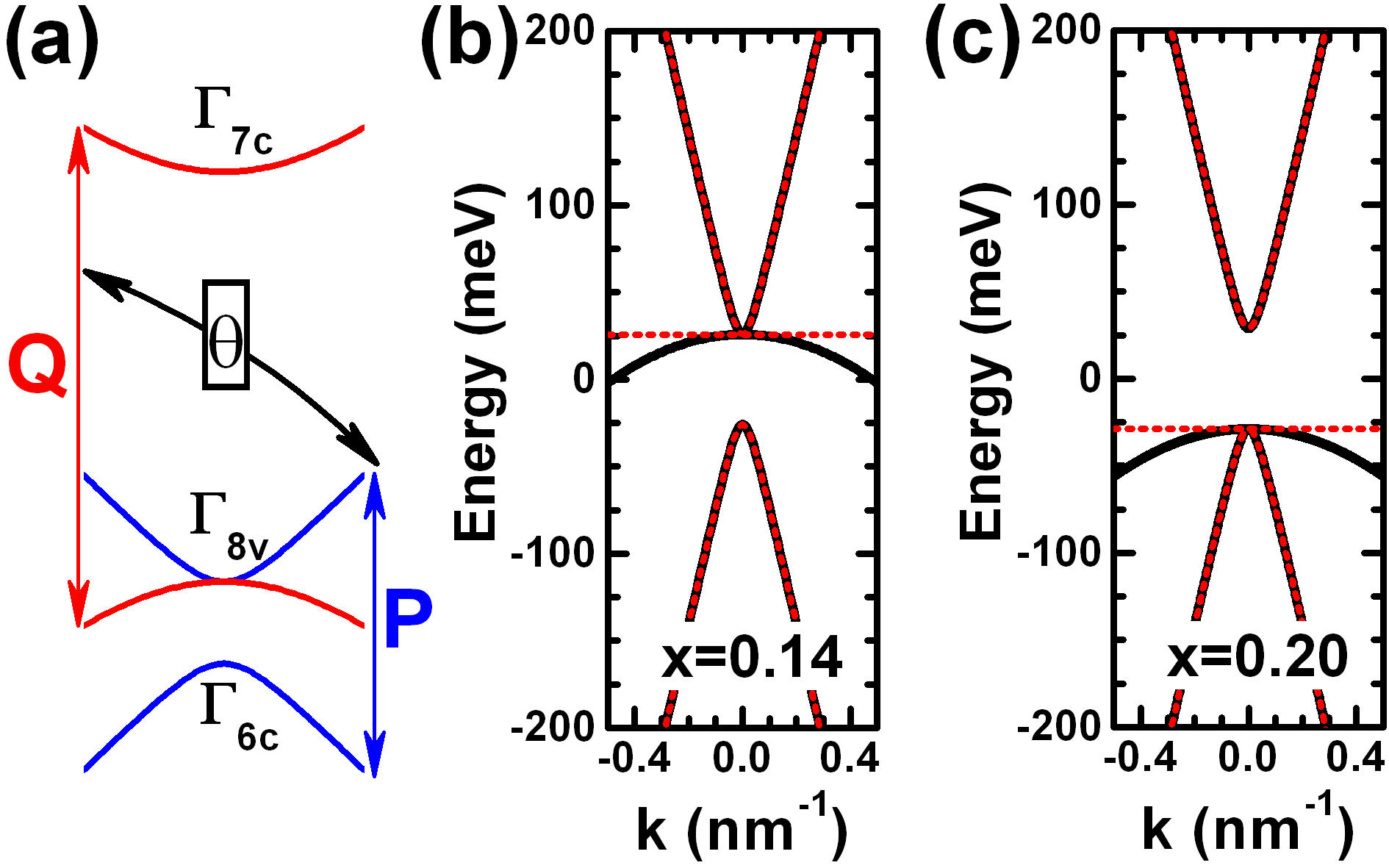} 
\caption{\label{Fig:1} (Color online) (a) Schematic representation of the (3+1)-band model
as two coupled Dirac systems described by ``hybridization angle'' $\theta$~\cite{q6} (for zinc-blende crystal, $\theta=\pi/3$). The ``effective speed of light'' in each system is determined by the corresponding momentum matrix element $P$ or $Q$. As shown in Appendix~\ref{sec:A2}, for narrow-gap HgCdTe crystals, one can assume $V=\sqrt{2/3}P=\sqrt{2/3}Q$.
(b,c) Band dispersion of Hg$_{1-x}$Cd$_{x}$Te bulk crystals in the three-band (in dotted red) and (3+1)-band (in solid black) models. For the latter case, the second conduction $\Gamma_{7c}$ band is high in energy lying beyond the figures' scale, and $\mathbf{k}$ is oriented along [100] crystallographic direction.}
\end{figure}

The first point is that the  $|\Gamma_{6c},{\pm}1/2\rangle$--$|\Gamma_{8v},{\pm}1/2\rangle$ and $|\Gamma_{7c},{\pm}1/2\rangle$--$|\Gamma_{8v},{\pm}3/2\rangle$ band pairs represents two Dirac subsystems (see Fig.~\ref{Fig:1}) each described by \emph{anisotropic} relativistic 3D Dirac Hamiltonian:
\begin{multline}
\label{L1eq:1}
\mathcal{\hat{H}}^{(1,2)}_{D}(\mathbf{k})=
C_{1,2}\mathcal{I}_4+M_{1,2}\alpha_0+\\
+V\cos{\theta}\left(k_{x}\alpha_{x}+k_{y}\alpha_{y}\right)+
Vk_z\alpha_z,
\end{multline}
where $\theta$ is ``hybridization angle'' origin from the mixing between light- and heavy-holes at non-zero quasimomentum~\cite{q6}, $\mathcal{I}_4$ is a $4\times4$ identity matrix and the $\alpha$ matrices are related to the Pauli matrices as
\begin{equation}
\label{L1eq:2}
\alpha_0=\begin{pmatrix}
\sigma_0 &0 \\
0 & -\sigma_0
\end{pmatrix},~~~~~
\alpha_{x,y,z}=\begin{pmatrix}
0 & \sigma_{x,y,z} \\
\sigma_{x,y,z} & 0
\end{pmatrix}.
\end{equation}
Here, the forth Pauli matrix $\sigma_0$ is defined as a $2\times2$ identity matrix. In Eq.~(\ref{L1eq:1}), the constants $C_{1,2}$ and mass parameters $M_{1,2}$ are defined by the energies of $|\Gamma_{6c},{\pm}1/2\rangle$, $|\Gamma_{8v},{\pm}1/2\rangle$, $|\Gamma_{7c},{\pm}1/2\rangle$ and $|\Gamma_{8v},{\pm}3/2\rangle$ bands:
\begin{eqnarray}
\label{L1eq:3}
C_1+M_1=E_{6c},~~~~~~~C_1-M_1=E_{8v},\nonumber\\
C_2+M_2=E_{7c},~~~~~~~C_2-M_2=E_{8v}.
\end{eqnarray}
The last expression determines the position of the heavy-hole band at $\mathbf{k}=0$.
For the case of uniaxially strained HgCdTe or Cd$_3$As$_2$, $C_2-M_2=E_{8v}+\delta_{\epsilon}$, where $\delta_{\epsilon}$ represents the gap between the light- and heavy-hole subbands~\cite{q5}.

The second point is that the two Dirac subsystems are coupled because of the inevitable hybridization between $|\Gamma_{8v},{\pm}1/2\rangle$ and $|\Gamma_{8v},{\pm}3/2\rangle$ bands in cubic semiconductors. By introducing the ``hybridization angle'' $\theta$ (see also Sec.~\ref{Sec:Problem}), the generalized (3+1)-band Hamiltonian is written as
\begin{multline}
\label{L1eq:4}
\mathcal{\hat{H}}^{(\mathbf{3+1})}_{8\times8}(\mathbf{k})= ~~~~~~~~\\
=\begin{pmatrix}
\mathcal{\hat{H}}^{(1)}_{D}(\mathbf{k}) & V\sin{\theta}\left(k_{x}\beta_{y}-k_{y}\beta_{x}\right) \\
V\sin{\theta}\left(k_{y}\beta_{x}-k_{x}\beta_{y}\right) & \mathcal{\hat{H}}^{(2)}_{D}(\mathbf{k})
\end{pmatrix},
\end{multline}
where we have introduced the $\beta$ matrices as
\begin{equation}
\label{L1eq:5}
\beta_0^{(z)}=\begin{pmatrix}
i\sigma_z &0 \\
0 & -i\sigma_z
\end{pmatrix},~~~~~
\beta_{x,y}=\begin{pmatrix}
0 & \sigma_{y,x} \\
-\sigma_{y,x} & 0
\end{pmatrix}.
\end{equation}
Note the indices order in the definition of $\beta_{x}$ and $\beta_{y}$.

The (3+1)-band Hamiltonian in Eq.~(\ref{L1eq:4}) shows that Kane fermions in HgCdTe can be indeed treated as a combination of two mutually hybridized Dirac particles, which is a direct proof of their pseudo-relativistic origin as spin-1/2 particles. This consideration is lacking in the treatment of Kane fermions with the three-band Hamiltonian $\mathcal{\hat{H}}_{6\times6}(\mathbf{k},\theta)$ in Eq.~(\ref{Keq:1}). In its turn, $\mathcal{\hat{H}}_{6\times6}(\mathbf{k},\theta)$ can be directly obtained from Eq.~(\ref{L1eq:4}) by projecting $\mathcal{\hat{H}}^{(\mathbf{3+1})}_{8\times8}(\mathbf{k})$ on the $\Gamma_{6c}+\Gamma_{8v}$ subspace in the limit of $E_{7c}-E_{8v}\rightarrow\infty$. Comparison of the energy dispersion for Hg$_{0.14}$Cd$_{0.86}$Te and Hg$_{0.20}$Cd$_{0.90}$Te bulk crystals ($C_1=0$, $\theta=\pi/3$) calculated within $\mathcal{\hat{H}}_{6\times6}(\mathbf{k},\theta)$ and $\mathcal{\hat{H}}^{(\mathbf{3+1})}_{8\times8}(\mathbf{k})$ is provided in Fig.~\ref{Fig:1}.
The band parameters for this case can be obtained on the basis of Appendix~\ref{sec:A2} and Ref.~\cite{q5}. Note that, in contrast to $\mathcal{\hat{H}}_{6\times6}(\mathbf{k},\theta)$, for which $\mathcal{\hat{H}}_{6\times6}(\mathbf{k},\pm\theta)$ and $\mathcal{\hat{H}}_{6\times6}(\mathbf{k},\pi/2\mp\theta)$ in the absence of magnetic field are related by unitary transformation, the extended (3+1)-band Hamiltonian $\mathcal{\hat{H}}^{(\mathbf{3+1})}_{8\times8}(\mathbf{k})$ does not hold this property.

Let us now dwell on several questions related to the (3+1)-band Hamiltonian. First, Eq.~(\ref{L1eq:4}) corresponds to the Cartesian axes $x$, $y$, $z$ oriented along [100], [010] and [001], respectively. For other orientations of the Cartesian axes, the form of $\mathcal{\hat{H}}^{(\mathbf{3+1})}_{8\times8}(\mathbf{k})$ differs from Eq.~(\ref{L1eq:4}) (see Appendix~\ref{sec:A1}). Secondly, one can see that $\mathcal{\hat{H}}^{(\mathbf{3+1})}_{8\times8}(\mathbf{k})$ does not hold the spherical rotational symmetry. This means that Landau levels of the system depend on the orientation of electric and magnetic fields with respect to the main crystallographic orientations. Further, we consider two different cases, in which Landau levels collapse under different conditions.

\subsection{\label{Sec:LorentzA}$\mathcal{B}\parallel[001]$ }
Let us now consider the behavior of Kane fermion in constant crossed electric and magnetic fields, when magnetic field is oriented along $[001]$ direction chosen as the $z$ axis, while the $x$ and $y$ axes are tilted at an angle $\phi$ from the [100] and [010] directions. In the Cartesian coordinate system, $\mathcal{\hat{H}}^{(\mathbf{3+1})}_{8\times8}(\mathbf{k})$ is written as (see Appendix~\ref{sec:A1}):
\begin{equation}
\label{L1eq:6}
{\mathcal{\hat{H}}}^{(\mathbf{3+1})}_{8\times8}({\mathbf{k}})=\begin{pmatrix}
\mathcal{\hat{H}}^{(1)}_{D}(\mathbf{k}) & V\sin{\theta}\mathcal{\hat{K}}(\mathbf{k}) \\
-V\sin{\theta}\mathcal{\hat{K}}(\mathbf{k}) & \mathcal{\hat{H}}^{(2)}_{D}(\mathbf{k})
\end{pmatrix},
\end{equation}
where $\mathcal{\hat{H}}^{(1,2)}_{D}(\mathbf{k})$ takes the form of Eq.~(\ref{L1eq:1}), while $\mathcal{\hat{K}}(\mathbf{k})$ is written as
\begin{equation}
\label{L1eq:7}
\mathcal{\hat{K}}(\mathbf{k})=\left(k_{x}\beta_{y}-k_{y}\beta_{x}\right)\cos{2\phi}-
\left(k_{x}\beta_{x}+k_{y}\beta_{y}\right)\sin{2\phi}.
\end{equation}

Assuming the orientation of electric field $\mathcal{E}$ in the $x$ direction and magnetic field $\mathcal{B}$ in the $z$ direction, one has to add the diagonal term $e\mathcal{E}{x}\mathcal{I}_8$ ($\mathcal{I}_8$ is a $8\times8$ identity matrix) to $\mathcal{\hat{H}}^{(\mathbf{3+1})}_{8\times8}(\mathbf{k})$ and simultaneously make the Peierls substitution:
\begin{equation}
\label{L1eq:6}
k_x{\rightarrow}\hat{k}_x=-i\dfrac{\partial}{\partial{x}},~~~~~~
k_y{\rightarrow}k_{y}+\dfrac{x}{a_B^2},
\end{equation}
where $a_B$ is the magnetic length ($a_B^2=c\hbar/e\mathcal{B}$, $c$ is the speed of light in vacuum). Then, introducing the ``time-momentum'' operator as $k_t=-i\hbar\partial/\partial{(V_{F}t)}$, the Schr\"{o}dinger equation for the Kane fermion in crossed fields is written as
\begin{equation}
\label{L1eq:7}
\left[\mathcal{\hat{H}}^{(\mathbf{3+1})}_{8\times8}(\hat{k}_x,k_y,k_z)+\left(V_{F}k_t+e\mathcal{E}{x}\right)\mathcal{I}_8\right]|\Psi_{8\times8}\rangle=0,
\end{equation}
where for the sake of brevity, we have introduced $V_F=V\cos{\theta}$.

In order to find the energies $E$ and the wave-function $|\Psi_{8\times8}\rangle$ of Kane fermion, it is convenient to boost to a frame of reference moving with a velocity perpendicular to $\mathcal{E}$ and $\mathcal{B}$. This is performed by means of a Lorentz boost in the $y$ direction on the space-time coordinate system
\begin{equation}
\label{L1eq:8}
\begin{pmatrix}
k_{t}' \\ k_{y}'
\end{pmatrix}=
\begin{pmatrix}
\cosh\alpha & \sinh\alpha \\
\sinh\alpha & \cosh\alpha
\end{pmatrix}
\begin{pmatrix}
k_{t} \\ k_{y}
\end{pmatrix}.
\end{equation}
Under the Lorentz boost, the wave function transforms, $|\Psi_{8\times8}'\rangle=\mathcal{N}{\cdot}L^{(y)}_{8\times8}(\alpha)|\Psi_{8\times8}\rangle$, where $L^{(y)}_{8\times8}(\alpha)=\sigma_0{\otimes}\exp\left(\alpha_y\cdot{\alpha}/{2}\right)$,
and $\mathcal{N}$ is a normalization constant required since $L^{(y)}_{8\times8}(\alpha)$ does not preserve the norm of the wave function.

Applying the above transformations and choosing
\begin{eqnarray}
\label{L1eq:9}
\tanh\alpha=-\beta=-\dfrac{c\hbar}{V}\dfrac{\mathcal{E}}{\mathcal{B}\cos{\theta}},\nonumber\\
\cosh\alpha=\gamma=\dfrac{1}{\sqrt{1-\beta^2}},~
\end{eqnarray}
the Schr\"{o}dinger equation in Eq.~(\ref{L1eq:7}), can be rewritten as
\begin{multline}
\label{L1eq:10}
\biggl[\begin{pmatrix}
\mathcal{\hat{H}}_{1}' & V_{F}\tan{\theta}\mathcal{\hat{K}}' \\
-V_{F}\tan{\theta}\mathcal{\hat{K}}' & \mathcal{\hat{H}}_{2}'
\end{pmatrix}+\\
+V_{F}k_t'
\begin{pmatrix}
\mathcal{I}_4 & -\beta\gamma\tan{\theta}\mathcal{\hat{K}}'_t \\
\beta\gamma\tan{\theta}\mathcal{\hat{K}}'_t & \mathcal{I}_4
\end{pmatrix}
\biggr]|\Psi_{8\times8}'\rangle=0,
\end{multline}
where $\mathcal{\hat{H}}_{1,2}'$ are the ``boosted'' Dirac Hamiltonians:
\begin{multline}
\label{L1eq:11}
\mathcal{\hat{H}}_{1,2}'=
{\gamma}C_{1,2}\left(\mathcal{I}_4-\beta\alpha_y\right)+M_{1,2}\alpha_0+\\
+V_{F}\hat{k}_{x}\alpha_{x}+V_{F}\left(\dfrac{x}{{\gamma}a_B^2}+k_{y}'\right)\alpha_{y}+
Vk_z\alpha_z,
\end{multline}
and
\begin{multline}
\label{L1eq:12}
\mathcal{\hat{K}}'=\left[\gamma\left(\beta_y-\beta\beta_0^{(z)}\right)\cos{2\phi}-\beta_x\sin{2\phi}\right]k_x-\\
-{\gamma}\left[\beta_{x}\cos{2\phi}+\gamma\left(\beta_y-\beta\beta_0^{(z)}\right)\sin{2\phi}\right]\left(\dfrac{x}{{\gamma}a_B^2}+k_{y}'\right),
\end{multline}
and
\begin{equation}
\label{L1eq:12a}
\mathcal{\hat{K}}'_t=\beta_{x}\cos{2\phi}+\gamma\left(\beta_y-\beta\beta_0^{(z)}\right)\sin{2\phi}.
\end{equation}
One can see that the Schr\"{o}dinger equation~(\ref{L1eq:10}) in the moving reference frame is described by the \emph{effective} magnetic field $\mathcal{B}'$ reduced to $\mathcal{B}'=\mathcal{B}/\gamma$ . Although it does not contain the electric field $\mathcal{E}$ explicitly, the terms proportional to $\beta$ and $\gamma$ are present in Eq.~(\ref{L1eq:10}). This means that, in contrast to Dirac fermion, the motion of Kane fermion in the moving reference frame is still dependent on the frame's velocity defined by the parameter $\beta$ (cf.~Refs~\cite{Flds3,Flds4,Flds25}).

Further, it is convenient to shift the origin of the coordinate $x$ as
\begin{equation}
\label{L1eq:13}
\tilde{x}=x+{\gamma}k_{y}'a_B^2-\dfrac{{\beta}{\gamma}^2a_B^2}{V_F}\dfrac{C_1+C_2}{2},
\end{equation}
that allows Eqs.~(\ref{L1eq:10})--(\ref{L1eq:12}) to be presented in a more symmetrical form
\begin{multline}
\label{L1eq:14}
\biggl[
\left(V_{F}k_t'+\mathcal{C}\gamma\right)
\begin{pmatrix}
\mathcal{I}_4 & -\beta\gamma\tan{\theta}\mathcal{\hat{K}}'_t \\
\beta\gamma\tan{\theta}\mathcal{\hat{K}}'_t & \mathcal{I}_4
\end{pmatrix}+\\
+\begin{pmatrix}
\mathcal{\tilde{H}}_{1}' & V_{F}\tan{\theta}\mathcal{\tilde{K}}' \\
-V_{F}\tan{\theta}\mathcal{\tilde{K}}' & \mathcal{\tilde{H}}_{2}'
\end{pmatrix}\biggr]|\Psi_{8\times8}'\rangle=0,
\end{multline}
where $\mathcal{C}=(C_1+C_2)/{2}$ and
\begin{multline}
\label{L1eq:15}
\mathcal{\tilde{H}}_{1,2}'=
\mp\gamma\dfrac{C_2-C_1}{2}\mathcal{I}_4+M_{1,2}\alpha_0+V_{F}\hat{k}_{x}\alpha_{x}+\\
+\dfrac{V_{F}}{{\gamma}a_B^2}\left(\tilde{x}{\pm}
\dfrac{\beta{\gamma}^{2}a_B^2}{V_{F}}\dfrac{C_2-C_1}{2}\right)\alpha_{y}+
Vk_z\alpha_z,
\end{multline}
and
\begin{multline}
\label{L1eq:16}
\mathcal{\hat{K}}'=\left[\gamma\left(\beta_y-\beta\beta_0^{(z)}\right)\cos{2\phi}-\beta_x\sin{2\phi}\right]k_x-\\
-{\gamma}\left[\beta_{x}\cos{2\phi}+\gamma\left(\beta_y-\beta\beta_0^{(z)}\right)\sin{2\phi}\right]\dfrac{\tilde{x}}{{\gamma}a_B^2}.
\end{multline}
In Eq.~(\ref{L1eq:15}), the upper signs correspond to the first Dirac subsystem, while the lower signs are for the second one.

Let us now consider two Dirac subsystems separately. Introducing the ladder operators $a_{1,2}$ and $a^{+}_{1,2}$
\begin{gather}
\label{L1eq:17}
a_{1,2}=\dfrac{1}{\sqrt{2}}\left(\xi{\pm}\xi_0+\dfrac{\partial}{\partial{\xi}}\right),\nonumber\\
a^{+}_{1,2}=\dfrac{1}{\sqrt{2}}\left(\xi{\pm}\xi_0-\dfrac{\partial}{\partial{\xi}}\right),
\end{gather}
where
\begin{eqnarray}
\label{L1eq:18}
\xi=\dfrac{(1-\beta^2)^{1/4}}{a_B}\tilde{x},~~~~~~~\nonumber\\
\xi_{0}=\dfrac{\beta{a_B}}{V_{F}(1-\beta^2)^{3/4}}\dfrac{C_2-C_1}{2},
\end{eqnarray}
each of the Hamiltonians $\mathcal{\tilde{H}}_{1,2}'$ is represented as
\begin{multline}
\label{L1eq:19}
\mathcal{\tilde{H}}_{1,2}'=
\mp\gamma\dfrac{C_2-C_1}{2}\mathcal{I}_4+M_{1,2}\alpha_0+Vk_z\alpha_z+
\\
+{i}\dfrac{\sqrt{2}V_{F}(1-\beta^2)^{1/4}}{a_B}\begin{pmatrix}
0 & 0 & 0 & -a_{1,2} \\
0 & 0 & a^{+}_{1,2} & 0 \\
0 & -a_{1,2} & 0 & 0 \\
a^{+}_{1,2} & 0 & 0 & 0
\end{pmatrix}.
\end{multline}
This form allows for an exact diagonalization of $\mathcal{\tilde{H}}_{1,2}'$ in terms of harmonic oscillator functions $F_n(\xi{\pm}\xi_0)$, where the positive and negative sign is respectively associated with the first and second type of the ladder operators. Here, $F_n(\xi){\equiv}0$ if $n<0$, while for $n\geq0$, it is written as
\begin{equation}
\label{L1eq:20}
F_n(\xi)=\dfrac{1}{\sqrt{2^{n}n!\sqrt{\pi}}}\mathcal{H}_{n}(\xi)e^{-\frac{\xi^2}{2}},
\end{equation}
where $\mathcal{H}_{n}(\xi)$ are the Hermite polynomials.

In view of the above, the wave-function at $\theta=0$ in Eq.~(\ref{L1eq:14}) has the form
\begin{multline}
\label{L1eq:21}
|\Psi_{8\times8}'\rangle_{\theta=0}=e^{-i\dfrac{(E_{\theta=0}'-\mathcal{C}\gamma)t'}{\hbar}}e^{ik_{y}'y'}e^{ik_{z}z}\times\\
\times\begin{pmatrix}
C^{(1)}_{n}F_{n}(\xi+\xi_0) \\
C^{(2)}_{n}F_{n+1}(\xi+\xi_0) \\
C^{(3)}_{n}F_{n}(\xi+\xi_0) \\
C^{(4)}_{n}F_{n+1}(\xi+\xi_0) \\
C^{(5)}_{n}F_{n}(\xi-\xi_0) \\
C^{(6)}_{n}F_{n+1}(\xi-\xi_0) \\
C^{(7)}_{n}F_{n}(\xi-\xi_0) \\
C^{(8)}_{n}F_{n+1}(\xi-\xi_0)
\end{pmatrix},
\end{multline}
where $n$ is Landau level index, $C^{(l)}_{n}$ ($l=1..8$) are the constants and
\begin{equation}
\label{L1eq:D}
E_{\theta=0}'=\mp\gamma\dfrac{C_2-C_1}{2}+\tau\sqrt{M_{1,2}^2+V^{2}k_{z}^{2}+\dfrac{2V^{2}(n+1)}{{\gamma}a_B^2}}
\end{equation}
with $\tau=+1$ and $\tau=-1$ for conduction and valence bands, respectively.

Now let us return to more the general case of Eq.~(\ref{L1eq:14}) with non-zero $\theta$.
Since the ``time-momentum'' operator $k_t'$ commutes with the Hamiltonian in Eq.~(\ref{L1eq:14}) , the wave function is presented in the similar form $|\Psi_{8\times8}'\rangle=\exp\left(-i(E'-\mathcal{C}\gamma)t'/\hbar\right)|\Phi_{8\times8}'\rangle$,
where $|\Phi_{8\times8}'\rangle$ is independent of $t'$ and obeys the equation
\begin{multline}
\label{L1eq:23}
\begin{pmatrix}
\mathcal{\tilde{H}}_{1}' & V_{F}\tan{\theta}\mathcal{\tilde{K}}' \\
-V_{F}\tan{\theta}\mathcal{\tilde{K}}' & \mathcal{\tilde{H}}_{2}'
\end{pmatrix}|\Phi_{8\times8}'\rangle=\\
=E'\begin{pmatrix}
\mathcal{I}_4 & -\beta\gamma\tan{\theta}\mathcal{\hat{K}}'_t \\
\beta\gamma\tan{\theta}\mathcal{\hat{K}}'_t & \mathcal{I}_4
\end{pmatrix}|\Phi_{8\times8}'\rangle.
\end{multline}
As seen, Eq.~(\ref{L1eq:23}) differs from conventional eigenvalue problem due to matrix coefficient at the energy $E'$. Since such matrix does not commute with the matrix in the left side, and the inverse matrix has a singularity at $\beta<1$, Eq.~(\ref{L1eq:23}) cannot be represented in the form of the eigenvalue problem. A similar type of mathematical problem has recently been considered in the context of Dirac systems~\cite{s3}.

In order to find solution of Eq.~(\ref{L1eq:23}), the function $|\Phi_{8\times8}'\rangle$ is convenient to expand in the complete basis of the wave-functions in Eq.~(\ref{L1eq:21}) found at $\theta=0$:
\begin{equation}
\label{L1eq:24}
|\Phi_{8\times8}'\rangle=e^{ik_{y}'y'}e^{ik_{z}z}\sum_{n=-1}^{\infty}\begin{pmatrix}
C^{(1)}_{n}F_{n}(\xi+\xi_0) \\
C^{(2)}_{n}F_{n+1}(\xi+\xi_0) \\
C^{(3)}_{n}F_{n}(\xi+\xi_0) \\
C^{(4)}_{n}F_{n+1}(\xi+\xi_0) \\
C^{(5)}_{n}F_{n}(\xi-\xi_0) \\
C^{(6)}_{n}F_{n+1}(\xi-\xi_0) \\
C^{(7)}_{n}F_{n}(\xi-\xi_0) \\
C^{(8)}_{n}F_{n+1}(\xi-\xi_0)
\end{pmatrix}.
\end{equation}
Note that $n$ ceases to be a good quantum number in contrast to case of $\theta=0$.

The present expansion leads to a matrix representation of Eq.~(\ref{L1eq:23}), where the energies and vectors $\hat{\mathcal{C}}_n=(C^{(1)}_n, ..., C^{(8)}_n)^{T}$ are found from the secular equation:
\begin{equation}
\label{L1eq:25}
\sum_{n=-1}^{\infty}\left(\mathcal{\hat{H}}_{mn}'+E'\mathcal{\hat{A}}_{mn}'\right)\hat{\mathcal{C}}_n=E'\hat{\mathcal{C}}_m,
\end{equation}
where $\mathcal{\hat{H}}_{nm}'$ and $\mathcal{\hat{A}}_{mn}'$ are $8\times8$ matrices written in the block form as
\begin{equation}
\label{L1eq:26}
\mathcal{\hat{H}}_{mn}'=\begin{pmatrix}
\mathcal{\hat{U}}^{(1)}_{mn} &
V\sin{\theta}\mathcal{\hat{R}}_{mn}(\xi_0,-\xi_0) \\
V\sin{\theta}\mathcal{\hat{R}}_{mn}(-\xi_0,\xi_0) & \mathcal{\hat{U}}^{(2)}_{mn}
\end{pmatrix},
\end{equation}
and
\begin{equation}
\label{L1eq:27}
\mathcal{\hat{A}}_{mn}'=\beta\gamma\tan{\theta}\begin{pmatrix}
0 & \mathcal{\hat{T}}_{mn}(\xi_0,-\xi_0) \\
-\mathcal{\hat{T}}_{mn}(-\xi_0,\xi_0) & 0
\end{pmatrix}.
\end{equation}
Here, $\mathcal{\hat{U}}^{(1,2)}_{mn}$ represent the expansion of the Dirac blocks $\mathcal{\tilde{H}}_{1,2}'$, while $\mathcal{\hat{R}}_{mn}(\pm\xi_0,\mp\xi_0)$ and $\mathcal{\hat{T}}_{mn}(\pm\xi_0,\mp\xi_0)$ arise due to the anti-diagonal blocks in Eq.~(\ref{L1eq:23}).

Note that all matrix elements of  $\mathcal{\hat{H}}_{mn}'$ and $\mathcal{\hat{A}}_{mn}'$ in Eq.~(\ref{L1eq:25}) are calculated analytically. Particularly, $\mathcal{\hat{U}}^{(1,2)}_{mn}$ are written as follows:
\begin{widetext}
\begin{multline}
\label{L1eq:28}
\mathcal{\hat{U}}^{(1,2)}_{mn}=
\begin{pmatrix}
M_{1,2}^{(\mp)}\mathcal{F}_{m}\mathcal{F}_{n} & 0 & Vk_{z}\mathcal{F}_{m}\mathcal{F}_{n} & 0 \\
0 & M_{1,2}^{(\mp)}\mathcal{F}_{m+1}\mathcal{F}_{n+1} & 0 & -Vk_{z}\mathcal{F}_{m+1}\mathcal{F}_{n+1} \\
Vk_{z}\mathcal{F}_{m}\mathcal{F}_{n} & 0 & -M_{1,2}^{(\pm)}\mathcal{F}_{m}\mathcal{F}_{n} & 0 \\
0 & -Vk_{z}\mathcal{F}_{m+1}\mathcal{F}_{n+1} & 0 & -M_{1,2}^{(\pm)}\mathcal{F}_{m+1}\mathcal{F}_{n+1}
\end{pmatrix}\delta_{m,n}+\\
+{i}\dfrac{\sqrt{2}V_{F}(1-\beta^2)^{1/4}}{a_B}\begin{pmatrix}
0 & 0 & 0 & -\sqrt{n+1}\mathcal{F}_{m}\mathcal{F}_{n+1} \\
0 & 0 & \sqrt{n+1}\mathcal{F}_{m+1}\mathcal{F}_{n} & 0 \\
0 & -\sqrt{n+1}\mathcal{F}_{m}\mathcal{F}_{n+1} & 0 & 0 \\
\sqrt{n+1}\mathcal{F}_{m+1}\mathcal{F}_{n} & 0 & 0 & 0
\end{pmatrix}\delta_{m,n},
\end{multline}
\end{widetext}
where $M_{1,2}^{(\pm)}=M_{1,2}\pm\gamma{(C_2-C_1)}/{2}$; $\delta_{m,n}$ is the Kronecker delta, while $\mathcal{F}_{n}=1$ for $n\geq0$ and $\mathcal{F}_{n}=0$ for negative $n$ values. In Eq.~(\ref{L1eq:28}), the upper and lower signs correspond to $\mathcal{\hat{U}}^{(1)}_{mn}$ and $\mathcal{\hat{U}}^{(2)}_{mn}$, respectively.

The matrices $\mathcal{\hat{R}}_{mn}(\pm\xi_0,\mp\xi_0)$ are rather cumbersome and not presented here. Their calculation is carried out in a trivial way, if one expresses
$\mathcal{\tilde{K}}'$ in Eq.~(\ref{L1eq:16}) either through the operators $a_{1}$, $a^{+}_{1}$
or $a_{2}$, $a^{+}_{2}$:
\begin{multline*}
\sqrt{\gamma}a_{B}\mathcal{\hat{K}}'=\pm{\gamma}\left[\beta_{x}\cos{2\phi}+\gamma\left(\beta_y-\beta\beta_0^{(z)}\right)\sin{2\phi}\right]\xi_0-\\
-{\gamma}\left[\beta_{x}\cos{2\phi}+\gamma\left(\beta_y-\beta\beta_0^{(z)}\right)\sin{2\phi}\right]\dfrac{a^{+}_{1,2}+a_{1,2}}{\sqrt{2}}+\\
+i\left[\gamma\left(\beta_y-\beta\beta_0^{(z)}\right)\cos{2\phi}-\beta_x\sin{2\phi}\right]\dfrac{a^{+}_{1,2}-a_{1,2}}{\sqrt{2}},
\end{multline*}
depending on whether the functions $F_{n}(\xi+\xi_0)$ or $F_{n}(\xi-\xi_0)$ appear on the left, respectively.

As a result, after the integration, in addition to the factors $\mathcal{F}_{n_1}$ and $\mathcal{F}_{n_2}$, \emph{all of the matrix elements} of $\mathcal{\hat{R}}_{mn}(\pm\xi_0,\mp\xi_0)$ will contain a factor $\Gamma_{n_{1},n_{2}}(\xi_{0},-\xi_{0})$
(instead of the Kronecker delta in Eq.~(\ref{L1eq:28}))
defined as
\begin{multline}
\label{L1eq:29}
\Gamma_{n_{1},n_{2}}(\xi_{1},\xi_{2})=\int\limits_{-\infty}^{+\infty}F_{n_1}(\xi+\xi_1)F_{n_2}(\xi+\xi_2)d\xi= \\ =\sqrt{\dfrac{2^{n-m}}{(n-m)!}}e^{-\delta^2}\mathcal{L}_{m}^{n-m}(2\delta^2)\times \\
\times\begin{cases}
\left(\dfrac{\xi_1-\xi_2}{2}\right)^{n_1-n_2} &\text{, $n_{1}{\geq}n_{2}$},\\
\left(\dfrac{\xi_2-\xi_1}{2}\right)^{n_2-n_1} &\text{, $n_{1}{<}n_{2}$},
\end{cases}
\end{multline}
where $n=\text{max}(n_1,n_2)$, $m=\text{min}(n_1,n_2)$ and $\delta=(\xi_2-\xi_1)/{2}$~\cite{Flds28}. 

The similar $\Gamma$-factors also arise in the calculations of the matrix elements of $\mathcal{\hat{A}}_{mn}'$. By introducing $a_{n_{1},n_{2}}(\xi_{1},\xi_{2})=\Gamma_{n_{1},n_{2}}(\xi_{1},\xi_{2})\mathcal{F}_{n_{1}}\mathcal{F}_{n_{2}}$,
one can write $\mathcal{\hat{T}}_{mn}(\xi_{1},\xi_{2})$ in Eq.~(\ref{L1eq:27}) as follows:
\begin{widetext}
\begin{equation}
\label{L1eq:30}
\mathcal{\hat{T}}_{mn}(\xi_{1},\xi_{2})=
i\begin{pmatrix}
-{\beta}a_{m,n}(\xi_{1},\xi_{2})\sin{2\phi} & 0 & 0 & -a_{m,n+1}(\xi_{1},\xi_{2})e^{2i\phi} \\
0 & {\beta}a_{m+1,n+1}(\xi_{1},\xi_{2})\sin{2\phi} & a_{m+1,n}(\xi_{1},\xi_{2})e^{-2i\phi} & 0 \\
0 & a_{m,n+1}(\xi_{1},\xi_{2})e^{2i\phi} & {\beta}a_{m,n}(\xi_{1},\xi_{2})\sin{2\phi} & 0 \\
-a_{m+1,n}(\xi_{1},\xi_{2})e^{-2i\phi} & 0 & 0 & -{\beta}a_{m+1,n+1}(\xi_{1},\xi_{2})\sin{2\phi}
\end{pmatrix},
\end{equation}
\end{widetext}

Thus, using of Eqs.~(\ref{L1eq:26}--\ref{L1eq:30}), the solution of the Schr\"{o}dinger equation in the boosted frame is reduced to the algebraic secular equation in Eq.~(\ref{L1eq:25}). The secular equation was solved by an iterative method, the $i$-step of which was reduced to the eigenvalue problem for the matrix with the energy $E_{i-1}'$ found at the previous iteration:
\begin{equation*}
\sum_{n=-1}^{N}\left(\mathcal{\hat{H}}_{mn}'+E_{(i-1)}'\mathcal{\hat{A}}_{mn}'\right)\hat{\mathcal{C}}_n=E_{(i)}'\hat{\mathcal{C}}_m.
\end{equation*}
The energy $E_{(0)}'$ for the zeroth iteration has been reduced to zero. Note that the solution of the eigenvalue problem at each iteration can be performed with any required accuracy by using the correspondingly truncated matrix based on the large number $N$ of terms involved in the expansion of $|\Phi_{8\times8}'\rangle$ in Eq.~(\ref{L1eq:26}). In Sec.~\ref{Sec:RandD}, we present the calculations performed for $N=50$ with the five iteration steps.

Knowing the energies $E'$ and wave-functions $|\Psi_{8\times8}'\rangle$ in the boosted frame, the ``physical'' energies $E$ and wave-function $|\Psi_{8\times8}\rangle$ in Eq.~(\ref{L1eq:7}) are found by means of the inverse boost transformation
\begin{equation}
\label{L1eq:32}
E=\dfrac{C_1+C_2}{2}+E'\sqrt{1-\beta^2}-{\beta}Vk_{y}\cos{\theta},
\end{equation}
\begin{equation}
\label{L1eq:33}
|\Psi_{8\times8}\rangle{\propto}e^{-i\dfrac{Et}{\hbar}}e^{ik_{y}y}e^{ik_{z}z}\hat{\mathbb{A}}\sum_{n=-1}^{\infty}\begin{pmatrix}
C^{(1)}_{n}F_{n}(\xi+\xi_0) \\
C^{(2)}_{n}F_{n+1}(\xi+\xi_0) \\
C^{(3)}_{n}F_{n}(\xi+\xi_0) \\
C^{(4)}_{n}F_{n+1}(\xi+\xi_0) \\
C^{(5)}_{n}F_{n}(\xi-\xi_0) \\
C^{(6)}_{n}F_{n+1}(\xi-\xi_0) \\
C^{(7)}_{n}F_{n}(\xi-\xi_0) \\
C^{(8)}_{n}F_{n+1}(\xi-\xi_0)
\end{pmatrix},
\end{equation}
\begin{eqnarray}
\label{L1eq:34}
\xi=\dfrac{(1-\beta^2)^{1/4}}{a_B}\left(x+\dfrac{a_B^{2}k_{y}}{1-\beta^2}+\dfrac{{\beta}a_B^2}{1-\beta^2}\dfrac{E-C_1-C_2}{V\cos{\theta}}\right),\nonumber\\
\xi_{0}=\dfrac{(1-\beta^2)^{1/4}}{a_B}\left(\dfrac{\beta{a_B^2}}{1-\beta^2}\dfrac{C_2-C_1}{2V\cos{\theta}}\right),~~~~~~~~~~~~~
\end{eqnarray}
where $\hat{\mathbb{A}}=\sigma_0{\otimes}\exp\left(-\alpha_y\cdot{\alpha}/{2}\right)$. Note that in the above equations, $\beta=(c\hbar\mathcal{E})/(V\mathcal{B}\cos{\theta})$, while $\alpha$ is defined by Eq.~(\ref{L1eq:9}). In the absence of electric field ($\beta=0$), the formulae above are reduced to the results of conventional Landau level calculations on the basis of non-isotropic Hamiltonian in Eq.~(\ref{L1eq:6}).

\subsection{\label{Sec:LorentzB} $\mathcal{E}\parallel[100]$ and $\mathcal{B}\parallel[010]$ }
Let us now consider another orientation of the electric and magnetic fields. Assuming $x\parallel[100]$, $y\parallel[00\bar{1}]$ and $z\parallel[010]$, and introducing additional $\beta$-matrices as
\begin{equation}
\label{L2eq:0}
\beta_{z}=\begin{pmatrix}
0 & \sigma_{z} \\
-\sigma_{z} & 0
\end{pmatrix},~~~~~
\beta_{x}^{(0)}=\begin{pmatrix}
i\sigma_{x} & 0 \\
0 & -i\sigma_{x}
\end{pmatrix},
\end{equation}
the (3+1)-band Hamiltonian has the form:
\begin{equation}
\label{L2eq:1}
{\mathcal{\hat{H}}}^{(\mathbf{3+1})}_{8\times8}({\mathbf{k}})=\begin{pmatrix}
\mathcal{\hat{H}}^{(1)}_{D}(\mathbf{k}) & V\sin{\theta}\mathcal{\hat{K}}(\mathbf{k}) \\
-V\sin{\theta}\mathcal{\hat{K}}(\mathbf{k}) & \mathcal{\hat{H}}^{(2)}_{D}(\mathbf{k})
\end{pmatrix},
\end{equation}
where $\mathcal{\hat{H}}^{(1,2)}_{D}(\mathbf{k})$ and $\mathcal{\hat{K}}(\mathbf{k})$ are written as
\begin{multline}
\label{L2eq:2}
\mathcal{\hat{H}}^{(1,2)}_{D}(\mathbf{k})=
C_{1,2}\mathcal{I}_4+M_{1,2}\alpha_0+V\cos{\theta}k_{x}\alpha_{x}\\
+Vk_{y}\alpha_{y}+V\cos{\theta}k_{z}\alpha_{z},
\end{multline}
and
\begin{equation}
\label{L2eq:3}
\mathcal{\hat{K}}(\mathbf{k})=k_{x}\beta_{y}-k_{z}\beta_{z}.
\end{equation}

Assuming the orientation of electric field $\mathcal{E}$ along the $x$ axis and magnetic field $\mathcal{B}$ along the $z$ axis and simultaneously making the Peierls substitution presented by Eq.~(\ref{L1eq:6}), we arrive at an equation similar to Eq.~(\ref{L1eq:7}). Then, after the Lorentz boost in the $y$ direction (see Eq.~(\ref{L1eq:8})) with the parameter $\alpha$ defined as
\begin{eqnarray}
\label{L2eq:4}
\tanh\alpha=-\delta=-\dfrac{c\hbar}{V}\dfrac{\mathcal{E}}{\mathcal{B}},\nonumber\\
\cosh\alpha=\tilde{\gamma}=\dfrac{1}{\sqrt{1-\delta^2}},
\end{eqnarray}
the Schr\"{o}dinger equation in the moving frame of reference is rewritten as
\begin{equation}
\label{L2eq:5}
\biggl[\begin{pmatrix}
\mathcal{\hat{H}}_{1}' & V\sin{\theta}\mathcal{\hat{K}}' \\
-V\sin{\theta}\mathcal{\hat{K}}' & \mathcal{\hat{H}}_{2}'
\end{pmatrix}+Vk_t'\mathcal{I}_8
\biggr]|\Psi_{8\times8}'\rangle=0,
\end{equation}
where $k_t'=-i\hbar\partial/\partial{(Vt')}$ is the ``time-momentum'' operator, $\mathcal{\hat{H}}_{1,2}'$ are the ``boosted'' Dirac Hamiltonians:
\begin{multline}
\label{L2eq:6}
\mathcal{\hat{H}}_{1,2}'=
\tilde{\gamma}C_{1,2}\left(\mathcal{I}_4-\delta\alpha_y\right)+M_{1,2}\alpha_0+V\cos{\theta}\hat{k}_{x}\alpha_{x}+\\
+V\left(\dfrac{x}{\tilde{\gamma}a_B^2}+k_{y}'\right)\alpha_{y}+
V\cos{\theta}k_z\alpha_z,
\end{multline}
and
\begin{equation}
\label{L2eq:7}
\mathcal{\hat{K}}'=\tilde{\gamma}\left(\beta_y-\delta\beta_0^{(z)}\right)k_x
-\tilde{\gamma}\left(\beta_z+\delta\beta_0^{(x)}\right)k_z.
\end{equation}
As seen from Eq.~(\ref{L2eq:6}), in the moving frame of reference the effective magnetic field $\mathcal{B}'$ is reduced as $\mathcal{B}'=\mathcal{B}/\tilde{\gamma}$ (cf.~Eqs.~(\ref{L1eq:11},\ref{L1eq:12}) in Sec.~\ref{Sec:LorentzA}). Similar to the case of $\mathcal{B}\parallel[010]$ considered previously, the Schr\"{o}dinger equation for the Kane fermion in the moving reference frame contains additional terms related to the frame's velocity defined by parameter $\delta$.

By analogy with Eq.~(\ref{L1eq:13}), it is also convenient here to shift the origin of the coordinate $x$ as
\begin{equation}
\label{L2eq:8}
\tilde{x}=x+\tilde{\gamma}k_{y}'a_B^2-\dfrac{\delta\tilde{\gamma}^2a_B^2}{V}\dfrac{C_1+C_2}{2}.
\end{equation}
The latter allows to rewrite Eqs.~(\ref{L2eq:5})--(\ref{L2eq:7}) in the form
\begin{equation}
\label{L2eq:9}
\biggl[\begin{pmatrix}
\mathcal{\tilde{H}}_{1}' & V\sin{\theta}\mathcal{\hat{K}}' \\
-V\sin{\theta}\mathcal{\hat{K}}' &\mathcal{\tilde{H}}_{2}'
\end{pmatrix}+\left(Vk_t'+\mathcal{C}\tilde{\gamma}\right)\mathcal{I}_8
\biggr]|\Psi_{8\times8}'\rangle=0,
\end{equation}
where $\mathcal{C}=(C_1+C_2)/{2}$ and
\begin{multline}
\label{L2eq:10}
\mathcal{\tilde{H}}_{1,2}'=
\mp\tilde{\gamma}\dfrac{C_2-C_1}{2}\mathcal{I}_4+M_{1,2}\alpha_0+V\cos{\theta}\hat{k}_{x}\alpha_{x}+\\
+\dfrac{V}{\tilde{\gamma}a_B^2}\left(\tilde{x}{\pm}
\dfrac{\delta\tilde{\gamma}^{2}a_B^2}{V}\dfrac{C_2-C_1}{2}\right)\alpha_{y}+
V\cos{\theta}k_z\alpha_z.
\end{multline}
Here, the upper signs correspond to the first Dirac subsystem, while the lower signs are for the second one. Note that the origin changing does not affect the form of the anti-diagonal blocks $\mathcal{\hat{K}}'$ in Eq.~(\ref{L2eq:10}).

Introducing the ladder operators $b_{1,2}$ and $b^{+}_{1,2}$
\begin{gather}
\label{L2eq:11}
b_{1,2}=\dfrac{1}{\sqrt{2}}\left(\xi{\pm}\xi_0+\dfrac{\partial}{\partial{\xi}}\right),\nonumber\\
b^{+}_{1,2}=\dfrac{1}{\sqrt{2}}\left(\xi{\pm}\xi_0-\dfrac{\partial}{\partial{\xi}}\right),
\end{gather}
where
\begin{eqnarray}
\label{L2eq:12}
\xi=\dfrac{(1-\delta^2)^{1/4}}{a_B\sqrt{\cos{\theta}}}\tilde{x},~~~~~~~~~~~~\nonumber\\
\xi_{0}=\dfrac{\delta{a_B}}{V\sqrt{\cos{\theta}}(1-\delta^2)^{3/4}}\dfrac{C_2-C_1}{2},
\end{eqnarray}
each of the Hamiltonians $\mathcal{\tilde{H}}_{1,2}'$ and $\mathcal{\hat{K}}'$ in Eq.~(\ref{L2eq:9}) are represented as
\begin{multline}
\label{L2eq:13}
\mathcal{\tilde{H}}_{1,2}'=
\mp\tilde{\gamma}\dfrac{C_2-C_1}{2}\mathcal{I}_4+M_{1,2}\alpha_0+V\cos{\theta}k_z\alpha_z+
\\
+{i}\dfrac{\sqrt{2}V\sqrt{\cos{\theta}}(1-\delta^2)^{1/4}}{a_B}\begin{pmatrix}
0 & 0 & 0 & -b_{1,2} \\
0 & 0 & b^{+}_{1,2} & 0 \\
0 & -b_{1,2} & 0 & 0 \\
b^{+}_{1,2} & 0 & 0 & 0
\end{pmatrix},
\end{multline}
and
\begin{multline}
\label{L2eq:14}
\mathcal{\hat{K}}'={i}\tilde{\gamma}\left(\beta_y-\delta\beta_0^{(z)}\right)\dfrac{(1-\delta^2)^{1/4}}{a_B\sqrt{\cos{\theta}}}\dfrac{b^{+}_{1,2}-b_{1,2}}{\sqrt{2}}-\\
-\tilde{\gamma}\left(\beta_z+\delta\beta_0^{(x)}\right)k_z.
\end{multline}

Then, representing the wave function in the form $|\Psi_{8\times8}'\rangle=\exp\left(-i(E'-\mathcal{C}\gamma)t'/\hbar\right)|\Phi_{8\times8}'\rangle$, and using the basis expansion of Eq.~(\ref{L1eq:24}) but with $\xi$ and $\xi_0$ defined above, the Schr\"{o}dinger equation~(\ref{L2eq:9}) in the moving reference frame is reduced to the eigenvalue problem (cf.~Eq.~(\ref{L1eq:25})):
\begin{equation}
\label{L2eq:15}
\sum_{n=-1}^{\infty}\mathcal{\hat{H}}_{mn}'\hat{\mathcal{C}}_n=E'\hat{\mathcal{C}}_m,
\end{equation}
where $\mathcal{\hat{H}}_{nm}'$ is $8\times8$ matrix written as follows:
\begin{equation}
\label{L2eq:16}
\mathcal{\hat{H}}_{mn}'=\begin{pmatrix}
\mathcal{\hat{U}}^{(1)}_{mn} &
V\sin{\theta}\mathcal{\hat{R}}_{mn}(\xi_0,-\xi_0) \\
V\sin{\theta}\mathcal{\hat{R}}_{mn}(-\xi_0,\xi_0) & \mathcal{\hat{U}}^{(2)}_{mn}
\end{pmatrix},
\end{equation}
where $\mathcal{\hat{U}}^{(1,2)}_{mn}$ represent the expansion of $\mathcal{\tilde{H}}_{1,2}'$ in Eq.~(\ref{L2eq:13}), while $\mathcal{\hat{R}}_{mn}(\pm\xi_0,\mp\xi_0)$ results from the anti-diagonal blocks $\mathcal{\hat{K}}'$ in Eq.~(\ref{L2eq:14}). In this section, we omit explicit expressions for $\mathcal{\hat{U}}^{(1,2)}_{mn}$ and $\mathcal{\hat{R}}_{mn}(\pm\xi_0,\mp\xi_0)$, which are not difficult to obtain. Note that $\mathcal{\hat{U}}^{(1,2)}_{mn}$ are proportional to the Kronecker delta $\delta_{m,n}$ (cf.~Eq.~(\ref{L1eq:28})), while all the matrix elements of $\mathcal{\hat{R}}_{mn}(\pm\xi_0,\mp\xi_0)$ contain the $\Gamma$-factors introduced by Eq.~(\ref{L1eq:29}).

In the absence of non-diagonal blocks in Eq.~(\ref{L2eq:16}), the eigenvalue problem in Eq.~(\ref{L2eq:15}) can be analytically solved resulting in
\begin{equation}
\label{L2eq:D}
E_{\theta=0}'=\mp\tilde{\gamma}\dfrac{C_2-C_1}{2}+\tau\sqrt{M_{1,2}^2+V^{2}k_{z}^{2}+\dfrac{2V^{2}(n+1)}{\tilde{\gamma}a_B^2}},
\end{equation}
where $n$ is the Landau level index; $\tau=+1$ and $\tau=-1$ for conduction and valence bands, respectively. At non-zero $\theta$, that solution of the eigenvalue problem can be obtained by numerical calculation with any required accuracy.

After knowing the energies $E'$ and wave-functions $|\Psi_{8\times8}'\rangle$ in the boosted frame, the energies $E$ and wave-function $|\Psi_{8\times8}\rangle$ of the (3+1)-band Hamiltonian in Eq.~(\ref{L2eq:1}) are found by means of the inverse boost transformation
\begin{equation}
\label{L2eq:17}
E=\dfrac{C_1+C_2}{2}+E'\sqrt{1-\delta^2}-{\delta}Vk_{y},
\end{equation}
\begin{equation}
\label{L2eq:18}
|\Psi_{8\times8}\rangle{\propto}e^{-i\dfrac{Et}{\hbar}}e^{ik_{y}y}e^{ik_{z}z}\hat{\mathbb{A}}\sum_{n=-1}^{\infty}\begin{pmatrix}
C^{(1)}_{n}F_{n}(\xi+\xi_0) \\
C^{(2)}_{n}F_{n+1}(\xi+\xi_0) \\
C^{(3)}_{n}F_{n}(\xi+\xi_0) \\
C^{(4)}_{n}F_{n+1}(\xi+\xi_0) \\
C^{(5)}_{n}F_{n}(\xi-\xi_0) \\
C^{(6)}_{n}F_{n+1}(\xi-\xi_0) \\
C^{(7)}_{n}F_{n}(\xi-\xi_0) \\
C^{(8)}_{n}F_{n+1}(\xi-\xi_0)
\end{pmatrix},
\end{equation}
\begin{eqnarray}
\label{L2eq:19}
\xi=\dfrac{(1-\delta^2)^{1/4}}{a_B\sqrt{\cos{\theta}}}
\left(x+\dfrac{a_B^{2}k_{y}}{1-\delta^2}+\dfrac{{\delta}a_B^2}{1-\delta^2}\dfrac{E-C_1-C_2}{V}\right),\nonumber\\
\xi_{0}=\dfrac{(1-\delta^2)^{1/4}}{a_B\sqrt{\cos{\theta}}}\left(\dfrac{\delta{a_B^2}}{1-\delta^2}\dfrac{C_2-C_1}{2V}\right),~~~~~~~~~~~
\end{eqnarray}
where $\hat{\mathbb{A}}=\sigma_0{\otimes}\exp\left(-\alpha_y\cdot{\alpha}/{2}\right)$. We remind that in the above equations, $\delta=(c\hbar\mathcal{E})/(V\mathcal{B})$, while $\alpha$ is defined by Eq.~(\ref{L2eq:4}).

\section{\label{Sec:RandD} Results and Discussion}
Let us now discuss some properties of Kane fermions directly seen from the (3+1)-band Hamiltonian in crossed electric and magnetic fields.
First, the structure of the (3+1)-band Hamiltonian proves the origin of the Kane fermions in HgCdTe as spin-1/2 particles. In this sense, the Kane fermion should be considered as a superposition of two Dirac particles of different masses hybridized in a special way. Interestingly, this situation is partially reminiscent of the case of Dirac fermion itself, which is also a superposition of two Weyl particles mutually hybridized  due to a finite rest mass. Note that the (3+1)-band representation of the Kane fermion is retained for any position of the $\Gamma_{7c}$, $\Gamma_{6c}$ and $\Gamma_{8v}$ bands (see Eq.~(\ref{L1eq:3})); therefore, it describes HgCdTe crystals with inverted and non-inverted band structures, as well as for Cd$_{3}$As$_{2}$~\cite{Flds32}.

As first shown by Aronov and Pikus~\cite{Flds3,Flds4}, the pseudo-relativistic character of Dirac fermions in semiconductors allows to apply the Lorentz boost to eliminate the electric field $\mathcal{E}'$ from the Hamiltonian in the reference frame moving with the drift velocity $V_d=c\mathcal{E}/\mathcal{B}$. Thus, the Dirac Hamiltonian in the moving frame includes only the effective magnetic field $\mathcal{B}'=\mathcal{B}\sqrt{1-\delta^2}$, where $\delta$ is the ratio of $V_d$ to ``effective speed of the light'' $\tilde{c}=V/{\hbar}$. Note that such an interpretation is compatible with the Lorentz transformation for electric and magnetic fields also described by the ``effective speed of the light''. Indeed, in the frame moving with velocity $\textbf{V}$, the electric $\mathbf{\mathcal{E}}$ and magnetic $\mathbf{\mathcal{B}}$ fields formally transform according to
\begin{eqnarray}
\label{L3eq:0aL1}
\mathbf{\mathcal{E}'}=\gamma_L\left(\mathbf{\mathcal{E}}+\dfrac{1}{c_L}\textbf{V}\times\mathbf{\mathcal{B}}\right)-\dfrac{\gamma_L^2}{\gamma_L+1}
\dfrac{\textbf{V}\left(\textbf{V}\cdot\mathbf{\mathcal{E}}\right)}{c_L^2},\nonumber\\
\mathbf{\mathcal{B}'}=\gamma_L\left(\mathbf{\mathcal{B}}-\dfrac{1}{c_L}\textbf{V}\times\mathbf{\mathcal{E}}\right)-\dfrac{\gamma_L^2}{\gamma_L+1}
\dfrac{\textbf{V}\left(\textbf{V}\cdot\mathbf{\mathcal{B}}\right)}{c_L^2},
\end{eqnarray}
where $c_L$ and $\gamma_L$ are related as $\gamma_L=1/\sqrt{1-\left|\textbf{V}\right|^2/c_L^2}$. It can be seen that this transformation leaves invariant the quantities
\begin{equation}
\label{L3eq:0aL2}
{\mathcal{B}'}^2-{\mathcal{E}'}^2={\mathcal{B}}^2-{\mathcal{E}}^2,~~~~~
(\mathcal{B}'\cdot\mathcal{E}')=(\mathcal{B}\cdot\mathcal{E}),
\end{equation}
for \emph{any values} of $c_L$. If $\mathcal{B}\perp\mathcal{E}$ and $\mathcal{B}>\mathcal{E}$ one can always eliminate the electric field by choosing a suitable moving coordinate system. For Dirac fermions, the transformations of the Hamiltonian and fields become compatible only if $c_L=\tilde{c}$.

For Kane fermions, the effective magnetic field $\mathcal{B}'$ in the moving frame, in which $\mathcal{E}'=0$, depends on the orientations of $\mathcal{E}$ and $\mathcal{B}$ with respect to the main crystallographic axes. Therefore, a compatibility of the (3+1)-band Hamiltonian with the fields' transformation in Eq.~(\ref{L3eq:0aL2}) takes place at the values of $c_L$ dependent on the crystallographic orientations of $\mathcal{E}$ and $\mathcal{B}$ in the original frame as well (see Appendix~\ref{sec:A3}). Particularly, for the two cases considered in Sec.~\ref{Sec:Lorentz}, $c_L=\tilde{c}\cos{\theta}$ for $\mathcal{B}\parallel[001]$, while for $\mathcal{B}\parallel[001]$ and $\mathcal{E}\parallel[100]$, $c_L=\tilde{c}$.

\begin{figure}
\includegraphics [width=1.0\columnwidth, keepaspectratio] {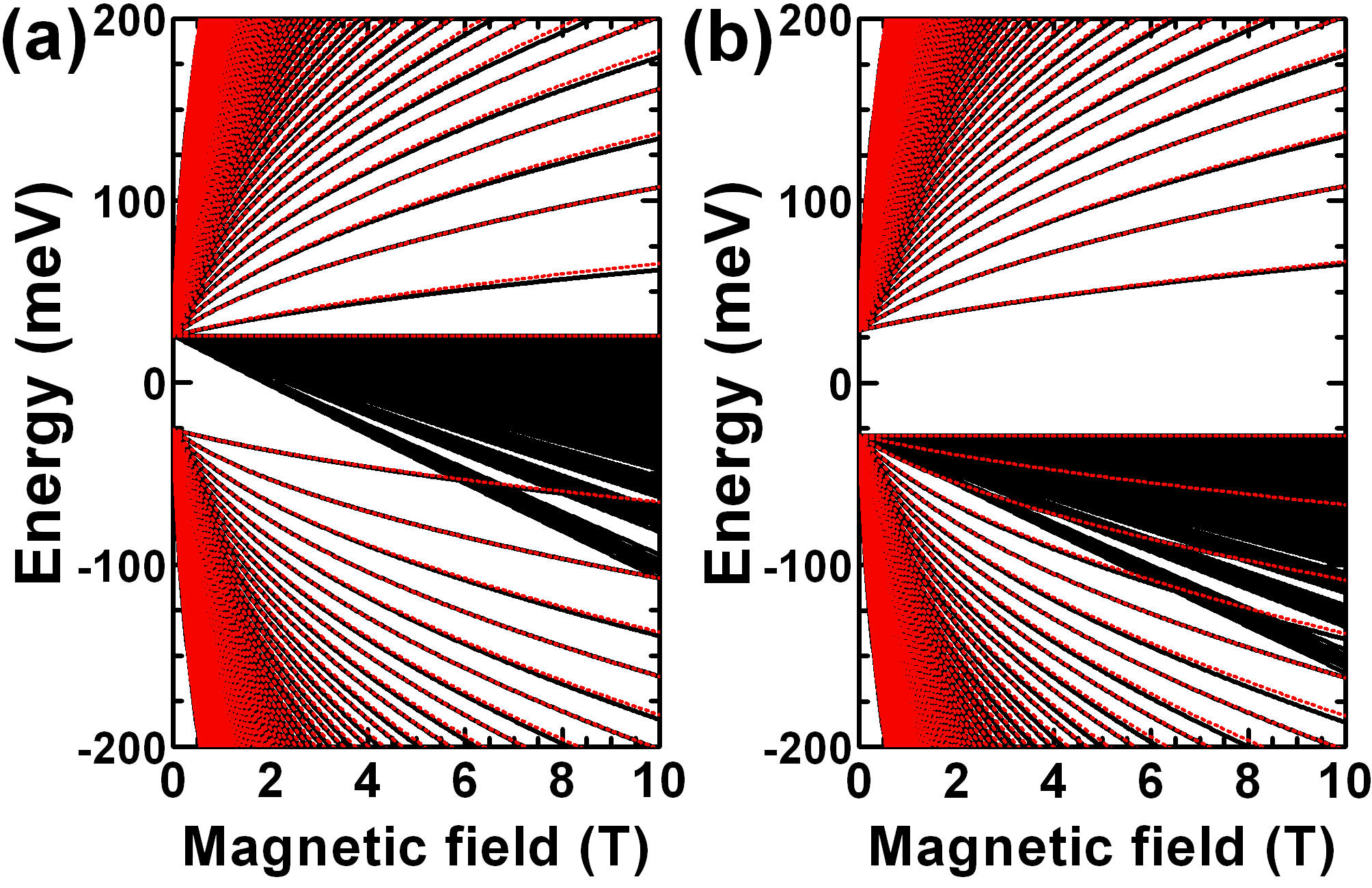} 
\caption{\label{Fig:2} (Color online) Landau levels in the absence of electric field in (a) Hg$_{0.14}$Cd$_{0.86}$Te and (b) Hg$_{0.20}$Cd$_{0.80}$Te bulk crystals with inverted and non-inverted band structures, respectively (cf.~Fig.~\ref{Fig:1}). The solid black curves are the calculations within the (3+1)-band model, while the dotted red curves represent the calculations based on Eq.~(\ref{Keq:5}). For both cases, $\mathcal{B}\parallel[001]$. The second conduction $\Gamma_{7c}$ band is high in energy lying beyond the figure scale.}
\end{figure}

Another difference between Kane fermions and Dirac particles is the breaking of spherical symmetry, which is a consequence of the symmetry of zinc blende crystals. The latter results in the fact that Landau level index, defined by the corresponding harmonic oscillator function in Eq.~(\ref{L1eq:20}), ceases to be a good quantum number that complicates the Landau level fan calculated within the (3+1)-band model. Figure~\ref{Fig:2} provides the energies of Landau levels for HgCdTe bulk crystals with inverted and non-inverted band structures in the absence of electric field ($\beta=0$) when magnetic field $\mathcal{B}$ is oriented along [001] crystallographic axis. In order to solve the eigenvalue problem arising from the secular equation~(\ref{L1eq:25}) at $\beta=0$, we have used the truncated matrix composed of $\mathcal{\hat{H}}_{mn}'$ with $N=50$ terms in the expansion of $|\Phi_{8\times8}'\rangle$ in Eq.~(\ref{L1eq:26}). It is seen that although the Landau levels for the light-hole and $\Gamma_{6c}$ bands are in good agreement with Eq.~(\ref{Keq:5}), the non-zero curvature of the heavy hole band in the absence of rotational symmetry in the plane perpendicular to the magnetic field leads to a non-equidistant spectrum of the Landau levels. Note that $\theta=\pi/3$ for zinc-blende HgCdTe and Cd$_{3}$As$_{2}$ crystals.

In the presence of electric field $\mathcal{E}$, the cubic symmetry leads to a strong dependence of the energy and the collapse of the Landau levels on the orientation of electric and magnetic field relative to the main axes of the crystal. This is well illustrated by the two cases considered in Sec.~\ref{Sec:Lorentz}, in which the collapse occurs when the drift velocity $V_d$ reaches $V\cos{\theta}/{\hbar}$ for $\mathcal{B}\parallel[001]$ and $V/{\hbar}$ for $\mathcal{B}\parallel[010]$, $\mathcal{E}\parallel[100]$. The former corresponds to only \emph{half} of the ``effective speed of the light'' $\tilde{c}=V/{\hbar}$, since $\theta=\pi/3$ for the Kane fermions. It can be shown that, for an arbitrary orientation of the magnetic field (and the electric field $\mathcal{E}\perp\mathcal{B}$), the Landau levels collapse when $V_d$ equals to $V^{*}_d$ lying between \emph{half} and \emph{whole} values of $V/{\hbar}$ (see Appendix~\ref{sec:A3}).

Let us now take a closer look at the Landau levels evolution in crossed electric and magnetic fields. As shown in Sec.~\ref{Sec:Lorentz}, in the presence of $\mathcal{E}$, the energies of Landau levels of Kane fermions are connected with the energies $E'$ and wave-functions $|\Psi_{8\times8}'\rangle$ of the ``boosted'' (3+1)-band Hamiltonian. The calculations in the moving frame, in their turn, are reduced to the solution of secular equations in the given expansion basis~(see Eq.~(\ref{L1eq:25}) and (\ref{L2eq:15})) composed of 4$\times$4 matrices describing the two Dirac subsystems. The two diagonal matrices in Eqs.~(\ref{L1eq:26}) and (\ref{L2eq:16}), namely $\mathcal{\hat{U}}^{(1,2)}_{mn}$, describe the evolution of each of the Dirac subsystem, while the anti-diagonal matrices $\mathcal{\hat{R}}_{mn}(\pm\xi_0,\mp\xi_0)$ (as well as $\mathcal{\hat{T}}_{mn}(\pm\xi_0,\mp\xi_0)$ in Eq.~(\ref{L1eq:27})) represent the mutual hybridization between the subsystems.

The most important point of the secular equations~(\ref{L1eq:25}) and (\ref{L2eq:15}) is that each of the matrix elements of $\mathcal{\hat{R}}_{mn}(\pm\xi_0,\mp\xi_0)$ and $\mathcal{\hat{T}}_{mn}(\pm\xi_0,\mp\xi_0)$ contains the corresponding $\Gamma$-factor.
As seen from Eq.~(\ref{L1eq:30}), $\Gamma_{n_{1},n_{2}}(\pm\xi_{0},\mp\xi_{0})$ is proportional to the Gaussian factor $\exp(-\xi_0^2)$ at any values of $n_1$ and $n_2$. The latter makes it possible to immediately understand the behavior of Landau levels in strong electric fields.

Particularly, for the case of $\mathcal{B}\parallel[001]$ (see Sec.~\ref{Sec:LorentzA}), the anti-diagonal blocks of $\mathcal{\hat{H}}_{mn}'$ in Eq.~(\ref{L1eq:26}) and $\mathcal{\hat{T}}_{mn}(\pm\xi_0,\mp\xi_0)$ in Eq.~(\ref{L1eq:27}) vanish at $\beta\rightarrow{1}$
due to $\xi_0\rightarrow{\infty}$ and
\begin{equation}
\label{L3eq:0a}
e^{-\xi_0^2}=
\exp\left[{-\dfrac{\beta^2{a_B^2}(C_2-C_1)^2}{4V^2\cos^2{\theta}(1-\beta^2)^{3/2}}}\right]\rightarrow{0}
\end{equation}
if $C_1{\neq}C_2$, which is always the case of HgCdTe and Cd$_{3}$As$_{2}$~\cite{Flds32} crystals. Since the Gaussian function $e^{-\xi_0^2}$ falls off very quickly, the anti-diagonal blocks $\mathcal{\hat{R}}_{mn}(\pm\xi_0,\mp\xi_0)$ and $\mathcal{\hat{T}}_{mn}(\pm\xi_0,\mp\xi_0)$ become negligibly small long before $\beta$ reaches $1$. In this case, the inverse Lorentz transformation gives the following Landau levels energies
\begin{multline}
\label{L3eq:1}
E_{\mathrm{Dirac}}^{[001]}=C_{1,2}-{\beta}Vk_{y}\cos{\theta}\pm\\
\pm\sqrt{1-\beta^2}\sqrt{M_{1,2}^2+V^{2}k_{z}^{2}+\sqrt{1-\beta^2}\dfrac{2V^{2}(n+1)\cos^2{\theta}}{a_B^2}},
\end{multline}
where $n\geq-1$ and $\beta={\hbar}V_d/(V\cos{\theta})=(c{\hbar}\mathcal{E})/(\mathcal{B}V\cos{\theta})$ (see Sec.~\ref{Sec:LorentzA}). Thus, Kane fermion in crossed electric and magnetic fields decays into two independent Dirac fermions by increasing the electric fields.

\begin{figure}
\includegraphics [width=1.0\columnwidth, keepaspectratio] {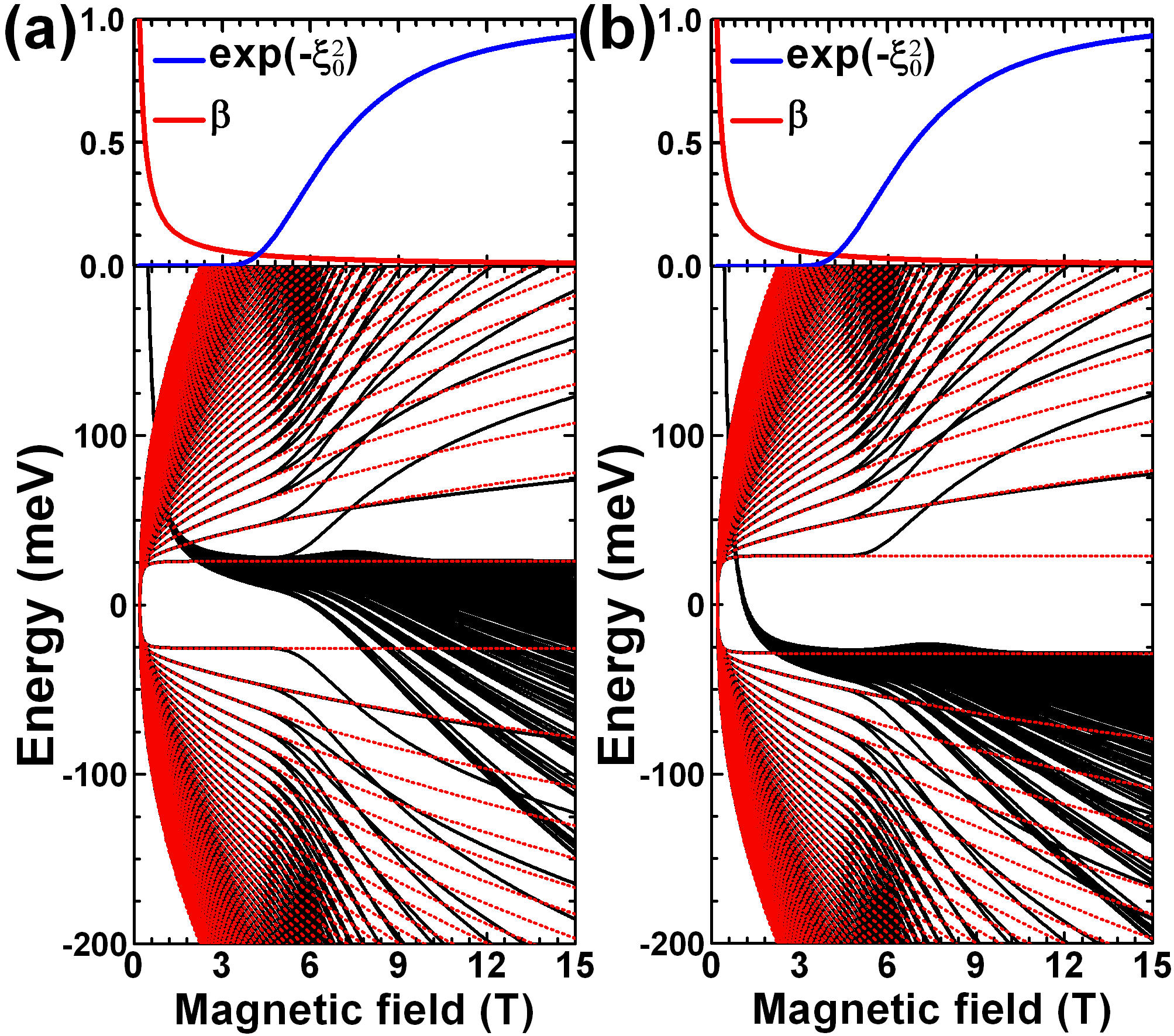} 
\caption{\label{Fig:3} (Color online) Landau levels at $\mathcal{E}=1000$~V/cm and $k_y=k_z=0$ in (a) Hg$_{0.14}$Cd$_{0.86}$Te and (b) Hg$_{0.20}$Cd$_{0.80}$Te bulk crystals for $\mathcal{B}\parallel[001]$ and $\mathcal{E}\parallel[100]$. The solid black curves are the calculations within the (3+1)-band model (see Sec.~\ref{Sec:LorentzA}), while the dotted red curves represent the calculations based on Eq.~(\ref{L3eq:1}). Here, the critical magnetic field $B_c=(c\hbar\mathcal{E})/(V\cos{\theta})$, at which Landau levels of Kane fermions ($\theta=\pi/3$) collapse, approximately equals to $0.19$~T. The upper panels represent $\beta$ and  $e^{-\xi_0^2}$ (where $\beta$ and $\xi_0$ are defined by Eq.~(\ref{L1eq:9}) and Eq.~(\ref{L1eq:18}), respectively) as functions of magnetic field.}
\end{figure}

Figure~\ref{Fig:3} provides the Landau levels calculations in crossed electric and magnetic fields in HgCdTe bulk crystals with inverted and non-inverted band structure for $\mathcal{B}\parallel[001]$ and $\mathcal{E}\parallel[100]$. The latter corresponds to $\phi=0$ in $\mathcal{\hat{K}}'_t$ and $\mathcal{\hat{K}}'$ defined by Eqs.~(\ref{L1eq:12a}) and (\ref{L1eq:16}), respectively. In order to solve the secular equation~(\ref{L1eq:25}), we applied the five-steps iteration procedure involving the truncated matrix composed of $\mathcal{\hat{H}}_{mn}'$ with $N=50$ terms in the expansion of $|\Phi_{8\times8}'\rangle$ in Eq.~(\ref{L1eq:26}). Then, solution of the secular equation~(\ref{L1eq:25}) was transformed into the energies of Landau levels by means of Eq.~(\ref{L1eq:32}).

The energy of the Landau levels in HgCdTe crystals is plotted in Fig.~\ref{Fig:3}, as function of magnetic field, in two different regimes. In small magnetic fields exceeding $B_c=(c\hbar\mathcal{E})/(V\cos{\theta})$ at which $\beta=1$, the Landau levels energies are reproduced by the picture involving two independent Dirac fermions in accordance with the conclusion made above. Above 3~T, the Dirac fermion regime, characterized by double-degenerated Landau levels (except the one at $n=-1$ in Eq~(\ref{L3eq:1})) progressively transforms into the Kane fermion picture characterized by strong hybridization between light- and heavy-hole bands. In high magnetic fields, corresponding to the small $\beta$ values, the Landau level energies asymptotically tend to the ones in the absence of electric fields.

\begin{figure}
\includegraphics [width=1.0\columnwidth, keepaspectratio] {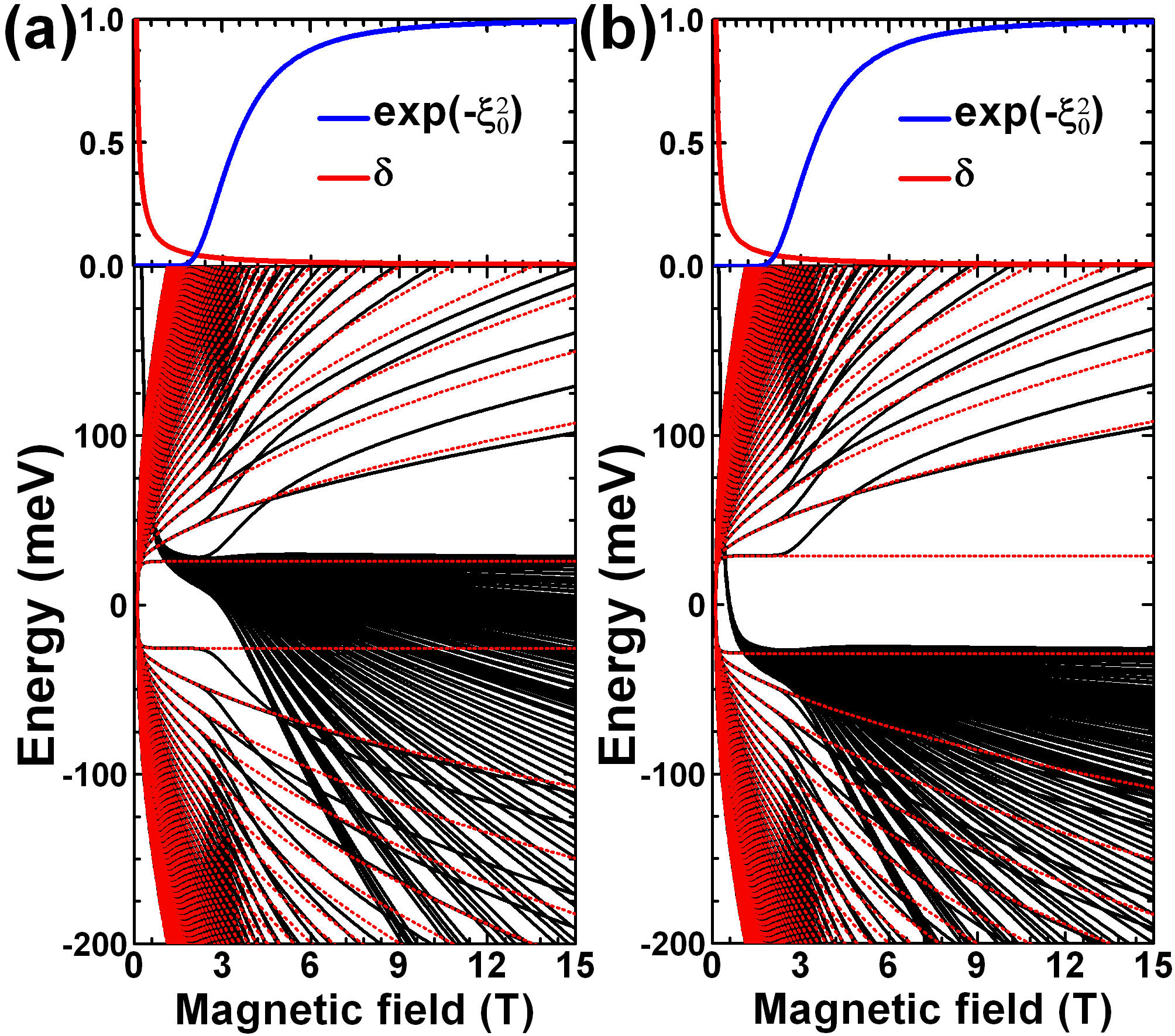} 
\caption{\label{Fig:4} (Color online) Landau levels at $\mathcal{E}=1000$~V/cm and $k_y=k_z=0$ in (a) Hg$_{0.14}$Cd$_{0.86}$Te and (b) Hg$_{0.20}$Cd$_{0.80}$Te bulk crystals for $\mathcal{B}\parallel[010]$ and $\mathcal{E}\parallel[100]$. The solid black curves are the calculations within the (3+1)-band model (see Sec.~\ref{Sec:LorentzB}), while the dotted red curves represent the calculations based on Eq.~(\ref{L3eq:2}). Here, the critical magnetic field $B_c=c\hbar\mathcal{E}/V$, at which Landau levels of Kane fermions ($\theta=\pi/3$) collapse, approximately equals to $0.095$~T. The upper panels represent $\delta$ and  $e^{-\xi_0^2}$ (where $\delta$ and $\xi_0$ are defined by Eq.~(\ref{L2eq:4}) and Eq.~(\ref{L2eq:12}), respectively) as functions of magnetic field.}
\end{figure}

A similar behavior of the Landau levels is observed for other orientations of the magnetic field, particularly for $\mathcal{B}\parallel[010]$ shown in Fig.~\ref{Fig:4}. As seen, the only qualitative difference between the Landau levels for $\mathcal{B}\parallel[010]$ and those for $\mathcal{B}\parallel[001]$ discussed above is the smaller range of magnetic fields, in which the Dirac fermion regime is established. This difference is caused by different definitions of $\xi_0$ for $\mathcal{B}\parallel[001]$ (see Eq.~(\ref{L1eq:18})) and $\mathcal{B}\parallel[010]$ (see Eq.~(\ref{L2eq:12})). In the latter case, the Gaussian factor $\exp(-\xi_0^2)$ determining the Landau levels evolution with magnetic field is written as

\begin{equation}
\label{L3eq:0b}
e^{-\xi_0^2}=
\exp\left[{-\dfrac{\delta^2{a_B^2}(C_2-C_1)^2}{4V^2\cos{\theta}(1-\delta^2)^{3/2}}}\right].
\end{equation}
As seen from Fig.~\ref{Fig:4}, the Landau levels energies in small magnetic fields (such that $\exp(-\xi_0^2)\rightarrow{0}$) are defined as
\begin{multline}
\label{L3eq:2}
E_{\mathrm{Dirac}}^{[010]}=C_{1,2}-{\delta}Vk_{y}\pm\\
\pm\sqrt{1-\delta^2}\sqrt{M_{1,2}^2+V^{2}k_{z}^{2}+\sqrt{1-\delta^2}\dfrac{2V^{2}(n+1)\cos{\theta}}{a_B^2}},
\end{multline}
where $n\geq-1$ and $\delta=(c\hbar\mathcal{E})/(V\mathcal{B})$ (see Sec.~\ref{Sec:LorentzB}).

Thus, we can conclude that the Kane fermion in crossed fields always decays into two independent Dirac particles, which is represented by a vanishing of the mixing between the light and heavy hole bands as the magnetic field decreases. The magnetic field range of such Dirac fermion regime is defined by the difference $C_2-C_1$ and the orientation of electric and magnetic fields with respect to the main crystallographic axes of the zinc-blende crystal. Note that $C_2-C_1$ is always non-zero due to the band positions in HgCdTe crystals (see~Eq.~(\ref{L1eq:3})). We also note that the conclusions above holds in the presence of a gap between the light- and heavy-hole subbands, which includes the uniaxially strained HgCdTe and Cd$_3$As$_2$~\cite{Flds32} bulk crystals. As seen from above, the decay of Kane fermion is absent if the (3+1)-band Hamiltonian preserves the particle-hole symmetry, i.e. at $C_1=C_2$.

The above conclusions remain valid even beyond the ``symmetric approximation'' of the (3+1)-band Hamiltonian, i.e. when two Dirac subsystems are characterized by different ``speeds of light'' $V_1/\hbar$ and $V_2/\hbar$ (see Appendix~\ref{sec:A2}). This arises if the inter-band momentum matrix elements $P$ for $\Gamma_{8v}$--$\Gamma_{6c}$ bands and $Q$ for $\Gamma_{8v}$--$\Gamma_{7c}$ bands differ significantly, that  takes place for instance in bulk InSb~\cite{q11}. The exact solution of the problem when $V_1{\neq}V_2$ is a rather cumbersome task and is beyond the scope of this work. Further, we only restrict ourselves to the qualitative algorithm description, following which one can accurately find the Landau levels beyond the ``symmetric approximation''.

If $V_1{\neq}V_2$, instead of the ``global'' Lorentz boost transformation, one must first find the Landau levels in the crossed fields for each of the Dirac subsystems separately. The solution to this problem, in turn, can be performed by means of two ``local'' Lorentz boost transformations, each characterized by its own speed of light. Then, by using the wave functions in a stationary frame of reference as the basis expansion, one can take into account the coupling between two Dirac subsystems and reduce the Landau levels calculations to the corresponding matrix secular equation. By analogy with the ``symmetric approximation'', the secular equation will contain the anti-diagonal blocks with the elements proportional to the novel $\Gamma$-factor (cf.~Eq.~(\ref{L1eq:29})) defined as
\begin{equation}
\label{L3eq:3}
\Gamma_{\lambda_{1},n_{1},\lambda_{2},n_{2}}(\xi_{1},\xi_{2})=\int\limits_{-\infty}^{+\infty}F_{n_1}(\lambda_{1}\xi+\xi_1)F_{n_2}(\lambda_{2}\xi+\xi_2)d\xi,
\end{equation}
where the presence of $\lambda_{1}$ and $\lambda_{2}$ is due to different ``speeds of light'' in the Dirac subsystems (see the definition of $\xi$ for $\mathcal{B}\parallel[001]$ in Eq.~(\ref{L1eq:18}) and for $\mathcal{B}\parallel[010]$ in Eq.~(\ref{L2eq:12})). Although the integral in Eq.~(\ref{L3eq:3}) cannot be calculated analytically, it can be shown that
\begin{equation}
\label{L3eq:4}
\Gamma_{\lambda_{1},n_{1},\lambda_{2},n_{2}}(\xi_{1},\xi_{2})\sim\exp\left[-\dfrac{(\lambda_{1}\xi_{2}+\lambda_{2}\xi_{1})^2}{2(\lambda_{1}^2+\lambda_{2}^2)}\right]
\end{equation}
for any values of $n_1$ and $n_2$. If $\xi_{1}$ or $\xi_{2}$ is large enough, which happens when the magnetic field approaches a critical value of the Landau levels collapse in one of the Dirac subsystems, $\Gamma_{\lambda_{1},n_{1},\lambda_{2},n_{2}}(\xi_{1},\xi_{2})$ vanishes. Under these conditions, the Kane fermion decays into two independent Dirac particles as in the symmetric case of $V_1=V_2$ discussed earlier.

Since the fermion decay into two Dirac particles in the (3+1)-band model occurs at any values of $M_2$ and $0<\theta<\pi/2$, one can also conclude that the flat band described by $\mathcal{\hat{H}}_{6\times6}(\mathbf{k},\theta)$ in Eq.~(\ref{Keq:1}) in the limiting case of $M_2\rightarrow{\infty}$ becomes fully decoupled from the conical bands when magnetic field approaches its critical value. This fact in a sense justifies the use of semiclassical approximation within the three-band model (see Sec.~\ref{Sec:Problem}), which ignores the inter-band coupling. Although the critical magnetic field calculated within the semiclassical approximation is independent from orientation of $\mathcal{B}$; this approach may however qualitatively describe the evolution of Landau levels if $\mathcal{E}$ is used as a fitting parameter. This particularly explains a good agreement between the Landau level transition energies calculated on the basis of Eq.~(\ref{Keq:6}) and experimental results on magneto-absorption of InSb observed by Zawadzki~\emph{et~al.}~\cite{Flds29}. Note that in Ref.~\cite{Flds29}, $\mathcal{E}\parallel[111]$, while the orientation of $\mathcal{B}\perp\mathcal{E}$ with respect to the main crystallographic axes were unknown that is different from two cases of $\mathcal{E}\parallel[100]$ considered in Sec.~\ref{Sec:Lorentz}.

\section{\label{Sec:Sum} Summary}
In conclusion, by taking into account an additional conduction $\Gamma_{7c}$ band resulting in finite curvature of the heavy-hole band in zinc-blende crystals,
we unequivocally identify the Kane fermions as complex spin-1/2 partiles composed of two \emph{mutually hybridized} Dirac fermions. The latter allows directly to apply the Lorentz transformation for the theoretical investigation of the Landau levels collapse of Kane fermions in crossed electric and magnetic fields. We have found that increasing of the electric field first leads to the decay of the Kane fermion into two independent Dirac particles. Then, the Landau levels of the decayed particles collapse when their drift velocities $V_d$ achieve $V^{*}_d$ lying between \emph{half} and \emph{whole} values of the ``effective speed of the light''. The latter strongly depends on the orientation of the electric and magnetic fields with respect to the main crystallographic axes of the zinc-blende crystals. This is a distinctive feature of Kane fermions, since the Landau levels of conventional Dirac fermions collapse occurs when $V_d$ approaches the ``effective speed of the light''. Our results pave the way for deep understanding of pseudo-relativistic effects in narrow-gap semiconductors arising in crossed electric and magnetic fields.

\begin{acknowledgments}
This work was partially supported by the Foundation for Polish Science: the IRAP program (Grant No. MAB/2018/9, project CENTERA), by CNRS through IRP ``TeraMIR'' and by the French Agence Nationale pour la Recherche (``Dirac 3D'' and ``Colector'' projects).
\end{acknowledgments}

\appendix
\section{\label{sec:A2} (3+1)-band model in cubic semiconductors}
The most accurate description of the band structure of diamond and zinc-blende semiconductors at the energies close to the fundamental gap can be performed by taking directly into consideration the $\Gamma_{8c}$, $\Gamma_{7c}$, $\Gamma_{6c}$, $\Gamma_{8v}$ and $\Gamma_{7v}$ bands, resulting in the 14$\times$14 \textbf{k$\cdot$p} Hamiltonian~\cite{q1,q2a,q2}.
In this Hamiltonian, the non-zero curvature of the heavy-hole band stems from the momentum matrix element $Q$ between the valence $\Gamma_{8v}+\Gamma_{7v}$ and conduction $\Gamma_{8c}+\Gamma_{7c}$ bands~\cite{q2a,q2}.

\begin{table*}[!]
\caption{\label{tab:1} Band parameters of bilk Hg$_{1-x}$Cd$_{x}$Te crystals at $T=2$~K in the absence of biaxial strain ($\delta_{\epsilon}=0$) used in the (3+1)-band \textbf{k$\cdot$p} model within the ``symmetric approximation''. Experimental values of $m^{\text{[100]}}_{hh}$ and $M_1$ were calculated on the basis of material parameters provided in Ref.~\cite{q5}. Other parameters were obtained on the basis of Eq.~(\ref{A2eq:5}) by assuming $P=Q$.}
\begin{ruledtabular}
\begin{tabular}{c|c|c|c|c|c|c|c|c}
x & ${m^{\text{[100]}}_{hh}/m_0}_{\text{exp}}$ & $P_{\text{exp}}$~(eV$\cdot${\AA}) & $E_{7c}-E_{8v}$~(eV) & $V_1=V_2$~(eV$\cdot${\AA}) & $C_1$~(eV) & $C_2$~(eV) & $M_1$~(eV) & $M_2$~(eV) \\
\hline
0.00 & 0.323 & 8.46~\cite{q5,s2} & $3.5$--$4.8$~\cite{q10,q12} & 6.91 & 0.000 & 2.173 & -0.151 & 2.021\\
0.10 & 0.334 & 8.46~\cite{q5,s2} & ----- & 6.91 & 0.000 & 2.156 & -0.062 & 2.094\\
0.14 & 0.339 & 8.46~\cite{q5,s2} & ----- & 6.91 & 0.000 & 2.150 & -0.026 & 2.124\\
$x_c\simeq$0.168 & 0.342 & 8.46~\cite{q5,s2} & ----- & 6.91 & 0.000 & 2.146 & 0 & 2.146\\
0.20 & 0.347 & 8.46~\cite{q5,s2} & ----- & 6.91 & 0.000 & 2.143 & 0.029 & 2.171\\
0.25 & 0.353 & 8.46~\cite{q5,s2} & ----- & 6.91 & 0.000 & 2.138 & 0.075 & 2.212\\
0.30 & 0.360 & 8.46~\cite{q5,s2} & ----- & 6.91 & 0.000 & 2.134 & 0.121 & 2.255\\
0.35 & 0.367 & 8.46~\cite{q5,s2} & ----- & 6.91 & 0.000 & 2.132 & 0.168 & 2.299
\end{tabular}
\end{ruledtabular}
\end{table*}

Before going further, we first note that the splitting at the $\Gamma$ point of the Brillouin zone between $\Gamma_{8c}$ and $\Gamma_{7c}$, as well as between $\Gamma_{8v}$ and $\Gamma_{7v}$ is caused by the spin-orbit interaction. The latter means that if one considers the case of large spin-orbit interaction, which is a good approximation for narrow-gap HgCdTe~\cite{s1,s2} and InSb~\cite{Flds29,Flds30} semiconductors, in addition to the low-lying valence $\Gamma_{7v}$ band, one should also exclude the high-lying conduction $\Gamma_{8c}$ band. Under these assumptions, the Hamiltonian for the $\Gamma_{7c}$, $\Gamma_{6c}$ and $\Gamma_{8v}$ bands can be presented in the block form:
\begin{equation}
\label{A2eq:1}
\mathcal{H}^{(\mathbf{3+1})}_{8\times8}(\mathbf{k})=\begin{pmatrix}
\mathcal{H}_{7c7c} & \mathcal{H}_{7c6c} & \mathcal{H}_{7c8v} \\
\mathcal{H}_{6c7c} & \mathcal{H}_{6c6c} & \mathcal{H}_{6c8v} \\
\mathcal{H}_{8v7c} & \mathcal{H}_{8v6c} & \mathcal{H}_{8v8v} \end{pmatrix}.
\end{equation}
Neglecting quadratic terms representing the remote bands contribution~\cite{q1,q2a,q2} and the small terms arising due to the absence of the inversion center in the unit cell of zinc-blende semiconductors~\cite{q2}, the blocks in Eq.~(\ref{A2eq:1}) are written as
\begin{gather}
\label{A2eq:2}
\mathcal{H}_{7c7c}=E_{7c}\sigma_0,\nonumber\\
\mathcal{H}_{7c6c}=\mathcal{H}^{\dag}_{6c7c}=0,\nonumber\\
\mathcal{H}_{7c8v}=\mathcal{H}_{8v7c}^{\dag}=-2Q\left(T_{yz}k_x+T_{zx}k_y+T_{xy}k_z\right),\nonumber\\
\mathcal{H}_{6c6c}=E_{6c}\sigma_0,\nonumber\\
\mathcal{H}_{6c8v}=\mathcal{H}_{8v6c}^{\dag}=\sqrt{3}P\left(T_{x}k_x+T_{y}k_y+T_{z}k_z\right),\nonumber\\
\mathcal{H}_{8v8v}=E_{8v}\begin{pmatrix}
\sigma_0 & 0 \\ 0 & \sigma_0\end{pmatrix},
\end{gather}
where $P$ is the momentum matrix element between the $\Gamma_{6c}$ and $\Gamma_{8v}$ bands, the symbol ``$\dag$'' represents Hermitian conjugation; $T_{a}$ and $T_{ab}$ are the matrices defined in Ref.~\cite{q3}; $E_{7c}$, $E_{6c}$ and $E_{8v}$ correspond to the energies of the $\Gamma_{7c}$, $\Gamma_{6c}$ and $\Gamma_{8v}$ bands, respectively. In Eq.~(\ref{A2eq:2}2),
the axes are assumed to be oriented as follows: $x\parallel[100]$, $y\parallel[010]$, and $z\parallel[001]$. Note that the given form of $\mathcal{H}_{8v8v}$ in Eq.~(\ref{A2eq:2}2) represents the bulk semiconductor in the absence of uniaxial strain. For the latter, one should introduce an additional gap at $\mathbf{k}=0$ between the light- and heavy-hole bands~\cite{q5}.

The Hamiltonian $\mathcal{H}^{(\mathbf{3+1})}_{8\times8}(\mathbf{k})$ in Eq.~(\ref{A2eq:1}) allows one to go beyond the flat-band approximation for the heavy holes (see Fig.~(\ref{Fig:1}) and obtain a more realistic description of the band structure for narrow-gap HgCdTe and InSb semiconductors at the energies close to the fundamental gap. One may argue that taking into account the high-lying conduction $\Gamma_{7c}$ band and simultaneously neglecting the valence $\Gamma_{7v}$ band is an excess of the accuracy of this model. However, the $\Gamma_{7v}$ band has no effect on the heavy-hole mass~\cite{q4}, while taking the $\Gamma_{7c}$ band into account yields the non-zero band curvature of the heavy holes. This can be directly demonstrated by projecting $\mathcal{H}^{(\mathbf{3+1})}_{8\times8}(\mathbf{k})$ on the $\Gamma_{6c}+\Gamma_{8v}$ subspace, which results in the square corrections to the three-band Hamiltonian arising in the diagonal $\mathcal{H}^{(\text{\textbf{3-band}})}_{8v8v}$ block:
\begin{multline}
\label{A2eq:3}
\mathcal{H}^{(\text{\textbf{3-band}})}_{8v8v}=E_{8v}\begin{pmatrix}
\sigma_0 & 0 \\ 0 & \sigma_0\end{pmatrix}+\\
+\dfrac{\hbar^2}{2m_0}\left[-\left(\gamma_1+\dfrac{5}{2}\gamma_2\right)\mathbf{k}^2+2\gamma_2\left(\mathbf{J}\cdot\mathbf{k}\right)^2\right]+\\
+\dfrac{2\hbar^2}{m_0}\left(\gamma_3-\gamma_2\right) \biggl[\left\{J_xJ_y\right\}k_xk_y
+\left\{J_yJ_z\right\}k_yk_z+\\
+\left\{J_zJ_x\right\}k_zk_x\biggr],
\end{multline}
where $m_0$ is free-electron mass, $\mathbf{J}$ is the vector composed of the matrices of the angular momentum $3/2$, $\left\{J_{a}J_{b}\right\}=(J_{a}J_{b}+J_{b}J_{a})/2$, and $\gamma_1$, $\gamma_2$, and $\gamma_3$ are the \emph{effective} Luttinger parameters for the three-band \textbf{k$\cdot$p} model defined as
\begin{gather}
\label{A2eq:4}
\gamma_1=\dfrac{1}{3}\dfrac{2m_0}{\hbar^2}\dfrac{Q^2}{E_{7c}-E_{8v}},\nonumber\\
\gamma_2=-\dfrac{1}{6}\dfrac{2m_0}{\hbar^2}\dfrac{Q^2}{E_{7c}-E_{8v}},\nonumber\\
\gamma_3=\dfrac{1}{6}\dfrac{2m_0}{\hbar^2}\dfrac{Q^2}{E_{7c}-E_{8v}}.
\end{gather}
As clear, Eq.~(\ref{A2eq:3}) represents nothing but the square corrections to the three-band \textbf{k$\cdot$p} Hamiltonian for the $\Gamma_{6c}$ and $\Gamma_{8v}$ bands, resulting in the non-zero curvature of the heavy-hole subband. The latter is characterized by anisotropic effective mass $m_{hh}$ defined by the Luttinger parameters~\cite{q7}. For instance, on the basis of the effective mass along [100] crystallographic direction
\begin{equation}
\label{A2eq:5}
\dfrac{m_0}{m^{\text{[100]}}_{hh}}=\gamma_1-2\gamma_2=\dfrac{2}{3}\dfrac{2m_0}{\hbar^2}\dfrac{Q^2}{E_{7c}-E_{8v}}. 
\end{equation}
one can override parameters $Q$ and $E_{7c}-E_{8v}$ by using the experimental values of $\gamma_1$ and $\gamma_2$.


The main advantage of $\mathcal{H}^{(\mathbf{3+1})}_{8\times8}(\mathbf{k})$ is its linearity in $\mathbf{k}$, which plays \emph{a key role in understanding of relativistic properties} of Kane fermions. By using the unitary transformation, Hamiltonian can be presented in the form of two coupled Dirac Hamiltonians:

\begin{widetext}
\begin{equation}
\label{A2eq:6}
\mathcal{\hat{H}}^{(\mathbf{3+1})}_{8\times8}(\mathbf{k})=
\begin{pmatrix}
C_1+M_1 & 0 & V_{1}k_{z} & \dfrac{V_{1}}{2}k_{-} & 0 & 0 & 0 & \dfrac{\sqrt{3}V_{1}}{2}k_{+} \\
0 & C_1+M_1 & \dfrac{V_{1}}{2}k_{+} & -V_{1}k_{z} & 0 & 0 & \dfrac{\sqrt{3}V_{1}}{2}k_{-} & 0 \\
V_{1}k_{z} & \dfrac{V_{1}}{2}k_{-} & C_1-M_1 & 0 & 0 & -\dfrac{\sqrt{3}V_{2}}{2}k_{+} & 0 & 0 \\
\dfrac{V_{1}}{2}k_{+} & -V_{1}k_{z} & 0 & C_1-M_1 & -\dfrac{\sqrt{3}V_{2}}{2}k_{-} & 0 & 0 & 0 \\
0 & 0 & 0 & -\dfrac{\sqrt{3}V_{2}}{2}k_{+} & C_2+M_2 & 0 & V_{2}k_{z} & \dfrac{V_{2}}{2}k_{-} \\
0 & 0 & -\dfrac{\sqrt{3}V_{2}}{2}k_{-} & 0 & 0 & C_2+M_2 & \dfrac{V_{2}}{2}k_{+} & -V_{2}k_{z} \\
0 & \dfrac{\sqrt{3}V_{1}}{2}k_{+} & 0 & 0 & V_{2}k_{z} & \dfrac{V_{2}}{2}k_{-} & C_2-M_2 & 0 \\
\dfrac{\sqrt{3}V_{1}}{2}k_{-} & 0 & 0 & 0 & \dfrac{V_{2}}{2}k_{+} & -V_{2}k_{z} & 0 & C_2-M_2
\end{pmatrix},
\end{equation}
where $k_{\pm}=k_x{\pm}ik_y$, $V_1=\sqrt{2/3}P$, $V_2=\sqrt{2/3}Q$ and $C_{1,2}$, $M_{1,2}$ are defined from the following equations:
\begin{gather}
\label{A2eq:7}
C_1+M_1=E_{6c},~~~~~~~
C_1-M_1=E_{8v},\nonumber\\
C_2+M_2=E_{7c},~~~~~~~
C_2-M_2=E_{8v}.
\end{gather}
The last expression determines the position of the heavy-hole band at $\mathbf{k}=0$. For the case of uniaxial strain or Cd$_3$As$_2$, $C_2-M_2=E_{8v}+\delta_{\epsilon}$, where $\delta_{\epsilon}$ represents the gap between the light- and heavy-hole subbands~\cite{q5}.

For better understanding the coupling origin between two $4\times4$ Dirac blocks, we further introduce ``hybridization angle'' $\theta$ describing the mixing between light- and heavy-hole branches in the $\Gamma_{8v}$ band~\cite{q6}. Thus, the Hamiltonian $\mathcal{\hat{H}}^{(\mathbf{3+1})}_{8\times8}(\mathbf{k})$ in Eq.~(\ref{A2eq:6}) is generalized as:
\begin{equation}
\label{A2eq:7}
\mathcal{\hat{H}}^{(\mathbf{3+1})}_{8\times8}(\mathbf{k})=
\begin{pmatrix}
C_1+M_1 & 0 & V_{1}k_{z} & V_{1}k_{-}\cos{\theta} & 0 & 0 & 0 & V_{1}k_{+}\sin{\theta} \\
0 & C_1+M_1 & V_{1}k_{+}\cos{\theta} & -V_{1}k_{z} & 0 & 0 & V_{1}k_{-}\sin{\theta} & 0 \\
V_{1}k_{z} & V_{1}k_{-}\cos{\theta} & C_1-M_1 & 0 & 0 & -V_{2}k_{+}\sin{\theta} & 0 & 0 \\
V_{1}k_{+}\cos{\theta} & -V_{1}k_{z} & 0 & C_1-M_1 & -V_{2}k_{-}\sin{\theta} & 0 & 0 & 0 \\
0 & 0 & 0 & -V_{2}k_{+}\sin{\theta} & C_2+M_2 & 0 & V_{2}k_{z} & V_{2}k_{-}\cos{\theta} \\
0 & 0 & -V_{2}k_{-}\sin{\theta} & 0 & 0 & C_2+M_2 & V_{2}k_{+}\cos{\theta} & -V_{2}k_{z} \\
0 & V_{1}k_{+}\sin{\theta} & 0 & 0 & V_{2}k_{z} & V_{2}k_{-}\cos{\theta} & C_2-M_2 & 0 \\
V_{1}k_{-}\sin{\theta} & 0 & 0 & 0 & V_{2}k_{+}\cos{\theta} & -V_{2}k_{z} & 0 & C_2-M_2
\end{pmatrix},
\end{equation}
\end{widetext}
where $\theta=\pi/3$ corresponds to the case of Kane fermions~\cite{q6} in narrow-gap HgCdTe and InSb semiconductors, while $\theta=0$ represents the case of two independent Dirac particles.


Finally, the Hamiltonian in Eq.~(\ref{A2eq:7}) can be further simplified for HgCdTe crystals by setting $V_{1}{\simeq}V_{2}$. As seen from Tab.~\ref{tab:1}, only the Luttinger parameters and the values of $P$ are well-known for HgCdTe semiconductors, while the scatter of the energies $E_{7c}-E_{8v}$ for HgTe is quite large in comparison with the band-gap $E_{6c}-E_{8v}=2M_1$~~\cite{q5}.
Converting the scatter range of $E_{7c}-E_{8v}$ into the range of the $Q$ values by means of Eq.~(\ref{A2eq:5}) with the known mass $m^{\text{[100]}}_{hh}$, one can find that $Q$ should change between $7.87$~eV$\cdot${\AA} and $9.22$~eV$\cdot${\AA} for HgTe. Since the $Q$ ranges overlap with the $P$ values in Tab.~\ref{tab:1}, one can indeed apply ``symmetric approximation'' for $\mathcal{\hat{H}}^{(\mathbf{3+1})}_{8\times8}(\mathbf{k})$ in narrow-gap HgCdTe semiconductors by setting $P{\simeq}Q$ and therefore $V_{1}{\simeq}V_{2}$.
Thus, in the calculations on the basis of (3+1)-band Hamiltonian, it is preferable to use $P=Q=8.46$~eV$\cdot${\AA}, resulting in $E_{7c}-E_{8v}=4.042$~eV for HgTe.
The values of $E_{7c}-E_{8v}$ for Hg$_{x}$Cd$_{1-x}$Te alloys can be calculated by assuming the linear variation of the Luttinger parameters with $x$~\cite{q5} and $P=Q=8.64$~eV$\cdot${\AA} independent of $x$~\cite{s2}.

In the ``symmetric approximation'', the Hamiltonian in Eq.~(\ref{A2eq:7}) can be written in more compact form:
\begin{multline}
\label{A2eq:8}
\mathcal{\hat{H}}^{(\mathbf{3+1})}_{8\times8}(\mathbf{k})= ~~~~~~~~\\
=\begin{pmatrix}
\mathcal{\hat{H}}^{(1)}_{D}(\mathbf{k}) & V\sin{\theta}\left(k_{x}\beta_{y}-k_{y}\beta_{x}\right) \\
V\sin{\theta}\left(k_{y}\beta_{x}-k_{x}\beta_{y}\right) & \mathcal{\hat{H}}^{(2)}_{D}(\mathbf{k})
\end{pmatrix},
\end{multline}
where $\mathcal{\hat{H}}^{(1,2)}_{D}(\mathbf{k})$ are
\begin{multline}
\label{A2eq:9}
\mathcal{\hat{H}}^{(1,2)}_{D}(\mathbf{k})=
C_{1,2}\mathcal{I}_4+M_{1,2}\alpha_0+\\
+V\cos{\theta}\left(k_{x}\alpha_{x}+k_{y}\alpha_{y}\right)+
Vk_z\alpha_z
\end{multline}
where $\mathcal{I}_4$ is a $4\times4$ identity matrix and the $\alpha$ and $\beta$ matrices are related to the Pauli matrices as
\begin{gather}
\label{A2eq:10}
\alpha_0=\begin{pmatrix}
\sigma_0 &0 \\
0 & -\sigma_0
\end{pmatrix},~~~~~~~
\alpha_{x,y,z}=\begin{pmatrix}
0 & \sigma_{x,y,z} \\
\sigma_{x,y,z} & 0
\end{pmatrix},~~~~~\nonumber\\
\beta_0^{(z)}=\begin{pmatrix}
i\sigma_z &0 \\
0 & -i\sigma_z
\end{pmatrix},~~~~~
\beta_{x,y}=\begin{pmatrix}
0 & \sigma_{y,x} \\
-\sigma_{y,x} & 0
\end{pmatrix}.
\end{gather}
The expediency of introducing the $\beta$-matrices in this form is due to their transformation upon Lorentz boost along the $y$ axis, $L^{(y)}_{4\times4}(\alpha)=\exp\left(\alpha_y\cdot{\alpha}/{2}\right)$:
\begin{gather}
\label{A2eq:11}
L^{(y)}_{4\times4}(\alpha)\beta_0^{(z)}L^{(y)}_{4\times4}(\alpha)=\beta_0^{(z)}\cosh\alpha+\beta_y\sinh\alpha,\nonumber\\
L^{(y)}_{4\times4}(\alpha)\beta_{y}L^{(y)}_{4\times4}(\alpha)=\beta_y\cosh\alpha+\beta_0^{(z)}\sinh\alpha,\nonumber\\
L^{(y)}_{4\times4}(\alpha)\beta_{x}L^{(y)}_{4\times4}(\alpha)=\beta_{x}.
\end{gather}
Note the indices order in the definition of $\beta_{x}$ and $\beta_{y}$.

\section{\label{sec:A1} (3+1)-band model for arbitrary orientations of electric and magnetic fields}
\begin{figure}
\includegraphics [width=0.94\columnwidth, keepaspectratio] {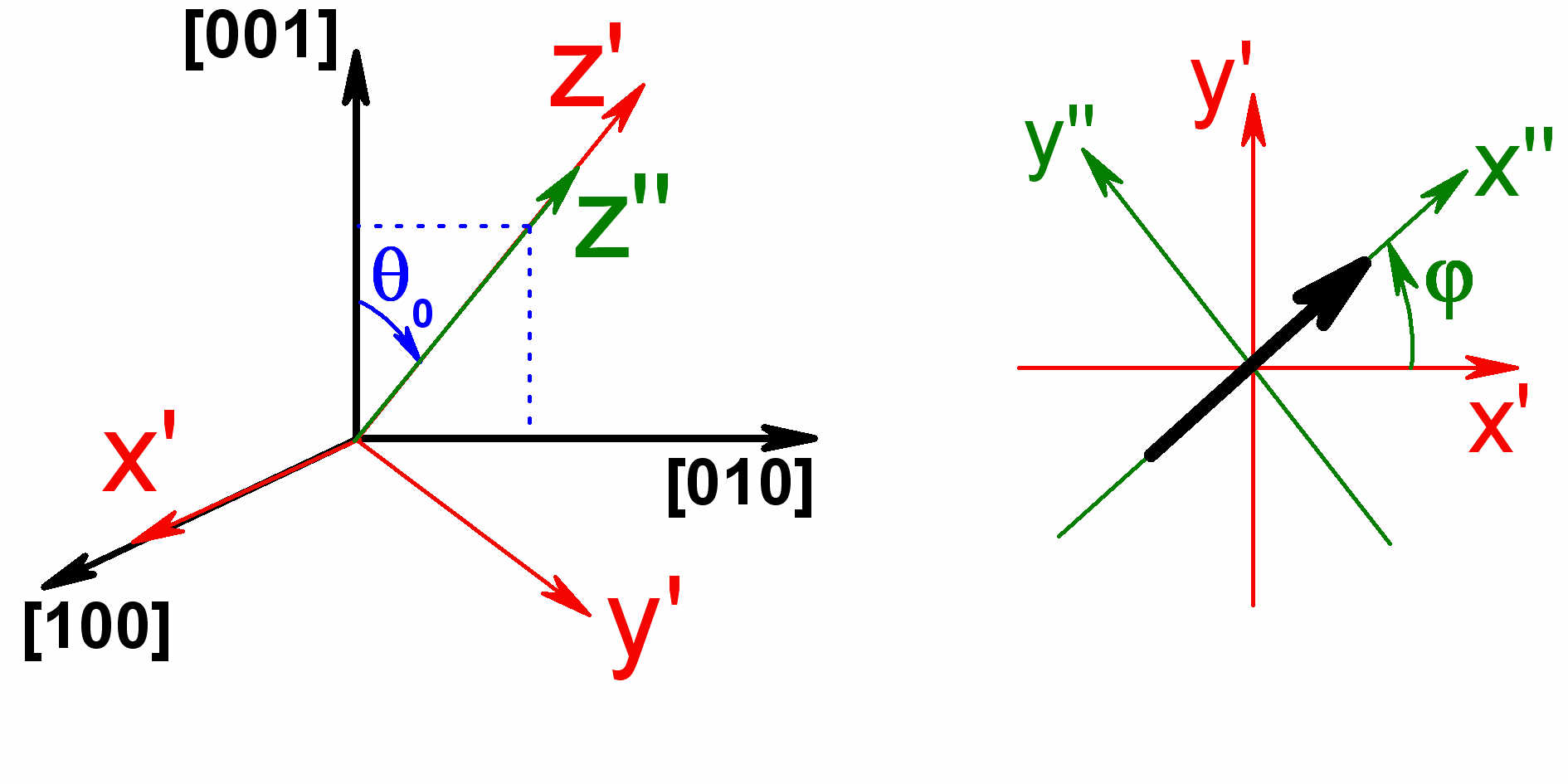} 
\caption{\label{Fig:Ap1} (Color online) Left panel: mutual orientation of two Cartesian axes ($x'$, $y'$, $z'$) and ($x''$, $y''$, $z''$) with respect to the main crystallographic axes. Right panel: the orientation of electric field $\mathcal{E}$ (shown by black arrow) in the Cartesian axes.}
\end{figure}

Up to now, the $x$, $y$ and $z$ axes are assumed to be oriented along the main crystallographic directions, namely [100] and [010] and [001], respectively. Further, we consider more general case when the magnetic field $\mathcal{B}$ is oriented along new $z'$ axis deflected at angle $\theta_0$ to the [001] direction, while electric field $\mathcal{E}$ is tilted at angle $\phi$ from the [100] direction as shown in Fig.~\ref{Fig:Ap1}.

To write the Hamiltonian in Eq.~(\ref{A2eq:8}) in the new coordinate system, one has to rotate the electron momentum around the [100] axis, and then around the new axis $z'$ in accordance with the transformations

\begin{equation}
\label{A1eq:1}
\begin{pmatrix}
k_{x} \\ k_{y} \\ k_{z}
\end{pmatrix}=
\begin{pmatrix}
1 & 0 & 0 \\
0 & \cos{\theta_0} & -\sin{\theta_0} \\
0 & \sin{\theta_0} & \cos{\theta_0}
\end{pmatrix}
\begin{pmatrix}
k_{x'} \\ k_{y'} \\ k_{z'}
\end{pmatrix}
\end{equation}
and
\begin{equation}
\label{A1eq:2}
\begin{pmatrix}
k_{x'} \\ k_{y'} \\ k_{z'}
\end{pmatrix}=
\begin{pmatrix}
\cos{\phi} & -\sin{\phi} & 0 \\
\sin{\phi} & \cos{\phi} & 0 \\
0 & 0 & 1
\end{pmatrix}
\begin{pmatrix}
k_{x''} \\ k_{y''} \\ k_{z''}
\end{pmatrix}.
\end{equation}

Simultaneously with the transition from ($k_{x}$, $k_{y}$, $k_{z}$) to ($k_{x''}$, $k_{y''}$, $k_{z''}$), one should also apply a unitary transformation to the Hamiltonian~(\ref{A2eq:8}):
\begin{equation}
\label{A1eq:3}
{\mathcal{\hat{H''}}}^{(\mathbf{3+1})}_{8\times8}({\mathbf{k}''})=
U^{-1}_z(\phi)U_x^{-1}(\theta_0)\mathcal{\hat{H}}^{(\mathbf{3+1})}_{8\times8}(\mathbf{k})U_x(\theta_0)U_z(\phi),
\end{equation}
where $U_x(\theta_0)$ and $U_z(\phi)$ are defined as
\begin{gather*}
U_x(\theta_0)=\mathcal{I}_4{\otimes}\exp\left(i\sigma_x\cdot{\theta_0}/{2}\right),\nonumber\\
U_z(\phi)=\mathcal{I}_4{\otimes}\exp\left(i\sigma_z\cdot{\phi}/{2}\right).
\end{gather*}

After all the calculations, the Hamiltonian ${\mathcal{\hat{H''}}}^{(\mathbf{3+1})}_{8\times8}({\mathbf{k}''})$ in the new coordinate system takes the form
\begin{equation}
\label{A1eq:4}
{\mathcal{\hat{H''}}}^{(\mathbf{3+1})}_{8\times8}({\mathbf{k}''})=\begin{pmatrix}
\mathcal{\hat{H''}}^{(1)}_{D}(\mathbf{k}'') & V\sin{\theta}\mathcal{\hat{K''}}(\mathbf{k}'') \\
-V\sin{\theta}\mathcal{\hat{K''}}(\mathbf{k}'') & \mathcal{\hat{H''}}^{(2)}_{D}(\mathbf{k''})
\end{pmatrix},
\end{equation}
where the blocks $\mathcal{\hat{H''}}^{(1,2)}_{D}(\mathbf{k}'')$ and $\mathcal{\hat{K''}}(\mathbf{k}'')$ are written as
\begin{widetext}
\begin{multline}
\label{A1eq:5}
\mathcal{\hat{H''}}^{(1,2)}_{D}(\mathbf{k}'')=
C_{1,2}\mathcal{I}_4+M_{1,2}\alpha_0+V\cos{\theta}\left(k_{x''}\alpha_{x''}+k_{y''}\alpha_{y''}\right)+
Vk_{z''}\alpha_{z''}+\\
+\left(1-\cos{\theta}\right)Vk_{x''}\left(\alpha_{x''}\sin^2{\theta_0}\sin^2{\phi}+\alpha_{y''}\dfrac{\sin^2{\theta_0}\sin{2\phi}}{2}
+\alpha_{z''}\dfrac{\sin{2\theta_0}\sin{\phi}}{2}\right)+\\
+\left(1-\cos{\theta}\right)Vk_{y''}\left(\alpha_{x''}\dfrac{\sin^2{\theta_0}\sin{2\phi}}{2}+
\alpha_{y''}\sin^2{\theta_0}\cos^2{\phi}+\alpha_{z''}\dfrac{\sin{2\theta_0}\cos{\phi}}{2}\right)+\\
+\left(1-\cos{\theta}\right)Vk_{z''}\left(\alpha_{x''}\dfrac{\sin{2\theta_0}\sin{\phi}}{2}+
\alpha_{y''}\dfrac{\sin{2\theta_0}\cos{\phi}}{2}-\alpha_{z''}\sin^2{\theta_0}\right)
\end{multline}
and
\begin{multline}
\label{A1eq:6}
\mathcal{\hat{K''}}(\mathbf{k}'')=
k_{x''}\left(-\beta_{x''}\dfrac{\left(1+\cos^2{\theta_0}\right)\sin{2\phi}}{2}
+\beta_{y''}\left(\cos^2{\phi}-\cos^2{\theta_0}\sin^2{\phi}\right)
+\beta_{z''}\dfrac{\sin{2\theta_0}\sin{\phi}}{2}\right)+\\
+k_{y''}\left(\beta_{x''}\left(\sin^2{\phi}-\cos^2{\theta_0}\cos^2{\phi}\right)
-\beta_{y''}\dfrac{\left(1+\cos^2{\theta_0}\right)\sin{2\phi}}{2}
+\beta_{z''}\dfrac{\sin{2\theta_0}\cos{\phi}}{2}\right)+\\
+k_{z''}\left(\beta_{x''}\dfrac{\sin{2\theta_0}\cos{\phi}}{2}+
\beta_{y''}\dfrac{\sin{2\theta_0}\sin{\phi}}{2}-\beta_{z''}\sin^2{\theta_0}\right),
\end{multline}
where we have additionally introduced $\beta_{z''}$ as
\begin{equation}
\label{A1eq:7}
\beta_{z''}=\begin{pmatrix}
0 & \sigma_{z''} \\
-\sigma_{z''} & 0
\end{pmatrix}.
\end{equation}
\end{widetext}
Further and in the main text, we omit the prime marks keeping in mind that orientation of new $x$, $y$ and $z$ axes does not coincide with the main crystallographic directions in the most general case.

If $z$ is oriented along the [001] direction ($\theta_0=0$, see Sec.~\ref{Sec:LorentzA}), $\mathcal{\hat{H}}^{(1,2)}_{D}(\mathbf{k})$ takes the form of Eq.~(\ref{A2eq:9}), while $\mathcal{\hat{K}}(\mathbf{k})$ is written as
\begin{equation}
\label{A1eq:8}
\mathcal{\hat{K}}(\mathbf{k})=\left(k_{x}\beta_{y}-k_{y}\beta_{x}\right)\cos{2\phi}-
\left(k_{x}\beta_{x}+k_{y}\beta_{y}\right)\sin{2\phi}.
\end{equation}
In this case, the $x$ and $y$ orientations relative to the main crystallographic directions are determined by the angle $\phi$ as shown in Fig.~\ref{Fig:Ap1}. For the particular case of $x\parallel[100]$, $y\parallel[00\bar{1}]$ and $z\parallel[010]$ ($\theta_0=\pi/2$ and $\phi=0$, see Sec.~\ref{Sec:LorentzB}),
$\mathcal{\hat{H}}^{(1,2)}_{D}(\mathbf{k})$ and $\mathcal{\hat{K}}(\mathbf{k})$ have the form
\begin{multline}
\label{A1eq:9}
\mathcal{\hat{H}}^{(1,2)}_{D}(\mathbf{k})=
C_{1,2}\mathcal{I}_4+M_{1,2}\alpha_0+V\cos{\theta}k_{x}\alpha_{x}\\
+Vk_{y}\alpha_{y}+V\cos{\theta}k_{z}\alpha_{z},
\end{multline}
and
\begin{equation}
\label{A1eq:10}
\mathcal{\hat{K}}(\mathbf{k})=k_{x}\beta_{y}-k_{z}\beta_{z}.
\end{equation}

\section{\label{sec:A3} Collapse of Landau levels at arbitrary orientations of electric and magnetic fields}
Let us now find the drift velocity $V_d$ for an arbitrary orientation of magnetic field (and electric field $\mathcal{E}\perp\mathcal{B}$), at which the collapse of the Landau levels occurs. Assuming the orientation of electric field $\mathcal{E}$ in the $x$ direction and magnetic field $\mathcal{B}$ in the $z$ direction, the Schr\"{o}dinger equation with the Hamiltonian~(\ref{A1eq:4}) characterized by the arbitrary orientation of the Cartesian axes with respect to the main crystallographic directions is written as
\begin{multline}
\label{A3eq:1}
\bigg[\mathcal{\hat{H}}^{(\mathbf{3+1})}_{8\times8}\left(\hat{k}_x,k_{y}+\dfrac{x}{a_B^2},k_z\right)+\\
+\left(-i\hbar\dfrac{\partial}{\partial{t}}+e\mathcal{E}{x}\right)\mathcal{I}_8\bigg]|\Psi_{8\times8}\rangle=0,
\end{multline}
where $\hat{k}_x=-i{\partial}/{\partial{x}}$.

As shown in Sec.~\ref{Sec:Lorentz}, the Landau level collapse is defined by the parameter $\alpha^{*}$ of the Lorentz boost along the $y$ axis. As clear, the elimination of electric field $\mathcal{E}$ in the moving reference frame is caused by the terms proportional to $k_{y}\alpha_{y}$ in the Dirac blocks~(\ref{A1eq:5}), namely by $\Lambda(\theta,\theta_0,\phi){V}k_{y}\alpha_{y}$, where
\begin{equation}
\label{A3eq:2}
\Lambda(\theta,\theta_0,\phi)=\cos{\theta}+\sin^2{\theta_0}\cos^2{\phi}\left(1-\cos{\theta}\right).
\end{equation}
Then, introducing the ``time-momentum'' operator ${k_t}$ as ${V}\Lambda(\theta,\theta_0,\phi){k_t}=-i\hbar\partial/\partial{t}$, one can show that the term ${k_t}\mathcal{I}_4+k_{y}\alpha_{y}$ in two Diagonal blocks of Eq.~(\ref{A3eq:1}) remains invariant upon the Lorentz transformation, i.e.
\begin{equation}
\label{A3eq:3}
\left({k_t}\mathcal{I}_4+k_{y}\alpha_{y}\right){\longrightarrow}\\
\left({k'_t}\mathcal{I}_4+k'_{y}\alpha_{y}\right).
\end{equation}
if $({k_t},{k_y})$ are connected with $({k'_t},{k'_y})$ by Eq.~(\ref{L1eq:8}).

On the contrary, the terms proportional to $x$ in the Diagonal blocks of Eq.~(\ref{A3eq:1})
\begin{equation*}
e\mathcal{E}{x}\mathcal{I}_4+\Lambda(\theta,\theta_0,\phi)V\dfrac{x}{a_B^2}\alpha_{y}
\end{equation*}
 are transformed as
\begin{multline}
\label{A3eq:4}
\left(\beta^*\mathcal{I}_4+\alpha_{y}\right)\rightarrow
\bigg[\left(\sinh{\alpha^*}+\beta^*\cosh{\alpha^*}\right)\mathcal{I}_4+\\
+\left(\cosh{\alpha^*}+\beta^*\sinh{\alpha^*}\right)\alpha_{y}\bigg],
\end{multline}
where $\beta^*$ is introduced as
\begin{equation}
\label{A3eq:5}
\beta^*=\dfrac{1}{\Lambda(\theta,\theta_0,\phi)}\dfrac{c\hbar}{V}\dfrac{\mathcal{E}}{\mathcal{B}}.
\end{equation}
Choosing $\tanh{\alpha^*}=-\beta^*$ and $\cosh{\alpha^*}=1/\sqrt{1-{\beta^{*}}^2}$, we completely remove the terms depending on $x$ from the main diagonal of the matrix Schr\"{o}dinger equation in the moving reference frame. The latter now contains the effective magnetic field $\mathcal{B}'=\mathcal{B}\sqrt{1-{\beta^{*}}^2}$ and additional terms dependent on $\beta^{*}$ (cf.~Eqs~(\ref{L1eq:10},\ref{L2eq:5})). Solving the Schr\"{o}dinger equation in the boosted frame, in the manner described in the main text, and then passing to the original reference frame, we can conclude that the collapse of the Landau levels occurs if $\beta^{*}=1$. This condition can be also represented in terms of drift velocity $V_d=c\mathcal{E}/\mathcal{B}$ as
\begin{equation}
\label{A3eq:6}
V_d=\dfrac{V}{\hbar}\left[\cos{\theta}+\sin^2{\theta_0}\cos^2{\phi}\left(1-\cos{\theta}\right)\right].
\end{equation}


\begin{thebibliography}{48}%
\makeatletter
\providecommand \@ifxundefined [1]{%
 \@ifx{#1\undefined}
}%
\providecommand \@ifnum [1]{%
 \ifnum #1\expandafter \@firstoftwo
 \else \expandafter \@secondoftwo
 \fi
}%
\providecommand \@ifx [1]{%
 \ifx #1\expandafter \@firstoftwo
 \else \expandafter \@secondoftwo
 \fi
}%
\providecommand \natexlab [1]{#1}%
\providecommand \enquote  [1]{``#1''}%
\providecommand \bibnamefont  [1]{#1}%
\providecommand \bibfnamefont [1]{#1}%
\providecommand \citenamefont [1]{#1}%
\providecommand \href@noop [0]{\@secondoftwo}%
\providecommand \href [0]{\begingroup \@sanitize@url \@href}%
\providecommand \@href[1]{\@@startlink{#1}\@@href}%
\providecommand \@@href[1]{\endgroup#1\@@endlink}%
\providecommand \@sanitize@url [0]{\catcode `\\12\catcode `\$12\catcode
  `\&12\catcode `\#12\catcode `\^12\catcode `\_12\catcode `\%12\relax}%
\providecommand \@@startlink[1]{}%
\providecommand \@@endlink[0]{}%
\providecommand \url  [0]{\begingroup\@sanitize@url \@url }%
\providecommand \@url [1]{\endgroup\@href {#1}{\urlprefix }}%
\providecommand \urlprefix  [0]{URL }%
\providecommand \Eprint [0]{\href }%
\providecommand \doibase [0]{http://dx.doi.org/}%
\providecommand \selectlanguage [0]{\@gobble}%
\providecommand \bibinfo  [0]{\@secondoftwo}%
\providecommand \bibfield  [0]{\@secondoftwo}%
\providecommand \translation [1]{[#1]}%
\providecommand \BibitemOpen [0]{}%
\providecommand \bibitemStop [0]{}%
\providecommand \bibitemNoStop [0]{.\EOS\space}%
\providecommand \EOS [0]{\spacefactor3000\relax}%
\providecommand \BibitemShut  [1]{\csname bibitem#1\endcsname}%
\let\auto@bib@innerbib\@empty
\bibitem [{\citenamefont {Keldysh}(1964)}]{Flds1}%
  \BibitemOpen
  \bibfield  {author} {\bibinfo {author} {\bibfnamefont {L.~V.}\ \bibnamefont
  {Keldysh}},\ }\href {http://jetp.ras.ru/cgi-bin/e/index/e/18/1/p253?a=list}
  {\bibfield  {journal} {\bibinfo  {journal} {Sov. Phys. JETP}\ }\textbf
  {\bibinfo {volume} {18}},\ \bibinfo {pages} {253} (\bibinfo {year}
  {1964})}\BibitemShut {NoStop}%
\bibitem [{\citenamefont {Wolff}(1964)}]{Flds2}%
  \BibitemOpen
  \bibfield  {author} {\bibinfo {author} {\bibfnamefont {P.~A.}\ \bibnamefont
  {Wolff}},\ }\href {\doibase https://doi.org/10.1016/0022-3697(64)90128-3}
  {\bibfield  {journal} {\bibinfo  {journal} {J. Phys. Chem. Solids}\ }\textbf
  {\bibinfo {volume} {25}},\ \bibinfo {pages} {1057} (\bibinfo {year}
  {1964})}\BibitemShut {NoStop}%
\bibitem [{\citenamefont {Novoselov}\ \emph {et~al.}(2005)\citenamefont
  {Novoselov}, \citenamefont {Geim}, \citenamefont {Morozov}, \citenamefont
  {Jiang}, \citenamefont {Katsnelson}, \citenamefont {Grigorieva},
  \citenamefont {Dubonos},\ and\ \citenamefont {Firsov}}]{Flds6}%
  \BibitemOpen
  \bibfield  {author} {\bibinfo {author} {\bibfnamefont {K.~S.}\ \bibnamefont
  {Novoselov}}, \bibinfo {author} {\bibfnamefont {A.~K.}\ \bibnamefont {Geim}},
  \bibinfo {author} {\bibfnamefont {S.~V.}\ \bibnamefont {Morozov}}, \bibinfo
  {author} {\bibfnamefont {D.}~\bibnamefont {Jiang}}, \bibinfo {author}
  {\bibfnamefont {M.~I.}\ \bibnamefont {Katsnelson}}, \bibinfo {author}
  {\bibfnamefont {I.~V.}\ \bibnamefont {Grigorieva}}, \bibinfo {author}
  {\bibfnamefont {S.~V.}\ \bibnamefont {Dubonos}}, \ and\ \bibinfo {author}
  {\bibfnamefont {A.~A.}\ \bibnamefont {Firsov}},\ }\href {\doibase
  10.1038/nature04233} {\bibfield  {journal} {\bibinfo  {journal} {Nature}\
  }\textbf {\bibinfo {volume} {438}},\ \bibinfo {pages} {197} (\bibinfo {year}
  {2005})}\BibitemShut {NoStop}%
\bibitem [{\citenamefont {Krishtopenko}\ \emph
  {et~al.}(2016{\natexlab{a}})\citenamefont {Krishtopenko}, \citenamefont
  {Knap},\ and\ \citenamefont {Teppe}}]{Flds7}%
  \BibitemOpen
  \bibfield  {author} {\bibinfo {author} {\bibfnamefont {S.~S.}\ \bibnamefont
  {Krishtopenko}}, \bibinfo {author} {\bibfnamefont {W.}~\bibnamefont {Knap}},
  \ and\ \bibinfo {author} {\bibfnamefont {F.}~\bibnamefont {Teppe}},\ }\href
  {\doibase 10.1038/srep30755} {\bibfield  {journal} {\bibinfo  {journal} {Sci.
  Rep.}\ }\textbf {\bibinfo {volume} {6}},\ \bibinfo {pages} {30755} (\bibinfo
  {year} {2016}{\natexlab{a}})}\BibitemShut {NoStop}%
\bibitem [{\citenamefont {Marcinkiewicz}\ \emph {et~al.}(2017)\citenamefont
  {Marcinkiewicz}, \citenamefont {Ruffenach}, \citenamefont {Krishtopenko},
  \citenamefont {Kadykov}, \citenamefont {Consejo}, \citenamefont {But},
  \citenamefont {Desrat}, \citenamefont {Knap}, \citenamefont {Torres},
  \citenamefont {Ikonnikov}, \citenamefont {Spirin}, \citenamefont {Morozov},
  \citenamefont {Gavrilenko}, \citenamefont {Mikhailov}, \citenamefont
  {Dvoretskii},\ and\ \citenamefont {Teppe}}]{Flds8}%
  \BibitemOpen
  \bibfield  {author} {\bibinfo {author} {\bibfnamefont {M.}~\bibnamefont
  {Marcinkiewicz}}, \bibinfo {author} {\bibfnamefont {S.}~\bibnamefont
  {Ruffenach}}, \bibinfo {author} {\bibfnamefont {S.~S.}\ \bibnamefont
  {Krishtopenko}}, \bibinfo {author} {\bibfnamefont {A.~M.}\ \bibnamefont
  {Kadykov}}, \bibinfo {author} {\bibfnamefont {C.}~\bibnamefont {Consejo}},
  \bibinfo {author} {\bibfnamefont {D.~B.}\ \bibnamefont {But}}, \bibinfo
  {author} {\bibfnamefont {W.}~\bibnamefont {Desrat}}, \bibinfo {author}
  {\bibfnamefont {W.}~\bibnamefont {Knap}}, \bibinfo {author} {\bibfnamefont
  {J.}~\bibnamefont {Torres}}, \bibinfo {author} {\bibfnamefont {A.~V.}\
  \bibnamefont {Ikonnikov}}, \bibinfo {author} {\bibfnamefont {K.~E.}\
  \bibnamefont {Spirin}}, \bibinfo {author} {\bibfnamefont {S.~V.}\
  \bibnamefont {Morozov}}, \bibinfo {author} {\bibfnamefont {V.~I.}\
  \bibnamefont {Gavrilenko}}, \bibinfo {author} {\bibfnamefont {N.~N.}\
  \bibnamefont {Mikhailov}}, \bibinfo {author} {\bibfnamefont {S.~A.}\
  \bibnamefont {Dvoretskii}}, \ and\ \bibinfo {author} {\bibfnamefont
  {F.}~\bibnamefont {Teppe}},\ }\href {\doibase 10.1103/PhysRevB.96.035405}
  {\bibfield  {journal} {\bibinfo  {journal} {Phys. Rev. B}\ }\textbf {\bibinfo
  {volume} {96}},\ \bibinfo {pages} {035405} (\bibinfo {year}
  {2017})}\BibitemShut {NoStop}%
\bibitem [{\citenamefont {Krishtopenko}\ \emph {et~al.}(2018)\citenamefont
  {Krishtopenko}, \citenamefont {Ruffenach}, \citenamefont {Gonzalez-Posada},
  \citenamefont {Boissier}, \citenamefont {Marcinkiewicz}, \citenamefont
  {Fadeev}, \citenamefont {Kadykov}, \citenamefont {Rumyantsev}, \citenamefont
  {Morozov}, \citenamefont {Gavrilenko}, \citenamefont {Consejo}, \citenamefont
  {Desrat}, \citenamefont {Jouault}, \citenamefont {Knap}, \citenamefont
  {Tourni\'e},\ and\ \citenamefont {Teppe}}]{Flds9b}%
  \BibitemOpen
  \bibfield  {author} {\bibinfo {author} {\bibfnamefont {S.~S.}\ \bibnamefont
  {Krishtopenko}}, \bibinfo {author} {\bibfnamefont {S.}~\bibnamefont
  {Ruffenach}}, \bibinfo {author} {\bibfnamefont {F.}~\bibnamefont
  {Gonzalez-Posada}}, \bibinfo {author} {\bibfnamefont {G.}~\bibnamefont
  {Boissier}}, \bibinfo {author} {\bibfnamefont {M.}~\bibnamefont
  {Marcinkiewicz}}, \bibinfo {author} {\bibfnamefont {M.~A.}\ \bibnamefont
  {Fadeev}}, \bibinfo {author} {\bibfnamefont {A.~M.}\ \bibnamefont {Kadykov}},
  \bibinfo {author} {\bibfnamefont {V.~V.}\ \bibnamefont {Rumyantsev}},
  \bibinfo {author} {\bibfnamefont {S.~V.}\ \bibnamefont {Morozov}}, \bibinfo
  {author} {\bibfnamefont {V.~I.}\ \bibnamefont {Gavrilenko}}, \bibinfo
  {author} {\bibfnamefont {C.}~\bibnamefont {Consejo}}, \bibinfo {author}
  {\bibfnamefont {W.}~\bibnamefont {Desrat}}, \bibinfo {author} {\bibfnamefont
  {B.}~\bibnamefont {Jouault}}, \bibinfo {author} {\bibfnamefont
  {W.}~\bibnamefont {Knap}}, \bibinfo {author} {\bibfnamefont {E.}~\bibnamefont
  {Tourni\'e}}, \ and\ \bibinfo {author} {\bibfnamefont {F.}~\bibnamefont
  {Teppe}},\ }\href {\doibase 10.1103/PhysRevB.97.245419} {\bibfield  {journal}
  {\bibinfo  {journal} {Phys. Rev. B}\ }\textbf {\bibinfo {volume} {97}},\
  \bibinfo {pages} {245419} (\bibinfo {year} {2018})}\BibitemShut {NoStop}%
\bibitem [{\citenamefont {Krishtopenko}\ and\ \citenamefont
  {Teppe}(2018)}]{Flds9}%
  \BibitemOpen
  \bibfield  {author} {\bibinfo {author} {\bibfnamefont {S.~S.}\ \bibnamefont
  {Krishtopenko}}\ and\ \bibinfo {author} {\bibfnamefont {F.}~\bibnamefont
  {Teppe}},\ }\href {\doibase 10.1126/sciadv.aap7529} {\bibfield  {journal}
  {\bibinfo  {journal} {Sci. Adv.}\ }\textbf {\bibinfo {volume} {4}},\ \bibinfo
  {pages} {eaap7529} (\bibinfo {year} {2018})}\BibitemShut {NoStop}%
\bibitem [{\citenamefont {Krishtopenko}\ \emph {et~al.}(2019)\citenamefont
  {Krishtopenko}, \citenamefont {Desrat}, \citenamefont {Spirin}, \citenamefont
  {Consejo}, \citenamefont {Ruffenach}, \citenamefont {Gonzalez-Posada},
  \citenamefont {Jouault}, \citenamefont {Knap}, \citenamefont {Maremyanin},
  \citenamefont {Gavrilenko}, \citenamefont {Boissier}, \citenamefont {Torres},
  \citenamefont {Zaknoune}, \citenamefont {Tourni\'e},\ and\ \citenamefont
  {Teppe}}]{Flds10}%
  \BibitemOpen
  \bibfield  {author} {\bibinfo {author} {\bibfnamefont {S.~S.}\ \bibnamefont
  {Krishtopenko}}, \bibinfo {author} {\bibfnamefont {W.}~\bibnamefont
  {Desrat}}, \bibinfo {author} {\bibfnamefont {K.~E.}\ \bibnamefont {Spirin}},
  \bibinfo {author} {\bibfnamefont {C.}~\bibnamefont {Consejo}}, \bibinfo
  {author} {\bibfnamefont {S.}~\bibnamefont {Ruffenach}}, \bibinfo {author}
  {\bibfnamefont {F.}~\bibnamefont {Gonzalez-Posada}}, \bibinfo {author}
  {\bibfnamefont {B.}~\bibnamefont {Jouault}}, \bibinfo {author} {\bibfnamefont
  {W.}~\bibnamefont {Knap}}, \bibinfo {author} {\bibfnamefont {K.~V.}\
  \bibnamefont {Maremyanin}}, \bibinfo {author} {\bibfnamefont {V.~I.}\
  \bibnamefont {Gavrilenko}}, \bibinfo {author} {\bibfnamefont
  {G.}~\bibnamefont {Boissier}}, \bibinfo {author} {\bibfnamefont
  {J.}~\bibnamefont {Torres}}, \bibinfo {author} {\bibfnamefont
  {M.}~\bibnamefont {Zaknoune}}, \bibinfo {author} {\bibfnamefont
  {E.}~\bibnamefont {Tourni\'e}}, \ and\ \bibinfo {author} {\bibfnamefont
  {F.}~\bibnamefont {Teppe}},\ }\href {\doibase 10.1103/PhysRevB.99.121405}
  {\bibfield  {journal} {\bibinfo  {journal} {Phys. Rev. B}\ }\textbf {\bibinfo
  {volume} {99}},\ \bibinfo {pages} {121405} (\bibinfo {year}
  {2019})}\BibitemShut {NoStop}%
\bibitem [{\citenamefont {Hsieh}\ \emph {et~al.}(2008)\citenamefont {Hsieh},
  \citenamefont {Qian}, \citenamefont {Wray}, \citenamefont {Xia},
  \citenamefont {Hor}, \citenamefont {Cava},\ and\ \citenamefont
  {Hasan}}]{Flds11}%
  \BibitemOpen
  \bibfield  {author} {\bibinfo {author} {\bibfnamefont {D.}~\bibnamefont
  {Hsieh}}, \bibinfo {author} {\bibfnamefont {D.}~\bibnamefont {Qian}},
  \bibinfo {author} {\bibfnamefont {L.}~\bibnamefont {Wray}}, \bibinfo {author}
  {\bibfnamefont {Y.}~\bibnamefont {Xia}}, \bibinfo {author} {\bibfnamefont
  {Y.~S.}\ \bibnamefont {Hor}}, \bibinfo {author} {\bibfnamefont {R.~J.}\
  \bibnamefont {Cava}}, \ and\ \bibinfo {author} {\bibfnamefont {M.~Z.}\
  \bibnamefont {Hasan}},\ }\href {\doibase 10.1038/nature06843} {\bibfield
  {journal} {\bibinfo  {journal} {Nature}\ }\textbf {\bibinfo {volume} {452}},\
  \bibinfo {pages} {970} (\bibinfo {year} {2008})}\BibitemShut {NoStop}%
\bibitem [{\citenamefont {Zhang}\ \emph {et~al.}(2009)\citenamefont {Zhang},
  \citenamefont {Liu}, \citenamefont {Qi}, \citenamefont {Dai}, \citenamefont
  {Fang},\ and\ \citenamefont {Zhang}}]{Flds12}%
  \BibitemOpen
  \bibfield  {author} {\bibinfo {author} {\bibfnamefont {H.}~\bibnamefont
  {Zhang}}, \bibinfo {author} {\bibfnamefont {C.-X.}\ \bibnamefont {Liu}},
  \bibinfo {author} {\bibfnamefont {X.-L.}\ \bibnamefont {Qi}}, \bibinfo
  {author} {\bibfnamefont {X.}~\bibnamefont {Dai}}, \bibinfo {author}
  {\bibfnamefont {Z.}~\bibnamefont {Fang}}, \ and\ \bibinfo {author}
  {\bibfnamefont {S.-C.}\ \bibnamefont {Zhang}},\ }\href {\doibase
  10.1038/nphys1270} {\bibfield  {journal} {\bibinfo  {journal} {Nat. Phys.}\
  }\textbf {\bibinfo {volume} {5}},\ \bibinfo {pages} {438} (\bibinfo {year}
  {2009})}\BibitemShut {NoStop}%
\bibitem [{\citenamefont {Assaf}\ \emph {et~al.}(2016)\citenamefont {Assaf},
  \citenamefont {Phuphachong}, \citenamefont {Volobuev}, \citenamefont
  {Inhofer}, \citenamefont {Bauer}, \citenamefont {Springholz}, \citenamefont
  {de~Vaulchier},\ and\ \citenamefont {Guldner}}]{Flds13}%
  \BibitemOpen
  \bibfield  {author} {\bibinfo {author} {\bibfnamefont {B.~A.}\ \bibnamefont
  {Assaf}}, \bibinfo {author} {\bibfnamefont {T.}~\bibnamefont {Phuphachong}},
  \bibinfo {author} {\bibfnamefont {V.~V.}\ \bibnamefont {Volobuev}}, \bibinfo
  {author} {\bibfnamefont {A.}~\bibnamefont {Inhofer}}, \bibinfo {author}
  {\bibfnamefont {G.}~\bibnamefont {Bauer}}, \bibinfo {author} {\bibfnamefont
  {G.}~\bibnamefont {Springholz}}, \bibinfo {author} {\bibfnamefont {L.~A.}\
  \bibnamefont {de~Vaulchier}}, \ and\ \bibinfo {author} {\bibfnamefont
  {Y.}~\bibnamefont {Guldner}},\ }\href {\doibase 10.1038/srep20323} {\bibfield
   {journal} {\bibinfo  {journal} {Sci. Rep.}\ }\textbf {\bibinfo {volume}
  {6}},\ \bibinfo {pages} {20323} (\bibinfo {year} {2016})}\BibitemShut
  {NoStop}%
\bibitem [{\citenamefont {Liu}\ \emph {et~al.}(2014{\natexlab{a}})\citenamefont
  {Liu}, \citenamefont {Zhou}, \citenamefont {Zhang}, \citenamefont {Wang},
  \citenamefont {Weng}, \citenamefont {Prabhakaran}, \citenamefont {Mo},
  \citenamefont {Shen}, \citenamefont {Fang}, \citenamefont {Dai},
  \citenamefont {Hussain},\ and\ \citenamefont {Chen}}]{Flds14}%
  \BibitemOpen
  \bibfield  {author} {\bibinfo {author} {\bibfnamefont {Z.~K.}\ \bibnamefont
  {Liu}}, \bibinfo {author} {\bibfnamefont {B.}~\bibnamefont {Zhou}}, \bibinfo
  {author} {\bibfnamefont {Y.}~\bibnamefont {Zhang}}, \bibinfo {author}
  {\bibfnamefont {Z.~J.}\ \bibnamefont {Wang}}, \bibinfo {author}
  {\bibfnamefont {H.~M.}\ \bibnamefont {Weng}}, \bibinfo {author}
  {\bibfnamefont {D.}~\bibnamefont {Prabhakaran}}, \bibinfo {author}
  {\bibfnamefont {S.-K.}\ \bibnamefont {Mo}}, \bibinfo {author} {\bibfnamefont
  {Z.~X.}\ \bibnamefont {Shen}}, \bibinfo {author} {\bibfnamefont
  {Z.}~\bibnamefont {Fang}}, \bibinfo {author} {\bibfnamefont {X.}~\bibnamefont
  {Dai}}, \bibinfo {author} {\bibfnamefont {Z.}~\bibnamefont {Hussain}}, \ and\
  \bibinfo {author} {\bibfnamefont {Y.~L.}\ \bibnamefont {Chen}},\ }\href
  {\doibase 10.1126/science.1245085} {\bibfield  {journal} {\bibinfo  {journal}
  {Science}\ }\textbf {\bibinfo {volume} {343}},\ \bibinfo {pages} {864}
  (\bibinfo {year} {2014}{\natexlab{a}})}\BibitemShut {NoStop}%
\bibitem [{\citenamefont {Liu}\ \emph {et~al.}(2014{\natexlab{b}})\citenamefont
  {Liu}, \citenamefont {Jiang}, \citenamefont {Zhou}, \citenamefont {Wang},
  \citenamefont {Zhang}, \citenamefont {Weng}, \citenamefont {Prabhakaran},
  \citenamefont {Mo}, \citenamefont {Peng}, \citenamefont {Dudin},
  \citenamefont {Kim}, \citenamefont {Hoesch}, \citenamefont {Fang},
  \citenamefont {Dai}, \citenamefont {Shen}, \citenamefont {Feng},
  \citenamefont {Hussain},\ and\ \citenamefont {Chen}}]{Flds15}%
  \BibitemOpen
  \bibfield  {author} {\bibinfo {author} {\bibfnamefont {Z.~K.}\ \bibnamefont
  {Liu}}, \bibinfo {author} {\bibfnamefont {J.}~\bibnamefont {Jiang}}, \bibinfo
  {author} {\bibfnamefont {B.}~\bibnamefont {Zhou}}, \bibinfo {author}
  {\bibfnamefont {Z.~J.}\ \bibnamefont {Wang}}, \bibinfo {author}
  {\bibfnamefont {Y.}~\bibnamefont {Zhang}}, \bibinfo {author} {\bibfnamefont
  {H.~M.}\ \bibnamefont {Weng}}, \bibinfo {author} {\bibfnamefont
  {D.}~\bibnamefont {Prabhakaran}}, \bibinfo {author} {\bibfnamefont {S.-K.}\
  \bibnamefont {Mo}}, \bibinfo {author} {\bibfnamefont {H.}~\bibnamefont
  {Peng}}, \bibinfo {author} {\bibfnamefont {P.}~\bibnamefont {Dudin}},
  \bibinfo {author} {\bibfnamefont {T.}~\bibnamefont {Kim}}, \bibinfo {author}
  {\bibfnamefont {M.}~\bibnamefont {Hoesch}}, \bibinfo {author} {\bibfnamefont
  {Z.}~\bibnamefont {Fang}}, \bibinfo {author} {\bibfnamefont {X.}~\bibnamefont
  {Dai}}, \bibinfo {author} {\bibfnamefont {Z.~X.}\ \bibnamefont {Shen}},
  \bibinfo {author} {\bibfnamefont {D.~L.}\ \bibnamefont {Feng}}, \bibinfo
  {author} {\bibfnamefont {Z.}~\bibnamefont {Hussain}}, \ and\ \bibinfo
  {author} {\bibfnamefont {Y.~L.}\ \bibnamefont {Chen}},\ }\href {\doibase
  10.1038/nmat3990} {\bibfield  {journal} {\bibinfo  {journal} {Nat. Mater.}\
  }\textbf {\bibinfo {volume} {13}},\ \bibinfo {pages} {677} (\bibinfo {year}
  {2014}{\natexlab{b}})}\BibitemShut {NoStop}%
\bibitem [{\citenamefont {Neupane}\ \emph {et~al.}(2014)\citenamefont
  {Neupane}, \citenamefont {Xu}, \citenamefont {Sankar}, \citenamefont
  {Alidoust}, \citenamefont {Bian}, \citenamefont {Liu}, \citenamefont
  {Belopolski}, \citenamefont {Chang}, \citenamefont {Jeng}, \citenamefont
  {Lin}, \citenamefont {Bansil}, \citenamefont {Chou},\ and\ \citenamefont
  {Hasan}}]{Flds16}%
  \BibitemOpen
  \bibfield  {author} {\bibinfo {author} {\bibfnamefont {M.}~\bibnamefont
  {Neupane}}, \bibinfo {author} {\bibfnamefont {S.-Y.}\ \bibnamefont {Xu}},
  \bibinfo {author} {\bibfnamefont {R.}~\bibnamefont {Sankar}}, \bibinfo
  {author} {\bibfnamefont {N.}~\bibnamefont {Alidoust}}, \bibinfo {author}
  {\bibfnamefont {G.}~\bibnamefont {Bian}}, \bibinfo {author} {\bibfnamefont
  {C.}~\bibnamefont {Liu}}, \bibinfo {author} {\bibfnamefont {I.}~\bibnamefont
  {Belopolski}}, \bibinfo {author} {\bibfnamefont {T.-R.}\ \bibnamefont
  {Chang}}, \bibinfo {author} {\bibfnamefont {H.-T.}\ \bibnamefont {Jeng}},
  \bibinfo {author} {\bibfnamefont {H.}~\bibnamefont {Lin}}, \bibinfo {author}
  {\bibfnamefont {A.}~\bibnamefont {Bansil}}, \bibinfo {author} {\bibfnamefont
  {F.}~\bibnamefont {Chou}}, \ and\ \bibinfo {author} {\bibfnamefont {M.~Z.}\
  \bibnamefont {Hasan}},\ }\href {\doibase 10.1038/ncomms4786} {\bibfield
  {journal} {\bibinfo  {journal} {Nat. Commun.}\ }\textbf {\bibinfo {volume}
  {5}},\ \bibinfo {pages} {3786} (\bibinfo {year} {2014})}\BibitemShut
  {NoStop}%
\bibitem [{\citenamefont {Desrat}\ \emph {et~al.}(2018)\citenamefont {Desrat},
  \citenamefont {Krishtopenko}, \citenamefont {Piot}, \citenamefont {Orlita},
  \citenamefont {Consejo}, \citenamefont {Ruffenach}, \citenamefont {Knap},
  \citenamefont {Nateprov}, \citenamefont {Arushanov},\ and\ \citenamefont
  {Teppe}}]{Flds17}%
  \BibitemOpen
  \bibfield  {author} {\bibinfo {author} {\bibfnamefont {W.}~\bibnamefont
  {Desrat}}, \bibinfo {author} {\bibfnamefont {S.~S.}\ \bibnamefont
  {Krishtopenko}}, \bibinfo {author} {\bibfnamefont {B.~A.}\ \bibnamefont
  {Piot}}, \bibinfo {author} {\bibfnamefont {M.}~\bibnamefont {Orlita}},
  \bibinfo {author} {\bibfnamefont {C.}~\bibnamefont {Consejo}}, \bibinfo
  {author} {\bibfnamefont {S.}~\bibnamefont {Ruffenach}}, \bibinfo {author}
  {\bibfnamefont {W.}~\bibnamefont {Knap}}, \bibinfo {author} {\bibfnamefont
  {A.}~\bibnamefont {Nateprov}}, \bibinfo {author} {\bibfnamefont
  {E.}~\bibnamefont {Arushanov}}, \ and\ \bibinfo {author} {\bibfnamefont
  {F.}~\bibnamefont {Teppe}},\ }\href {\doibase 10.1103/PhysRevB.97.245203}
  {\bibfield  {journal} {\bibinfo  {journal} {Phys. Rev. B}\ }\textbf {\bibinfo
  {volume} {97}},\ \bibinfo {pages} {245203} (\bibinfo {year}
  {2018})}\BibitemShut {NoStop}%
\bibitem [{\citenamefont {Lv}\ \emph {et~al.}(2015)\citenamefont {Lv},
  \citenamefont {Xu}, \citenamefont {Weng}, \citenamefont {Ma}, \citenamefont
  {Richard}, \citenamefont {Huang}, \citenamefont {Zhao}, \citenamefont {Chen},
  \citenamefont {Matt}, \citenamefont {Bisti}, \citenamefont {Strocov},
  \citenamefont {Mesot}, \citenamefont {Fang}, \citenamefont {Dai},
  \citenamefont {Qian}, \citenamefont {Shi},\ and\ \citenamefont
  {Ding}}]{Flds18}%
  \BibitemOpen
  \bibfield  {author} {\bibinfo {author} {\bibfnamefont {B.~Q.}\ \bibnamefont
  {Lv}}, \bibinfo {author} {\bibfnamefont {N.}~\bibnamefont {Xu}}, \bibinfo
  {author} {\bibfnamefont {H.~M.}\ \bibnamefont {Weng}}, \bibinfo {author}
  {\bibfnamefont {J.~Z.}\ \bibnamefont {Ma}}, \bibinfo {author} {\bibfnamefont
  {P.}~\bibnamefont {Richard}}, \bibinfo {author} {\bibfnamefont {X.~C.}\
  \bibnamefont {Huang}}, \bibinfo {author} {\bibfnamefont {L.~X.}\ \bibnamefont
  {Zhao}}, \bibinfo {author} {\bibfnamefont {G.~F.}\ \bibnamefont {Chen}},
  \bibinfo {author} {\bibfnamefont {C.~E.}\ \bibnamefont {Matt}}, \bibinfo
  {author} {\bibfnamefont {F.}~\bibnamefont {Bisti}}, \bibinfo {author}
  {\bibfnamefont {V.~N.}\ \bibnamefont {Strocov}}, \bibinfo {author}
  {\bibfnamefont {J.}~\bibnamefont {Mesot}}, \bibinfo {author} {\bibfnamefont
  {Z.}~\bibnamefont {Fang}}, \bibinfo {author} {\bibfnamefont {X.}~\bibnamefont
  {Dai}}, \bibinfo {author} {\bibfnamefont {T.}~\bibnamefont {Qian}}, \bibinfo
  {author} {\bibfnamefont {M.}~\bibnamefont {Shi}}, \ and\ \bibinfo {author}
  {\bibfnamefont {H.}~\bibnamefont {Ding}},\ }\href {\doibase
  10.1038/nphys3426} {\bibfield  {journal} {\bibinfo  {journal} {Nat. Phys.}\
  }\textbf {\bibinfo {volume} {11}},\ \bibinfo {pages} {724} (\bibinfo {year}
  {2015})}\BibitemShut {NoStop}%
\bibitem [{\citenamefont {Yang}\ \emph {et~al.}(2015)\citenamefont {Yang},
  \citenamefont {Liu}, \citenamefont {Sun}, \citenamefont {Peng}, \citenamefont
  {Yang}, \citenamefont {Zhang}, \citenamefont {Zhou}, \citenamefont {Zhang},
  \citenamefont {Guo}, \citenamefont {Rahn}, \citenamefont {Prabhakaran},
  \citenamefont {Hussain}, \citenamefont {Mo}, \citenamefont {Felser},
  \citenamefont {Yan},\ and\ \citenamefont {Chen}}]{Flds19}%
  \BibitemOpen
  \bibfield  {author} {\bibinfo {author} {\bibfnamefont {L.~X.}\ \bibnamefont
  {Yang}}, \bibinfo {author} {\bibfnamefont {Z.~K.}\ \bibnamefont {Liu}},
  \bibinfo {author} {\bibfnamefont {Y.}~\bibnamefont {Sun}}, \bibinfo {author}
  {\bibfnamefont {H.}~\bibnamefont {Peng}}, \bibinfo {author} {\bibfnamefont
  {H.~F.}\ \bibnamefont {Yang}}, \bibinfo {author} {\bibfnamefont
  {T.}~\bibnamefont {Zhang}}, \bibinfo {author} {\bibfnamefont
  {B.}~\bibnamefont {Zhou}}, \bibinfo {author} {\bibfnamefont {Y.}~\bibnamefont
  {Zhang}}, \bibinfo {author} {\bibfnamefont {Y.~F.}\ \bibnamefont {Guo}},
  \bibinfo {author} {\bibfnamefont {M.}~\bibnamefont {Rahn}}, \bibinfo {author}
  {\bibfnamefont {D.}~\bibnamefont {Prabhakaran}}, \bibinfo {author}
  {\bibfnamefont {Z.}~\bibnamefont {Hussain}}, \bibinfo {author} {\bibfnamefont
  {S.-K.}\ \bibnamefont {Mo}}, \bibinfo {author} {\bibfnamefont
  {C.}~\bibnamefont {Felser}}, \bibinfo {author} {\bibfnamefont
  {B.}~\bibnamefont {Yan}}, \ and\ \bibinfo {author} {\bibfnamefont {Y.~L.}\
  \bibnamefont {Chen}},\ }\href {\doibase 10.1038/nphys3425} {\bibfield
  {journal} {\bibinfo  {journal} {Nat. Phys.}\ }\textbf {\bibinfo {volume}
  {11}},\ \bibinfo {pages} {728} (\bibinfo {year} {2015})}\BibitemShut
  {NoStop}%
\bibitem [{\citenamefont {Xu}\ \emph {et~al.}(2015)\citenamefont {Xu},
  \citenamefont {Alidoust}, \citenamefont {Belopolski}, \citenamefont {Yuan},
  \citenamefont {Bian}, \citenamefont {Chang}, \citenamefont {Zheng},
  \citenamefont {Strocov}, \citenamefont {Sanchez}, \citenamefont {Chang},
  \citenamefont {Zhang}, \citenamefont {Mou}, \citenamefont {Wu}, \citenamefont
  {Huang}, \citenamefont {Lee}, \citenamefont {Huang}, \citenamefont {Wang},
  \citenamefont {Bansil}, \citenamefont {Jeng}, \citenamefont {Neupert},
  \citenamefont {Kaminski}, \citenamefont {Lin}, \citenamefont {Jia},\ and\
  \citenamefont {Hasan}}]{Flds20}%
  \BibitemOpen
  \bibfield  {author} {\bibinfo {author} {\bibfnamefont {S.-Y.}\ \bibnamefont
  {Xu}}, \bibinfo {author} {\bibfnamefont {N.}~\bibnamefont {Alidoust}},
  \bibinfo {author} {\bibfnamefont {I.}~\bibnamefont {Belopolski}}, \bibinfo
  {author} {\bibfnamefont {Z.}~\bibnamefont {Yuan}}, \bibinfo {author}
  {\bibfnamefont {G.}~\bibnamefont {Bian}}, \bibinfo {author} {\bibfnamefont
  {T.-R.}\ \bibnamefont {Chang}}, \bibinfo {author} {\bibfnamefont
  {H.}~\bibnamefont {Zheng}}, \bibinfo {author} {\bibfnamefont {V.~N.}\
  \bibnamefont {Strocov}}, \bibinfo {author} {\bibfnamefont {D.~S.}\
  \bibnamefont {Sanchez}}, \bibinfo {author} {\bibfnamefont {G.}~\bibnamefont
  {Chang}}, \bibinfo {author} {\bibfnamefont {C.}~\bibnamefont {Zhang}},
  \bibinfo {author} {\bibfnamefont {D.}~\bibnamefont {Mou}}, \bibinfo {author}
  {\bibfnamefont {Y.}~\bibnamefont {Wu}}, \bibinfo {author} {\bibfnamefont
  {L.}~\bibnamefont {Huang}}, \bibinfo {author} {\bibfnamefont {C.-C.}\
  \bibnamefont {Lee}}, \bibinfo {author} {\bibfnamefont {S.-M.}\ \bibnamefont
  {Huang}}, \bibinfo {author} {\bibfnamefont {B.}~\bibnamefont {Wang}},
  \bibinfo {author} {\bibfnamefont {A.}~\bibnamefont {Bansil}}, \bibinfo
  {author} {\bibfnamefont {H.-T.}\ \bibnamefont {Jeng}}, \bibinfo {author}
  {\bibfnamefont {T.}~\bibnamefont {Neupert}}, \bibinfo {author} {\bibfnamefont
  {A.}~\bibnamefont {Kaminski}}, \bibinfo {author} {\bibfnamefont
  {H.}~\bibnamefont {Lin}}, \bibinfo {author} {\bibfnamefont {S.}~\bibnamefont
  {Jia}}, \ and\ \bibinfo {author} {\bibfnamefont {M.~Z.}\ \bibnamefont
  {Hasan}},\ }\href {\doibase 10.1038/nphys3437} {\bibfield  {journal}
  {\bibinfo  {journal} {Nat. Phys.}\ }\textbf {\bibinfo {volume} {11}},\
  \bibinfo {pages} {748} (\bibinfo {year} {2015})}\BibitemShut {NoStop}%
\bibitem [{\citenamefont {Aronov}\ and\ \citenamefont
  {Pikus}(1967{\natexlab{a}})}]{Flds3}%
  \BibitemOpen
  \bibfield  {author} {\bibinfo {author} {\bibfnamefont {A.~G.}\ \bibnamefont
  {Aronov}}\ and\ \bibinfo {author} {\bibfnamefont {G.~E.}\ \bibnamefont
  {Pikus}},\ }\href {http://jetp.ras.ru/cgi-bin/e/index/e/24/1/p188?a=list}
  {\bibfield  {journal} {\bibinfo  {journal} {Sov. Phys. JETP}\ }\textbf
  {\bibinfo {volume} {24}},\ \bibinfo {pages} {188} (\bibinfo {year}
  {1967}{\natexlab{a}})}\BibitemShut {NoStop}%
\bibitem [{\citenamefont {Aronov}\ and\ \citenamefont
  {Pikus}(1967{\natexlab{b}})}]{Flds4}%
  \BibitemOpen
  \bibfield  {author} {\bibinfo {author} {\bibfnamefont {A.~G.}\ \bibnamefont
  {Aronov}}\ and\ \bibinfo {author} {\bibfnamefont {G.~E.}\ \bibnamefont
  {Pikus}},\ }\href {http://jetp.ras.ru/cgi-bin/e/index/e/24/2/p339?a=list}
  {\bibfield  {journal} {\bibinfo  {journal} {Sov. Phys. JETP}\ }\textbf
  {\bibinfo {volume} {24}},\ \bibinfo {pages} {339} (\bibinfo {year}
  {1967}{\natexlab{b}})}\BibitemShut {NoStop}%
\bibitem [{\citenamefont {Zawadzki}\ and\ \citenamefont {Lax}(1966)}]{Flds5}%
  \BibitemOpen
  \bibfield  {author} {\bibinfo {author} {\bibfnamefont {W.}~\bibnamefont
  {Zawadzki}}\ and\ \bibinfo {author} {\bibfnamefont {B.}~\bibnamefont {Lax}},\
  }\href {\doibase 10.1103/PhysRevLett.16.1001} {\bibfield  {journal} {\bibinfo
   {journal} {Phys. Rev. Lett.}\ }\textbf {\bibinfo {volume} {16}},\ \bibinfo
  {pages} {1001} (\bibinfo {year} {1966})}\BibitemShut {NoStop}%
\bibitem [{\citenamefont {Weiler}\ \emph {et~al.}(1967)\citenamefont {Weiler},
  \citenamefont {Zawadzki},\ and\ \citenamefont {Lax}}]{Flds5b}%
  \BibitemOpen
  \bibfield  {author} {\bibinfo {author} {\bibfnamefont {M.~H.}\ \bibnamefont
  {Weiler}}, \bibinfo {author} {\bibfnamefont {W.}~\bibnamefont {Zawadzki}}, \
  and\ \bibinfo {author} {\bibfnamefont {B.}~\bibnamefont {Lax}},\ }\href
  {\doibase 10.1103/PhysRev.163.733} {\bibfield  {journal} {\bibinfo  {journal}
  {Phys. Rev.}\ }\textbf {\bibinfo {volume} {163}},\ \bibinfo {pages} {733}
  (\bibinfo {year} {1967})}\BibitemShut {NoStop}%
\bibitem [{\citenamefont {Lukose}\ \emph {et~al.}(2007)\citenamefont {Lukose},
  \citenamefont {Shankar},\ and\ \citenamefont {Baskaran}}]{Flds25}%
  \BibitemOpen
  \bibfield  {author} {\bibinfo {author} {\bibfnamefont {V.}~\bibnamefont
  {Lukose}}, \bibinfo {author} {\bibfnamefont {R.}~\bibnamefont {Shankar}}, \
  and\ \bibinfo {author} {\bibfnamefont {G.}~\bibnamefont {Baskaran}},\ }\href
  {\doibase 10.1103/PhysRevLett.98.116802} {\bibfield  {journal} {\bibinfo
  {journal} {Phys. Rev. Lett.}\ }\textbf {\bibinfo {volume} {98}},\ \bibinfo
  {pages} {116802} (\bibinfo {year} {2007})}\BibitemShut {NoStop}%
\bibitem [{\citenamefont {Burkov}\ \emph {et~al.}(2011)\citenamefont {Burkov},
  \citenamefont {Hook},\ and\ \citenamefont {Balents}}]{Flds21}%
  \BibitemOpen
  \bibfield  {author} {\bibinfo {author} {\bibfnamefont {A.~A.}\ \bibnamefont
  {Burkov}}, \bibinfo {author} {\bibfnamefont {M.~D.}\ \bibnamefont {Hook}}, \
  and\ \bibinfo {author} {\bibfnamefont {L.}~\bibnamefont {Balents}},\ }\href
  {\doibase 10.1103/PhysRevB.84.235126} {\bibfield  {journal} {\bibinfo
  {journal} {Phys. Rev. B}\ }\textbf {\bibinfo {volume} {84}},\ \bibinfo
  {pages} {235126} (\bibinfo {year} {2011})}\BibitemShut {NoStop}%
\bibitem [{\citenamefont {Soluyanov}\ \emph {et~al.}(2015)\citenamefont
  {Soluyanov}, \citenamefont {Gresch}, \citenamefont {Wang}, \citenamefont
  {Wu}, \citenamefont {Troyer}, \citenamefont {Dai},\ and\ \citenamefont
  {Bernevig}}]{Flds22}%
  \BibitemOpen
  \bibfield  {author} {\bibinfo {author} {\bibfnamefont {A.~A.}\ \bibnamefont
  {Soluyanov}}, \bibinfo {author} {\bibfnamefont {D.}~\bibnamefont {Gresch}},
  \bibinfo {author} {\bibfnamefont {Z.}~\bibnamefont {Wang}}, \bibinfo {author}
  {\bibfnamefont {Q.}~\bibnamefont {Wu}}, \bibinfo {author} {\bibfnamefont
  {M.}~\bibnamefont {Troyer}}, \bibinfo {author} {\bibfnamefont
  {X.}~\bibnamefont {Dai}}, \ and\ \bibinfo {author} {\bibfnamefont {B.~A.}\
  \bibnamefont {Bernevig}},\ }\href {\doibase 10.1038/nature15768} {\bibfield
  {journal} {\bibinfo  {journal} {Nature}\ }\textbf {\bibinfo {volume} {527}},\
  \bibinfo {pages} {495} (\bibinfo {year} {2015})}\BibitemShut {NoStop}%
\bibitem [{\citenamefont {Wieder}\ \emph {et~al.}(2016)\citenamefont {Wieder},
  \citenamefont {Kim}, \citenamefont {Rappe},\ and\ \citenamefont
  {Kane}}]{Flds23}%
  \BibitemOpen
  \bibfield  {author} {\bibinfo {author} {\bibfnamefont {B.~J.}\ \bibnamefont
  {Wieder}}, \bibinfo {author} {\bibfnamefont {Y.}~\bibnamefont {Kim}},
  \bibinfo {author} {\bibfnamefont {A.~M.}\ \bibnamefont {Rappe}}, \ and\
  \bibinfo {author} {\bibfnamefont {C.~L.}\ \bibnamefont {Kane}},\ }\href
  {\doibase 10.1103/PhysRevLett.116.186402} {\bibfield  {journal} {\bibinfo
  {journal} {Phys. Rev. Lett.}\ }\textbf {\bibinfo {volume} {116}},\ \bibinfo
  {pages} {186402} (\bibinfo {year} {2016})}\BibitemShut {NoStop}%
\bibitem [{\citenamefont {Bradlyn}\ \emph {et~al.}(2016)\citenamefont
  {Bradlyn}, \citenamefont {Cano}, \citenamefont {Wang}, \citenamefont
  {Vergniory}, \citenamefont {Felser}, \citenamefont {Cava},\ and\
  \citenamefont {Bernevig}}]{Flds24}%
  \BibitemOpen
  \bibfield  {author} {\bibinfo {author} {\bibfnamefont {B.}~\bibnamefont
  {Bradlyn}}, \bibinfo {author} {\bibfnamefont {J.}~\bibnamefont {Cano}},
  \bibinfo {author} {\bibfnamefont {Z.}~\bibnamefont {Wang}}, \bibinfo {author}
  {\bibfnamefont {M.~G.}\ \bibnamefont {Vergniory}}, \bibinfo {author}
  {\bibfnamefont {C.}~\bibnamefont {Felser}}, \bibinfo {author} {\bibfnamefont
  {R.~J.}\ \bibnamefont {Cava}}, \ and\ \bibinfo {author} {\bibfnamefont
  {B.~A.}\ \bibnamefont {Bernevig}},\ }\href {\doibase 10.1126/science.aaf5037}
  {\bibfield  {journal} {\bibinfo  {journal} {Science}\ }\textbf {\bibinfo
  {volume} {353}},\ \bibinfo {pages} {aaf5037} (\bibinfo {year}
  {2016})}\BibitemShut {NoStop}%
\bibitem [{\citenamefont {Orlita}\ \emph {et~al.}(2014)\citenamefont {Orlita},
  \citenamefont {Basko}, \citenamefont {Zholudev}, \citenamefont {Teppe},
  \citenamefont {Knap}, \citenamefont {Gavrilenko}, \citenamefont {Mikhailov},
  \citenamefont {Dvoretskii}, \citenamefont {Neugebauer}, \citenamefont
  {Faugeras}, \citenamefont {Barra}, \citenamefont {Martinez},\ and\
  \citenamefont {Potemski}}]{s1}%
  \BibitemOpen
  \bibfield  {author} {\bibinfo {author} {\bibfnamefont {M.}~\bibnamefont
  {Orlita}}, \bibinfo {author} {\bibfnamefont {D.~M.}\ \bibnamefont {Basko}},
  \bibinfo {author} {\bibfnamefont {M.~S.}\ \bibnamefont {Zholudev}}, \bibinfo
  {author} {\bibfnamefont {F.}~\bibnamefont {Teppe}}, \bibinfo {author}
  {\bibfnamefont {W.}~\bibnamefont {Knap}}, \bibinfo {author} {\bibfnamefont
  {V.~I.}\ \bibnamefont {Gavrilenko}}, \bibinfo {author} {\bibfnamefont
  {N.~N.}\ \bibnamefont {Mikhailov}}, \bibinfo {author} {\bibfnamefont {S.~A.}\
  \bibnamefont {Dvoretskii}}, \bibinfo {author} {\bibfnamefont
  {P.}~\bibnamefont {Neugebauer}}, \bibinfo {author} {\bibfnamefont
  {C.}~\bibnamefont {Faugeras}}, \bibinfo {author} {\bibfnamefont {A.-L.}\
  \bibnamefont {Barra}}, \bibinfo {author} {\bibfnamefont {G.}~\bibnamefont
  {Martinez}}, \ and\ \bibinfo {author} {\bibfnamefont {M.}~\bibnamefont
  {Potemski}},\ }\href {\doibase 10.1038/nphys2857} {\bibfield  {journal}
  {\bibinfo  {journal} {Nature Phys.}\ }\textbf {\bibinfo {volume} {10}},\
  \bibinfo {pages} {233} (\bibinfo {year} {2014})}\BibitemShut {NoStop}%
\bibitem [{\citenamefont {Teppe}\ \emph {et~al.}(2016)\citenamefont {Teppe},
  \citenamefont {Marcinkiewicz}, \citenamefont {Krishtopenko}, \citenamefont
  {Ruffenach}, \citenamefont {Consejo}, \citenamefont {Kadykov}, \citenamefont
  {Desrat}, \citenamefont {But}, \citenamefont {Knap}, \citenamefont {Ludwig},
  \citenamefont {Moon}, \citenamefont {Smirnov}, \citenamefont {Orlita},
  \citenamefont {Jiang}, \citenamefont {Morozov}, \citenamefont {Gavrilenko},
  \citenamefont {Mikhailov},\ and\ \citenamefont {Dvoretskii}}]{s2}%
  \BibitemOpen
  \bibfield  {author} {\bibinfo {author} {\bibfnamefont {F.}~\bibnamefont
  {Teppe}}, \bibinfo {author} {\bibfnamefont {M.}~\bibnamefont
  {Marcinkiewicz}}, \bibinfo {author} {\bibfnamefont {S.~S.}\ \bibnamefont
  {Krishtopenko}}, \bibinfo {author} {\bibfnamefont {S.}~\bibnamefont
  {Ruffenach}}, \bibinfo {author} {\bibfnamefont {C.}~\bibnamefont {Consejo}},
  \bibinfo {author} {\bibfnamefont {A.~M.}\ \bibnamefont {Kadykov}}, \bibinfo
  {author} {\bibfnamefont {W.}~\bibnamefont {Desrat}}, \bibinfo {author}
  {\bibfnamefont {D.}~\bibnamefont {But}}, \bibinfo {author} {\bibfnamefont
  {W.}~\bibnamefont {Knap}}, \bibinfo {author} {\bibfnamefont {J.}~\bibnamefont
  {Ludwig}}, \bibinfo {author} {\bibfnamefont {S.}~\bibnamefont {Moon}},
  \bibinfo {author} {\bibfnamefont {D.}~\bibnamefont {Smirnov}}, \bibinfo
  {author} {\bibfnamefont {M.}~\bibnamefont {Orlita}}, \bibinfo {author}
  {\bibfnamefont {Z.}~\bibnamefont {Jiang}}, \bibinfo {author} {\bibfnamefont
  {S.~V.}\ \bibnamefont {Morozov}}, \bibinfo {author} {\bibfnamefont
  {V.}~\bibnamefont {Gavrilenko}}, \bibinfo {author} {\bibfnamefont {N.~N.}\
  \bibnamefont {Mikhailov}}, \ and\ \bibinfo {author} {\bibfnamefont {S.~A.}\
  \bibnamefont {Dvoretskii}},\ }\href {\doibase 10.1038/ncomms12576} {\bibfield
   {journal} {\bibinfo  {journal} {Nat. Commun.}\ }\textbf {\bibinfo {volume}
  {7}},\ \bibinfo {pages} {12576} (\bibinfo {year} {2016})}\BibitemShut
  {NoStop}%
\bibitem [{\citenamefont {Malcolm}\ and\ \citenamefont {Nicol}(2015)}]{Flds26}%
  \BibitemOpen
  \bibfield  {author} {\bibinfo {author} {\bibfnamefont {J.~D.}\ \bibnamefont
  {Malcolm}}\ and\ \bibinfo {author} {\bibfnamefont {E.~J.}\ \bibnamefont
  {Nicol}},\ }\href {\doibase 10.1103/PhysRevB.92.035118} {\bibfield  {journal}
  {\bibinfo  {journal} {Phys. Rev. B}\ }\textbf {\bibinfo {volume} {92}},\
  \bibinfo {pages} {035118} (\bibinfo {year} {2015})}\BibitemShut {NoStop}%
\bibitem [{\citenamefont {Raoux}\ \emph {et~al.}(2014)\citenamefont {Raoux},
  \citenamefont {Morigi}, \citenamefont {Fuchs}, \citenamefont {Pi\'echon},\
  and\ \citenamefont {Montambaux}}]{Flds27}%
  \BibitemOpen
  \bibfield  {author} {\bibinfo {author} {\bibfnamefont {A.}~\bibnamefont
  {Raoux}}, \bibinfo {author} {\bibfnamefont {M.}~\bibnamefont {Morigi}},
  \bibinfo {author} {\bibfnamefont {J.-N.}\ \bibnamefont {Fuchs}}, \bibinfo
  {author} {\bibfnamefont {F.}~\bibnamefont {Pi\'echon}}, \ and\ \bibinfo
  {author} {\bibfnamefont {G.}~\bibnamefont {Montambaux}},\ }\href {\doibase
  10.1103/PhysRevLett.112.026402} {\bibfield  {journal} {\bibinfo  {journal}
  {Phys. Rev. Lett.}\ }\textbf {\bibinfo {volume} {112}},\ \bibinfo {pages}
  {026402} (\bibinfo {year} {2014})}\BibitemShut {NoStop}%
\bibitem [{\citenamefont {Krishtopenko}\ \emph {et~al.}(2020)\citenamefont
  {Krishtopenko}, \citenamefont {Antezza},\ and\ \citenamefont {Teppe}}]{q6}%
  \BibitemOpen
  \bibfield  {author} {\bibinfo {author} {\bibfnamefont {S.~S.}\ \bibnamefont
  {Krishtopenko}}, \bibinfo {author} {\bibfnamefont {M.}~\bibnamefont
  {Antezza}}, \ and\ \bibinfo {author} {\bibfnamefont {F.}~\bibnamefont
  {Teppe}},\ }\href {\doibase 10.1088/1361-648x/ab6741} {\bibfield  {journal}
  {\bibinfo  {journal} {J. Phys.: Condens. Matter}\ }\textbf {\bibinfo {volume}
  {32}},\ \bibinfo {pages} {165501} (\bibinfo {year} {2020})}\BibitemShut
  {NoStop}%
\bibitem [{\citenamefont {Zawadzki}\ \emph {et~al.}(1985)\citenamefont
  {Zawadzki}, \citenamefont {Klahn},\ and\ \citenamefont {Merkt}}]{Flds29}%
  \BibitemOpen
  \bibfield  {author} {\bibinfo {author} {\bibfnamefont {W.}~\bibnamefont
  {Zawadzki}}, \bibinfo {author} {\bibfnamefont {S.}~\bibnamefont {Klahn}}, \
  and\ \bibinfo {author} {\bibfnamefont {U.}~\bibnamefont {Merkt}},\ }\href
  {\doibase 10.1103/PhysRevLett.55.983} {\bibfield  {journal} {\bibinfo
  {journal} {Phys. Rev. Lett.}\ }\textbf {\bibinfo {volume} {55}},\ \bibinfo
  {pages} {983} (\bibinfo {year} {1985})}\BibitemShut {NoStop}%
\bibitem [{\citenamefont {Zawadzki}\ \emph {et~al.}(1986)\citenamefont
  {Zawadzki}, \citenamefont {Klahn},\ and\ \citenamefont {Merkt}}]{Flds30}%
  \BibitemOpen
  \bibfield  {author} {\bibinfo {author} {\bibfnamefont {W.}~\bibnamefont
  {Zawadzki}}, \bibinfo {author} {\bibfnamefont {S.}~\bibnamefont {Klahn}}, \
  and\ \bibinfo {author} {\bibfnamefont {U.}~\bibnamefont {Merkt}},\ }\href
  {\doibase 10.1103/PhysRevB.33.6916} {\bibfield  {journal} {\bibinfo
  {journal} {Phys. Rev. B}\ }\textbf {\bibinfo {volume} {33}},\ \bibinfo
  {pages} {6916} (\bibinfo {year} {1986})}\BibitemShut {NoStop}%
\bibitem [{\citenamefont {Lifshitz}\ and\ \citenamefont
  {Kaganov}(1960)}]{Flds31}%
  \BibitemOpen
  \bibfield  {author} {\bibinfo {author} {\bibfnamefont {I.~M.}\ \bibnamefont
  {Lifshitz}}\ and\ \bibinfo {author} {\bibfnamefont {M.~I.}\ \bibnamefont
  {Kaganov}},\ }\href {\doibase 10.1070/PU1960v002n06ABEH003183} {\bibfield
  {journal} {\bibinfo  {journal} {Sov. Phys. Usp.}\ }\textbf {\bibinfo {volume}
  {2}},\ \bibinfo {pages} {831} (\bibinfo {year} {1960})}\BibitemShut {NoStop}%
\bibitem [{\citenamefont {R\"{o}ssler}(1984)}]{q1}%
  \BibitemOpen
  \bibfield  {author} {\bibinfo {author} {\bibfnamefont {U.}~\bibnamefont
  {R\"{o}ssler}},\ }\href {\doibase
  https://doi.org/10.1016/0038-1098(84)90299-0} {\bibfield  {journal} {\bibinfo
   {journal} {Solid State Commun.}\ }\textbf {\bibinfo {volume} {49}},\
  \bibinfo {pages} {943} (\bibinfo {year} {1984})}\BibitemShut {NoStop}%
\bibitem [{\citenamefont {Pfeffer}\ and\ \citenamefont {Zawadzki}(1990)}]{q2a}%
  \BibitemOpen
  \bibfield  {author} {\bibinfo {author} {\bibfnamefont {P.}~\bibnamefont
  {Pfeffer}}\ and\ \bibinfo {author} {\bibfnamefont {W.}~\bibnamefont
  {Zawadzki}},\ }\href {\doibase 10.1103/PhysRevB.41.1561} {\bibfield
  {journal} {\bibinfo  {journal} {Phys. Rev. B}\ }\textbf {\bibinfo {volume}
  {41}},\ \bibinfo {pages} {1561} (\bibinfo {year} {1990})}\BibitemShut
  {NoStop}%
\bibitem [{\citenamefont {Mayer}\ and\ \citenamefont {R\"{o}ssler}(1991)}]{q2}%
  \BibitemOpen
  \bibfield  {author} {\bibinfo {author} {\bibfnamefont {H.}~\bibnamefont
  {Mayer}}\ and\ \bibinfo {author} {\bibfnamefont {U.}~\bibnamefont
  {R\"{o}ssler}},\ }\href {\doibase 10.1103/PhysRevB.44.9048} {\bibfield
  {journal} {\bibinfo  {journal} {Phys. Rev. B}\ }\textbf {\bibinfo {volume}
  {44}},\ \bibinfo {pages} {9048} (\bibinfo {year} {1991})}\BibitemShut
  {NoStop}%
\bibitem [{\citenamefont {Krishtopenko}\ \emph
  {et~al.}(2016{\natexlab{b}})\citenamefont {Krishtopenko}, \citenamefont
  {Yahniuk}, \citenamefont {But}, \citenamefont {Gavrilenko}, \citenamefont
  {Knap},\ and\ \citenamefont {Teppe}}]{q5}%
  \BibitemOpen
  \bibfield  {author} {\bibinfo {author} {\bibfnamefont {S.~S.}\ \bibnamefont
  {Krishtopenko}}, \bibinfo {author} {\bibfnamefont {I.}~\bibnamefont
  {Yahniuk}}, \bibinfo {author} {\bibfnamefont {D.~B.}\ \bibnamefont {But}},
  \bibinfo {author} {\bibfnamefont {V.~I.}\ \bibnamefont {Gavrilenko}},
  \bibinfo {author} {\bibfnamefont {W.}~\bibnamefont {Knap}}, \ and\ \bibinfo
  {author} {\bibfnamefont {F.}~\bibnamefont {Teppe}},\ }\href {\doibase
  10.1103/PhysRevB.94.245402} {\bibfield  {journal} {\bibinfo  {journal} {Phys.
  Rev. B}\ }\textbf {\bibinfo {volume} {94}},\ \bibinfo {pages} {245402}
  (\bibinfo {year} {2016}{\natexlab{b}})}\BibitemShut {NoStop}%
\bibitem [{\citenamefont {{Guti\'{e}rrez Jim\'{e}nez}}\ and\ \citenamefont
  {Torba}(2020)}]{s3}%
  \BibitemOpen
  \bibfield  {author} {\bibinfo {author} {\bibfnamefont {N.}~\bibnamefont
  {{Guti\'{e}rrez Jim\'{e}nez}}}\ and\ \bibinfo {author} {\bibfnamefont
  {S.~M.}\ \bibnamefont {Torba}},\ }\href {\doibase
  https://doi.org/10.1016/j.amc.2019.124911} {\bibfield  {journal} {\bibinfo
  {journal} {Appl. Math. Comput.}\ }\textbf {\bibinfo {volume} {370}},\
  \bibinfo {pages} {124911} (\bibinfo {year} {2020})}\BibitemShut {NoStop}%
\bibitem [{\citenamefont {Gradshteyn}\ and\ \citenamefont
  {Ryzhik}(1980)}]{Flds28}%
  \BibitemOpen
  \bibfield  {author} {\bibinfo {author} {\bibfnamefont {I.}~\bibnamefont
  {Gradshteyn}}\ and\ \bibinfo {author} {\bibfnamefont {I.}~\bibnamefont
  {Ryzhik}},\ }\href {https://doi.org/10.1016/B978-0-12-294760-5.50019-2}
  {\bibfield  {journal} {\bibinfo  {journal} {\emph{Table of Integrals, Series,
  and Products}}\ } (\bibinfo {year} {Elsevier--Academic Press
  1980})}\BibitemShut {NoStop}%
\bibitem [{\citenamefont {Akrap}\ \emph {et~al.}(2016)\citenamefont {Akrap},
  \citenamefont {Hakl}, \citenamefont {Tchoumakov}, \citenamefont {Crassee},
  \citenamefont {Kuba}, \citenamefont {Goerbig}, \citenamefont {Homes},
  \citenamefont {Caha}, \citenamefont {Nov\'ak}, \citenamefont {Teppe},
  \citenamefont {Desrat}, \citenamefont {Koohpayeh}, \citenamefont {Wu},
  \citenamefont {Armitage}, \citenamefont {Nateprov}, \citenamefont
  {Arushanov}, \citenamefont {Gibson}, \citenamefont {Cava}, \citenamefont
  {van~der Marel}, \citenamefont {Piot}, \citenamefont {Faugeras},
  \citenamefont {Martinez}, \citenamefont {Potemski},\ and\ \citenamefont
  {Orlita}}]{Flds32}%
  \BibitemOpen
  \bibfield  {author} {\bibinfo {author} {\bibfnamefont {A.}~\bibnamefont
  {Akrap}}, \bibinfo {author} {\bibfnamefont {M.}~\bibnamefont {Hakl}},
  \bibinfo {author} {\bibfnamefont {S.}~\bibnamefont {Tchoumakov}}, \bibinfo
  {author} {\bibfnamefont {I.}~\bibnamefont {Crassee}}, \bibinfo {author}
  {\bibfnamefont {J.}~\bibnamefont {Kuba}}, \bibinfo {author} {\bibfnamefont
  {M.~O.}\ \bibnamefont {Goerbig}}, \bibinfo {author} {\bibfnamefont {C.~C.}\
  \bibnamefont {Homes}}, \bibinfo {author} {\bibfnamefont {O.}~\bibnamefont
  {Caha}}, \bibinfo {author} {\bibfnamefont {J.}~\bibnamefont {Nov\'ak}},
  \bibinfo {author} {\bibfnamefont {F.}~\bibnamefont {Teppe}}, \bibinfo
  {author} {\bibfnamefont {W.}~\bibnamefont {Desrat}}, \bibinfo {author}
  {\bibfnamefont {S.}~\bibnamefont {Koohpayeh}}, \bibinfo {author}
  {\bibfnamefont {L.}~\bibnamefont {Wu}}, \bibinfo {author} {\bibfnamefont
  {N.~P.}\ \bibnamefont {Armitage}}, \bibinfo {author} {\bibfnamefont
  {A.}~\bibnamefont {Nateprov}}, \bibinfo {author} {\bibfnamefont
  {E.}~\bibnamefont {Arushanov}}, \bibinfo {author} {\bibfnamefont {Q.~D.}\
  \bibnamefont {Gibson}}, \bibinfo {author} {\bibfnamefont {R.~J.}\
  \bibnamefont {Cava}}, \bibinfo {author} {\bibfnamefont {D.}~\bibnamefont
  {van~der Marel}}, \bibinfo {author} {\bibfnamefont {B.~A.}\ \bibnamefont
  {Piot}}, \bibinfo {author} {\bibfnamefont {C.}~\bibnamefont {Faugeras}},
  \bibinfo {author} {\bibfnamefont {G.}~\bibnamefont {Martinez}}, \bibinfo
  {author} {\bibfnamefont {M.}~\bibnamefont {Potemski}}, \ and\ \bibinfo
  {author} {\bibfnamefont {M.}~\bibnamefont {Orlita}},\ }\href {\doibase
  10.1103/PhysRevLett.117.136401} {\bibfield  {journal} {\bibinfo  {journal}
  {Phys. Rev. Lett.}\ }\textbf {\bibinfo {volume} {117}},\ \bibinfo {pages}
  {136401} (\bibinfo {year} {2016})}\BibitemShut {NoStop}%
\bibitem [{\citenamefont {Winkler}(2003)}]{q11}%
  \BibitemOpen
  \bibfield  {author} {\bibinfo {author} {\bibfnamefont {R.}~\bibnamefont
  {Winkler}},\ }\href {https://doi.org/10.1007/b13586} {\bibfield  {journal}
  {\bibinfo  {journal} {\emph{Spin-Orbit Coupling Effects in Two-Dimensional
  Electron and Hole Systems}}\ } (\bibinfo {year} {Springer-Verlag Berlin
  Heidelberg 2003})}\BibitemShut {NoStop}%
\bibitem [{q10(2004)}]{q10}%
  \BibitemOpen
  \href {https://doi.org/10.1007/978-3-642-18865-7} {\bibfield  {journal}
  {\bibinfo  {journal} {\emph{Semiconductors:~Data~Handbook} edited by O.
  Madelung}\ } (\bibinfo {year} {Springer-Verlag, Berlin, 2004})}\BibitemShut
  {NoStop}%
\bibitem [{\citenamefont {Svane}\ \emph {et~al.}(2011)\citenamefont {Svane},
  \citenamefont {Christensen}, \citenamefont {Cardona}, \citenamefont
  {Chantis}, \citenamefont {van Schilfgaarde},\ and\ \citenamefont
  {Kotani}}]{q12}%
  \BibitemOpen
  \bibfield  {author} {\bibinfo {author} {\bibfnamefont {A.}~\bibnamefont
  {Svane}}, \bibinfo {author} {\bibfnamefont {N.~E.}\ \bibnamefont
  {Christensen}}, \bibinfo {author} {\bibfnamefont {M.}~\bibnamefont
  {Cardona}}, \bibinfo {author} {\bibfnamefont {A.~N.}\ \bibnamefont
  {Chantis}}, \bibinfo {author} {\bibfnamefont {M.}~\bibnamefont {van
  Schilfgaarde}}, \ and\ \bibinfo {author} {\bibfnamefont {T.}~\bibnamefont
  {Kotani}},\ }\href {\doibase 10.1103/PhysRevB.84.205205} {\bibfield
  {journal} {\bibinfo  {journal} {Phys. Rev. B}\ }\textbf {\bibinfo {volume}
  {84}},\ \bibinfo {pages} {205205} (\bibinfo {year} {2011})}\BibitemShut
  {NoStop}%
\bibitem [{\citenamefont {Trebin}\ \emph {et~al.}(1979)\citenamefont {Trebin},
  \citenamefont {R\"{o}ssler},\ and\ \citenamefont {Ranvaud}}]{q3}%
  \BibitemOpen
  \bibfield  {author} {\bibinfo {author} {\bibfnamefont {H.~R.}\ \bibnamefont
  {Trebin}}, \bibinfo {author} {\bibfnamefont {U.}~\bibnamefont {R\"{o}ssler}},
  \ and\ \bibinfo {author} {\bibfnamefont {R.}~\bibnamefont {Ranvaud}},\ }\href
  {\doibase 10.1103/PhysRevB.20.686} {\bibfield  {journal} {\bibinfo  {journal}
  {Phys. Rev. B}\ }\textbf {\bibinfo {volume} {20}},\ \bibinfo {pages} {686}
  (\bibinfo {year} {1979})}\BibitemShut {NoStop}%
\bibitem [{\citenamefont {Kane}(1957)}]{q4}%
  \BibitemOpen
  \bibfield  {author} {\bibinfo {author} {\bibfnamefont {E.~O.}\ \bibnamefont
  {Kane}},\ }\href {\doibase https://doi.org/10.1016/0022-3697(57)90013-6}
  {\bibfield  {journal} {\bibinfo  {journal} {J. Phys. Chem. Solids}\ }\textbf
  {\bibinfo {volume} {1}},\ \bibinfo {pages} {249} (\bibinfo {year}
  {1957})}\BibitemShut {NoStop}%
\bibitem [{\citenamefont {Luttinger}\ and\ \citenamefont {Kohn}(1955)}]{q7}%
  \BibitemOpen
  \bibfield  {author} {\bibinfo {author} {\bibfnamefont {J.~M.}\ \bibnamefont
  {Luttinger}}\ and\ \bibinfo {author} {\bibfnamefont {W.}~\bibnamefont
  {Kohn}},\ }\href {\doibase 10.1103/PhysRev.97.869} {\bibfield  {journal}
  {\bibinfo  {journal} {Phys. Rev.}\ }\textbf {\bibinfo {volume} {97}},\
  \bibinfo {pages} {869} (\bibinfo {year} {1955})}\BibitemShut {NoStop}%
\end{thebibliography}

\end{document}